\documentclass[manuscript]{aastex}

\usepackage[varg]{txfonts}
\usepackage{enumerate}
\usepackage{enumitem}
\usepackage{float}
\usepackage{array}
\usepackage[english]{babel}
\usepackage{amsmath}

\usepackage{natbib}
\defcitealias{wahlberg14}{WJJ}
\defcitealias{bukhari15}{Paper I}

\newcommand{\unit}[1]{\,\mathrm{#1}}

\makeatletter
\newcommand{\thickhline}{%
	\noalign {\ifnum 0=`}\fi \hrule height 1.2pt
	\futurelet \reserved@a \@xhline
}
\newcolumntype{"}{@{\hskip\tabcolsep\vrule width 1.2pt\hskip\tabcolsep}}
\makeatother

\begin{document}


\title{The role of pebble fragmentation in planetesimal formation \\ I. Experimental study}


\author{M. Bukhari Syed}
\affil{Institut f\"ur Geophysik und extraterrestrische Physik, Technische Universit\"at zu Braunschweig, Mendelssohnstr. 3, 38106 Braunschweig, Germany}

\author{J. Blum}
\affil{Institut f\"ur Geophysik und extraterrestrische Physik, Technische Universit\"at zu Braunschweig, Mendelssohnstr. 3, 38106 Braunschweig, Germany}

\author{K. Wahlberg Jansson}
\affil{Lund Observatory, Department of Astronomy and Theoretical Physics, Lund University, Box 43, SE-221 00 Lund, Sweden}

\and

\author{A. Johansen}
\affil{Lund Observatory, Department of Astronomy and Theoretical Physics, Lund University, Box 43, SE-221 00 Lund, Sweden}

%\author{}
%\altaffilmark{4,5}
%\affil{Department of Astronomy and Theoretical Physics, Lund University}
%\email{aastex-help@aas.org}

%\and

%\author{R. J. Hanisch\altaffilmark{5}}
%\affil{Space Telescope Science Institute, Baltimore, MD 21218}


%\altaffiltext{1}{Visiting Astronomer, Cerro Tololo Inter-American Observatory. CTIO is operated by AURA, Inc.\ under contract to the National ScienceFoundation.}
%\altaffiltext{2}{Society of Fellows, Harvard University.}
%\altaffiltext{3}{present address: Center for Astrophysics,
%    60 Garden Street, Cambridge, MA 02138}
%\altaffiltext{4}{Visiting Programmer, Space Telescope Science Institute}
%\altaffiltext{5}{Patron, Alonso's Bar and Grill}


\begin{abstract}

\noindent Previous work on protoplanetary dust growth shows halt at centimeter sizes owing to the occurrence of bouncing at velocities of $\stackrel{>}{\sim} 0.1 \unit{m~s^{-1}}$ and fragmentation at velocities $\stackrel{>}{\sim} 1 \unit{m~s^{-1}}$. To overcome these barriers, spatial concentration of cm-sized dust pebbles and subsequent gravitational collapse have been proposed. However, numerical investigations have shown that dust aggregates may undergo fragmentation during the gravitational collapse phase. This fragmentation in turn changes the size distribution of the solids and thus must be taken into account in order to understand the properties of the planetesimals that form. To explore the fate of dust pebbles undergoing fragmenting collisions, we conducted laboratory experiments on dust-aggregate collisions with a focus on establishing a collision model for this stage of planetesimal formation. In our experiments, we analysed collisions of dust aggregates with masses between 1.4 g and 180 g, mass ratios between target and projectile from 125 to 1 at a fixed porosity of 65\%, within the velocity range of 1.5\textendash8.7 $\unit{m~s^{-1}}$, at low atmospheric pressure of $\sim 10^{-3}$ mbar and in free-fall conditions. We derived the mass of the largest fragment, the fragment size/mass distribution, and the efficiency of mass transfer as a function of collision velocity and projectile/target aggregate size. Moreover, we give recipes for an easy-to-use fragmentation and mass-transfer model for further use in modeling work. In a companion paper, we utilize the experimental findings and the derived dust-aggregate collision model to investigate the fate of dust pebbles during gravitational collapse.

\end{abstract}





\section{\label{sect:intro}Introduction}
Over the past decade, a significant amount of work on protoplanetary dust growth has been contributed by modeller and experimenters, which has significantly advanced our understanding about the formation of planetesimals. In the field of planetesimal formation, broad consensus has been reached on the pre-gravitational dust-growth regime in which micrometer-sized dust grains grow to at least centimetre sizes by sticking collisions in protoplanetary discs. Based upon the first complete laboratory-based dust-aggregate collision model by \citet{GuettlerEtal:2010}, \citet{ZsomEtal:2010a} showed that dust aggregates experience a bouncing barrier when they reach millimetre sizes, which limits growth and leads to relatively compact pebble-sized dust aggregates with volume filling factors of $\phi \sim 0.4$ (i.e., 60\% porosity). 

The further growth from pebbles to planetesimals faces severe obstacles by the absence of direct hit-and-stick processes \citep{GuettlerEtal:2010}, the onset of fragmentation in collisions between dust aggregates of similar size around $\sim 1 \unit{m~s^{-1}}$ \citep{GuettlerEtal:2010} and the strong influence of radial drift, which leads to the rapid depletion of boulders around 1 m in size at 1 AU \citep{Weidenschilling:1977a}. This halt of growth at pebble sizes is in agreement with observations, which show the presence of mm-cm-sized dust particles in protoplanetary disks (see \citet{Testi:2014} for a review). However, \citet{OkuzumiEtal:2012} have shown that under very favourable conditions (sub-micrometer-sized water-ice particles), direct coagulation into planetesimals is feasible. As we are interested in a more generic formation scenario that is less restricted in terms of grain size, particle material and location in the protoplanetary disk, we hereafter do assume that the growth pathway demonstrated by \citet{OkuzumiEtal:2012} is not feasible for micron-sized or warm dust particles.

Two competing models of planetesimal formation in the presence of the above-mentioned obstacles have been developed in the past years. Based upon an extensive body of laboratory work on mass transfer in high-velocity collisions between dust aggregates of dissimilar masses \citep{WurmEtal:2005a,TeiserWurm:2009a,GuettlerEtal:2010,Kotheetal:2010,Teiseretal:2011a,DeckersTeiser:2014}, \citet{WindmarkEtal:2012a}, \citet{WindmarkEtal:2012b} and \citet{Garaudetal:2013} describe the direct collisional formation of planetesimals, ignoring particle transport by radial drift. Although mass transfer in the process of fragmentation of the smaller projectile aggregate during an impact into the larger target aggregate has been clearly proven to exist, the formation of planetesimals of kilometre sizes or larger by this process faces severe problems, such as the rather large time scales required \citep{JohansenEtal:2014}, the role of counter-acting erosion \citep{SchraeplerBulm:2011}, and fragmentation in collisions between similar-sized planetesimals.

A planetesimal-formation model relying on particle concentration and self-gravity has been proposed by \citet{JohansenEtal:2007}, who showed that the streaming instability, first described by \citet{YoudinGoodman:2005}, is capable of concentrating pebble-sized dust aggregates such that planetesimals can directly form by gravitational instability. Since then, this formation scenario has been refined and proven capable of forming planetesimals of up to several 100 km in size from dust aggregates with Stokes numbers in the range St$\sim 0.01$ to St$\sim 1$ within the radii 1 to 10 AU  \citep{BaiStone:2010b,JohansenEtal:2012,CarreraEtal:2015}. Here, the Stokes number is defined as the ratio between the gas-grain coupling time and the inverse Keplerian frequency \citep{Cuzzi:1993}. At 1 AU, this range in Stokes numbers corresponds to cm- to m-sized dust aggregates in a minimum mass solar nebula model. As the formation of dust aggregates at the upper end of the size range faces the above-mentioned drift and fragmentation problems, this planetesimal-formation scenario is likely to operate with pebble-sized rather than boulder-sized dust particles. One of the main issues with previous studies on planetesimal formation via gravitational collapse is the use of inert dust, i.e. dust agglomerates were indestructible. However, numerical simulations predict that they collide with rather high velocities, typically a few $\unit{m~s^{-1}}$ according to \citet{JohansenEtal:2009}. At these velocities, aggregates are supposed to fragment as shown in the model of \citet{GuettlerEtal:2010}.

 %% One of the main unsolved issues with the gravitational-instability formation of planetesimals is the fact that previous studies used inert dust, i.e. dust agglomerates were indestructible, although numerical simulations predict that they collide with rather high velocities, typically a few $\unit{m~s^{-1}}$ according to \citet{JohansenEtal:2009}, at which they are supposed to fragment following the model by \citet{GuettlerEtal:2010}.	

Additionally, during the gravitational collapse of the pebble clouds, speeds high enough for fragmentation can be reached for planetesimals above a few 10 km in size \citep{Wahlberg:2014}. This fragmentation changes the size distribution of the pebbles and thus influences the porosity and packing of the planetesimal that forms.

Nevertheless, \citet{Skorov:2012}, \citet{BlumEtal:2014} and \citet{BlumEtal:2015} have shown that the dust activity of comets as they approach the Sun (as well as the low mass density and thermal conductivity) can only be explained by the gravitational instability scenario of planetesimal formation, due to the resulting low tensile strengths of the accreted dust pebbles.

In this paper, we will present new experimental work on the collision behavior of cm-sized dust aggregates in the velocity range up to $8.7 \unit{m~s^{-1}}$ for mass ratios between target and projectile agglomerates of 1 to 125. These results will be used in the companion paper (Wahlberg Jansson et al. 2016; hereafter Paper II) to simulate how fragmentation affects the gravitational collapse phase and the interior structure of planetesimals that form by gravitational instability.

In Section \ref{sect:exp}, we introduce our new experimental setup. Section \ref{sect:samples} describes the sample preparation and sample properties. In Section \ref{sect:results}, the results and analyses of our experiments are presented. Based upon these results, in Section \ref{sect:fragmod} we propose a simple empirical model to describe the general outcome in aggregate-aggregate collisions, which will be applied in Paper II. Section \ref{sect:simul} describes briefly how we use the new data to better describe the collapse of a pebble cloud. In Section \ref{sect:conclusions}, we conclude our work and discuss its astrophysical implications.

\section{\label{sect:exp}Experimental Setup}
\begin{figure}[htp]
\centering
\includegraphics[width=0.6\textwidth]{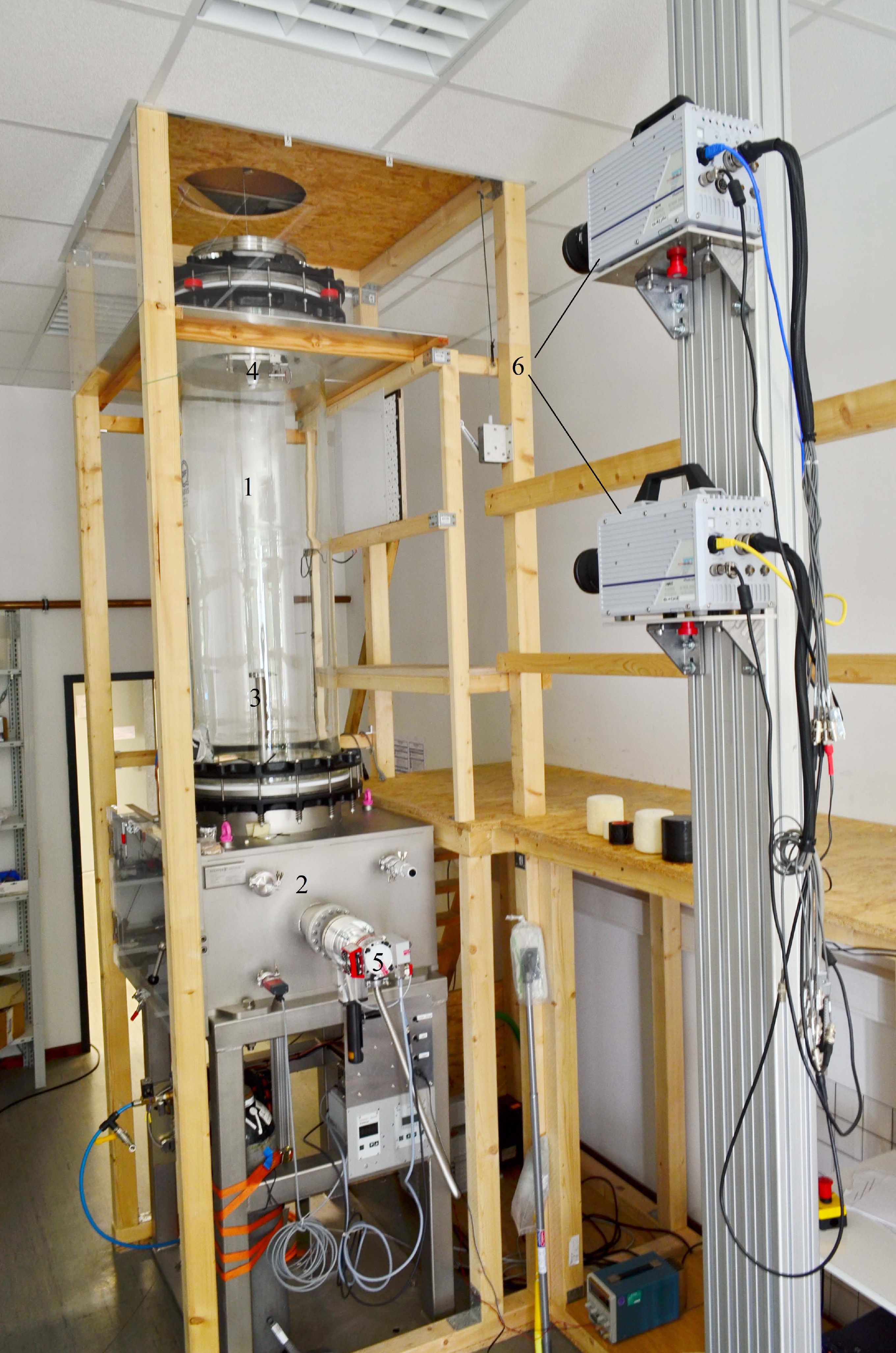}
\caption{Picture of the drop-tower setup for the investigation of aggregate-aggregate collisions. The marked components are (1) glass vacuum tube, (2) vacuum chamber, (3) shaft of pneumatic accelerator, (4) target release mechanism, (5) turbo pump, (6) high-speed cameras.
\label{fig:DrpTwr}}
\end{figure}

Regardless of whether the formation of planetesimals occurs by the process of mass transfer or through gravitational collapse of dust pebbles, cm-sized dust aggregates that collide with velocities in the range of 1-10 $\unit{m~s^{-1}}$ play a crucial role. To analyse the collision outcome in this parameter range, we designed a new experimental setup, which is shown in Figure \ref{fig:DrpTwr}.

The heart of the experimental setup is a vacuum glass cylinder (labeled (1) in Figure \ref{fig:DrpTwr}), which has a length of 150 cm and a diameter of 50 cm and is mounted on top of a steel vacuum chamber (2). Inside this chamber, the projectile dust aggregate is placed on a sample holder, which is attached to a pneumatic accelerator (3). The pneumatic accelerator is connected to a pressurized gas bottle filled with nitrogen gas. The target dust aggregate is loaded on top of a double-winged trap-door release mechanism (4), which is adjusted at the top of the glass cylinder. The trap-door release system consists of solenoid magnets and eddy-current brakes, designed to release the target aggregate into a rotation-free free-fall. Bright-field illumination is accomplished by an LED panel and the colliding dust aggregates are imaged by two Megapixel high-speed cameras (6) operated at 7,500 frames per second. The setup and its operation (with the exception of the pneumatic accelerator) are extensively described in \citet{BlumEtal:2014}.

\begin{figure}[h!]
\begin{center}
\includegraphics[width=0.5\textwidth]{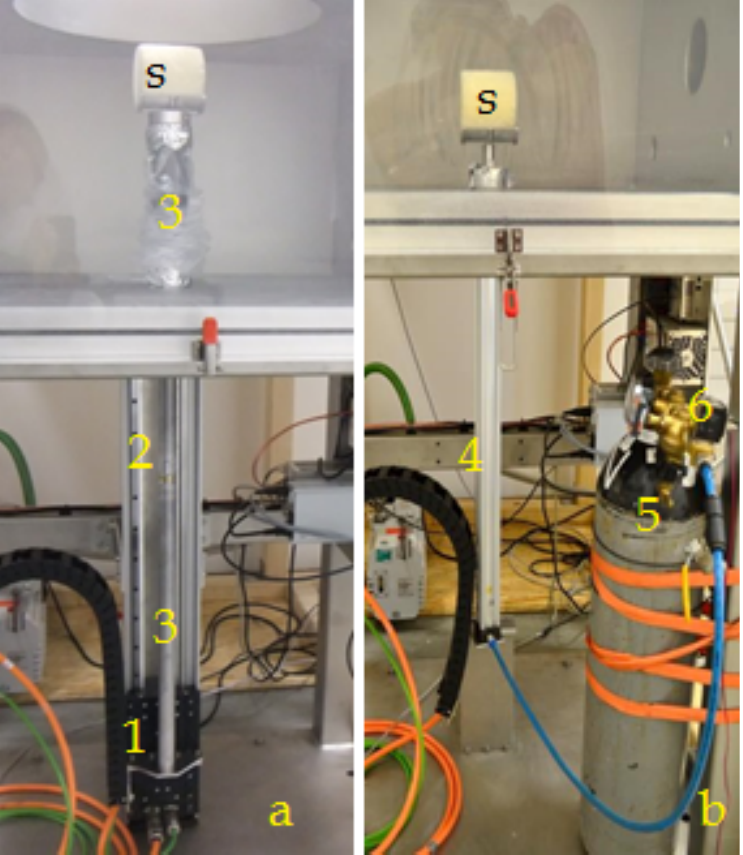}
\caption{The electromagnetic accelerator (a) and the pneumatic accelerator (b), each holding a projectile (labeled S). The labeled components of the electromagnetic accelerator are (1) sledge, (2) track, (3) shaft, and of the pneumatic accelerator (4) shaft, (5) pressurised gas bottle, (6) solenoid valve. \label{fig:Accelerators}}
			\end{center}
\end{figure}

Some of the experimental results presented in Section \ref{sect:results} were conducted by using an electromagnetic accelerator, which is also described in \citet{BlumEtal:2014} and shown in Figure \ref{fig:Accelerators}a. It consists of a sledge (labeled 1 in Figure \ref{fig:Accelerators}a), which is electromagnetically guided over a track of 1040 mm (2). The shaft (3), which operates partially in air and partially in vacuum, is fitted to the sledge that remains outside the vacuum chamber. With the manufacturer-provided software, one can easily adjust the desired acceleration and reach a final velocity up to $5 \unit{m~s^{-1}}$. However, for some applications the selected parameters (e.g., motor currents for the desired velocity) were not effective when the device was subjected to vacuum. High-vacuum conditions inside the vacuum chamber unintentionally accelerated the shaft, which required an unnecessarily high current for deceleration. This, in turn, induced undesirable jitter motion and resulted in pre-collision cracks within the projectile aggregate and complete fragmentation when the projectile was less massive. Thus, the success rate of launching an intact projectile was too low for an efficient experimentation. Hence, we replaced the electromagnetic by a pneumatic accelerator. Figure \ref{fig:Accelerators}b shows the pneumatic accelerator carrying a simulant aggregate (labeled S). The pneumatic accelerator consists of a simple shaft housed in an aluminium casing (4) and is connected to a pressurized gas bottle (5). Upon opening a solenoid valve (6), the gas bottle delivers a pressure of 4-5 bar to the shaft, sufficient to gently accelerate the projectile aggregate to a final velocity of up to $\sim$ 5-6 $\unit{m~s^{-1}}$. Test experiments showed that the pneumatic accelerator does not induce cracks or fragmentation into the fragile dust aggregates.

\section{\label{sect:samples}Sample Preparation and Sample Properties}

\subsection{Preparation of cm-sized dust aggregates}
The dust material used in this study is silicon dioxide ($\unit{SiO_{2}}$). According to the manufacturer (Sigma-Aldrich), the dust is 99 \% pure, consists of irregular grains of size 0.5-10 $\unit{\mu m}$ (approximately 80\% of the grains being between 1 and 5 $\mu$m in diameter) and possesses a material density of $\rho_{\unit{SiO_2}} = 2.60 \unit{g~cm^{-3}}$. This dust analog material has been widely used in laboratory experiments before \citep{Blum:2006,Beitzetal:2012a,SchraeplerEtal:2012,DeckersTeiser:2013,DeckersTeiser:2014} so that the results published here can be related to earlier work.

However, as the dust provided by the manufacturer possesses a lumpy structure, it requires some processing before being used for cm-sized dust aggregates, because we require the aggregates to be as homogeneous as possible. To remove the lumps, the dust powder was first sifted using an electrically vibrated sieve with 500 $\unit{\mu m}$ mesh size. The mass of the sieved dust was carefully measured per desired fill factor and volume and then poured into a respective mold for further compression, which was done manually and for 5-cm-sized aggregates hydraulically. The resulting aggregates possess a cylindrical shape with lengths equalling diameters, both ranging between 1 cm and 5 cm. Details of the dust processing have been already published in \citet{BlumEtal:2014}.

\subsection{Properties of the dust aggregates}
The volume filling factor $\phi$ is defined as the ratio of the overall density of a dust aggregate and the material density of the dust particles. In this study, we fixed the volume filling factor to $\phi = 0.35$, because previous work has shown that this is close to the expected value in the bouncing regime \citep{WeidlingEtal:2009,ZsomEtal:2010a}. Since the volume filling factor is one of the critical parameters in defining the collision outcome, it was carefully controlled throughout the dust processing. To investigate whether the compressed dust cylinders were homogeneous with a volume filling factor of $\phi = 0.35$ throughout their volume, X-ray tomography (XRT) measurements were performed on selected dust aggregates.

\begin{figure}[h!]
\includegraphics[width= 170mm,  height = 80mm]{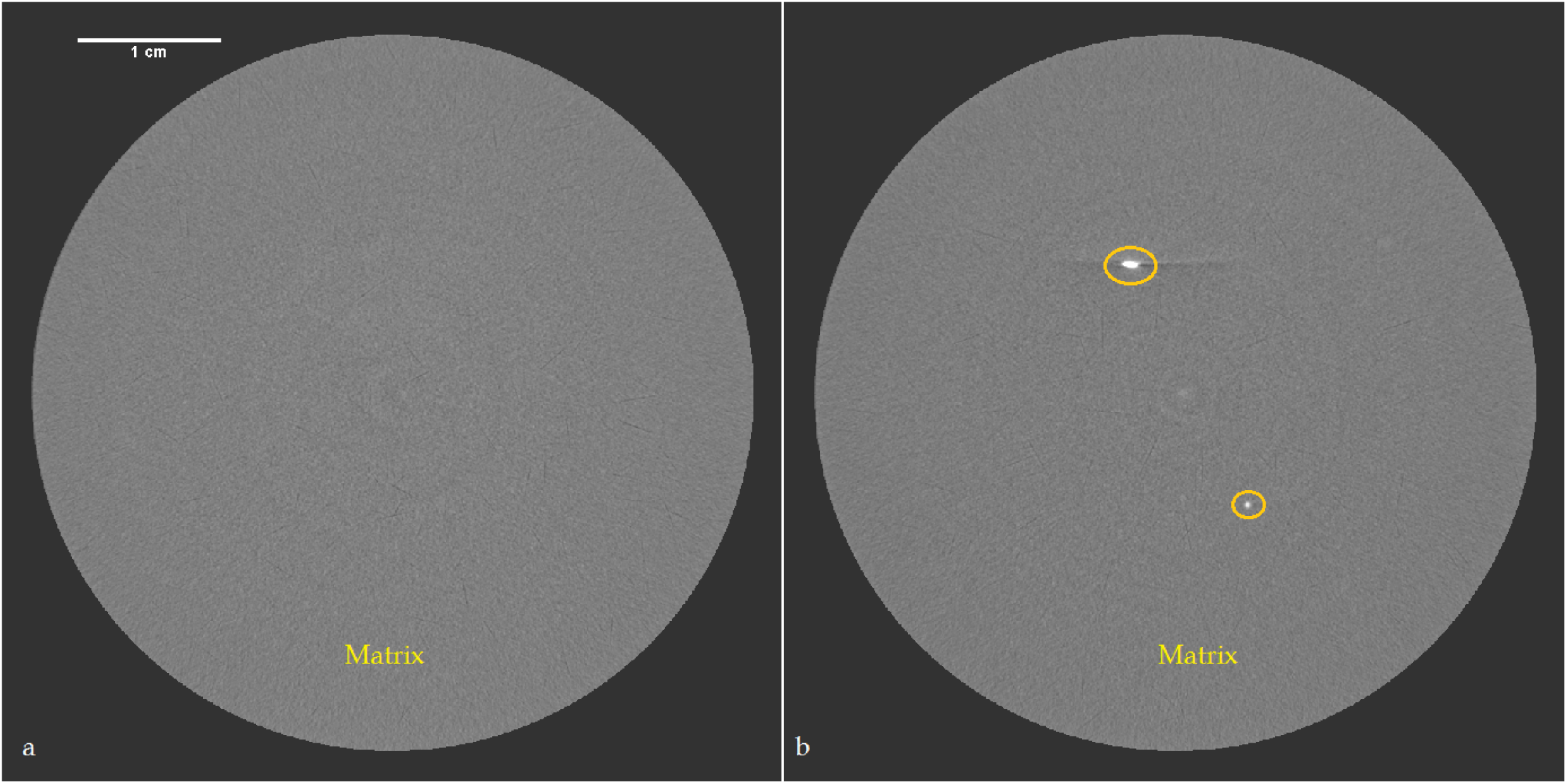}
	\caption{XRT reconstructions of a 5-cm-sized dust aggregate. (a) A cross-sectional view of the sample matrix is shown, which exhibits a very homogeneous volume filling factor (i.e., grey scale). However, in (b) some bright spots of high volume filling factor are visible, which are randomly scattered over the volume of the sample.
\label{fig:XRT-1}}
\end{figure}

Figure \ref{fig:XRT-1} shows reconstructed slices of the XRT analysis of a 5-cm-sized dust aggregate. The homogeneous grey matrix area possesses a standard deviation, over the full 5 cm length, of 5 to 6\%, which translates into $\Delta \phi = \pm 0.02$ relative to its mean value $\bar{\phi} = 0.35$. However, one occasionally can also find bright spots of dense regions, encircled in Figure \ref{fig:XRT-1}b, which occur sporadically throughout the cylindric dust sample. The mean volume filling factor of these dense spots is $\phi \sim 0.57$ and reaches up to $\phi = 0.68$. However, the fractional volume occupied by the dense spots is $< 10^{-5}$ so that their influence on the collision behaviour of the dust aggregates is negligible.

\begin{figure}[h!]
\includegraphics[width= 170mm]{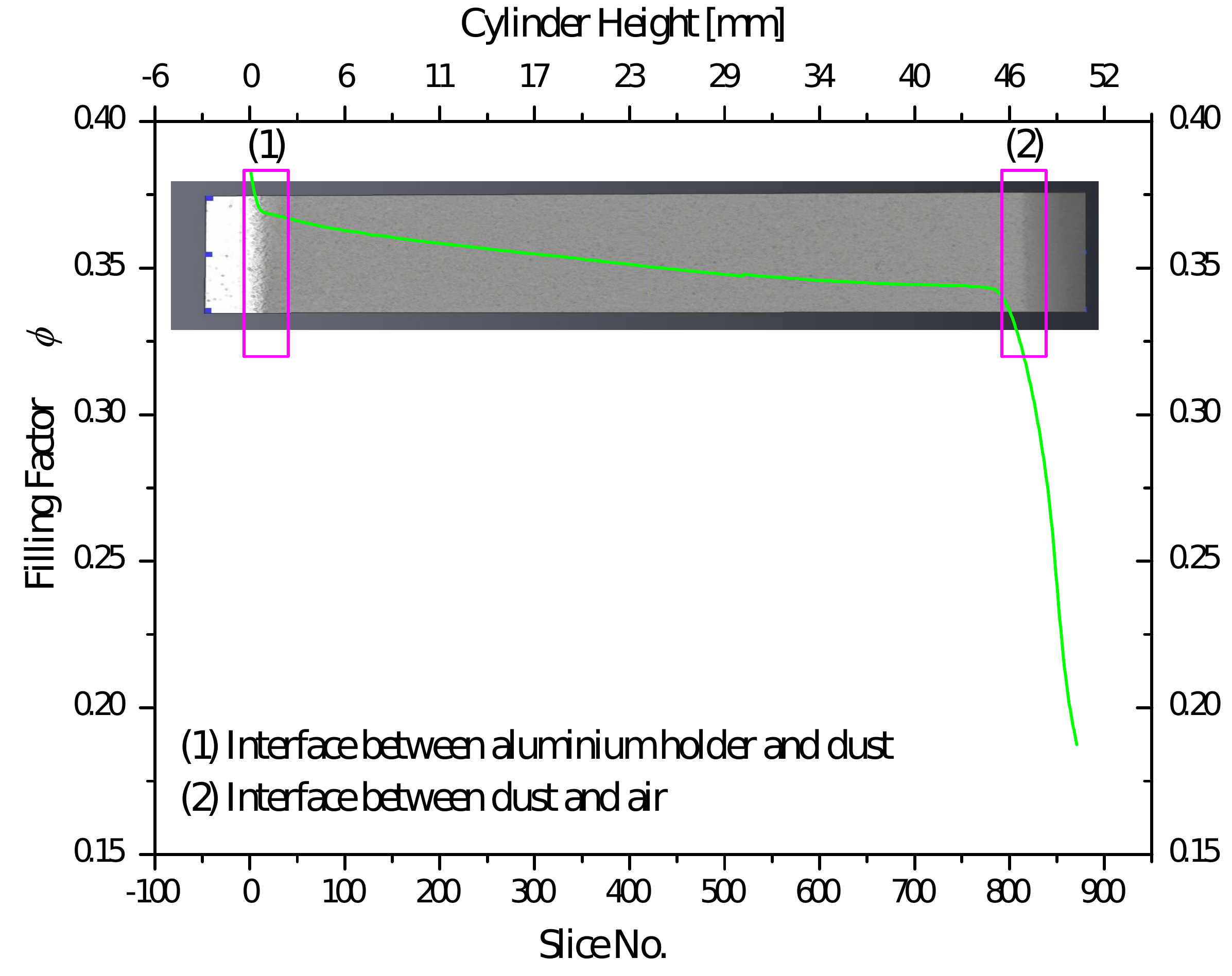}
	\caption{The volume filling factor of a 5-cm-sized dust aggregate along its symmetry axis, averaged over slices of $\sim 57 ~ \rm \mu m$ thickness. The inset shows an X-ray image of the sample. Highlighted with boxes are the interfaces between the solid aluminium holder and the dust aggregate (1) and between the dust aggregate and air (2).
	\label{fig:xrt}}
\end{figure}

Figure \ref{fig:xrt} shows the global volume filling factor profile of an entire 5-cm-sized dust aggregate, which is very similar to the profile showed in \citet{SchraeplerEtal:2012}. The slightly higher volume filling factor in the first 15 slices (or 0.9 mm from the bottom of the cylinder) is an artefact due to the reflection of x-rays from the aluminium sample holder on which the aggregate was placed. The declining tail, starting from slice $\sim 800$ (or 46 mm from the bottom) is assumed to be the reflection from the air-material interface. Between slice 15 and 800, the slice-averaged volume filling factor slightly decreases from $\phi = 0.37$ to $\phi = 0.33$. This decline is caused by the unidirectional compression of the sample \citep{beitzetal2013}.

\section{\label{sect:results}Data Analysis and Experimental Results}
We performed 142 individual aggregate-aggregate collisions in the following 8 series (projectile diameter/height -- target diameter/height): 1 cm -- 1 cm, 1 cm -- 2 cm, 1 cm -- 2.6 cm, 1 cm -- 5 cm, 2 cm -- 2 cm, 2 cm -- 5 cm, 3.5 cm -- 5 cm, and 5 cm -- 5 cm. Table \ref{tab:parameter} summarizes all collision parameters investigated here.

\begin{table}[h!]
	\caption{Experimental parameters of the aggregate-aggregate collisions investigated in this study. Projectile and target aggregate possess cylindrical shape of equal height and diameter and a fixed volume filling factor of $\phi = 0.35$. CF stands for catastrophic fragmentation in which the target and projectile aggregates both fragment. FM stands for fragmentation with mass transfer in which only the projectile fragments and transfers part of its mass to the intact target. $p_{\unit{sur}}$ is the probability with which the target survives the impact intact for collision velocities $v < v_{\unit{sur}}$, and $v_{\unit{sur}}$ is the highest velocity for which the target survives the impact velocity. The asterisks indicate that mass transfer was observed in a few events only, see Figure \ref{fig:MuPlt} below.	 \label{tab:parameter}	}
	
	\begin{center}
		\begin{tabular}{ |l|l|l|l|l|l|l| }	
			\hline
			No. & projectile size & No. of  & Velocity range & Outcome & $v_{\unit{sur}}$ & $p_{\unit{sur}}$ \\
			& -- target size & collisions & $\unit{m~s^{-1}}$ & & $\unit{m~s^{-1}}$  \\
			\hline
			1 & 1.0 cm -- 1.0 cm & 13 & 2.1 -- 5.3 & CF+FM* & 3.1 & $0.2 \pm 0.14$ \\
			2 & 2.0 cm -- 2.0 cm & 18 & 2.0 -- 6.0 & CF+FM* & 2.6 & $0.14 \pm 0.14$\\
			3 & 5.0 cm -- 5.0 cm & 21 & 2.0 -- 6.2 & CF  & $< 2$ & 0 \\
			4 & 1.0 cm -- 2.0 cm & 10 & 2.2 -- 7.7 & CF & $< 2.1$ & 0\\
			5 & 1.0 cm -- 2.6 cm & 19 & 2.5 -- 5.7 & CF+FM & 4.5 & $0.53 \pm 0.19$ \\
			6 & 1.0 cm -- 5.0 cm & 21 & 3.6 -- 8.7 & CF+FM & $ \geq 8.4$ & $0.80 \pm 0.20$ \\
			7 & 2.0 cm -- 5.0 cm & 24 & 1.6 -- 7.1 & CF+FM & 4.6 & $0.52 \pm 0.17$\\
			8 & 3.5 cm -- 5.0 cm & 16 & 1.5 -- 4.4 & CF & $< 1.5$ & 0\\
			\hline
		\end{tabular}
	\end{center}
\end{table}

\begin{figure}[h!]
	\begin{center}
		\includegraphics[width= 120mm,  height = 100mm]{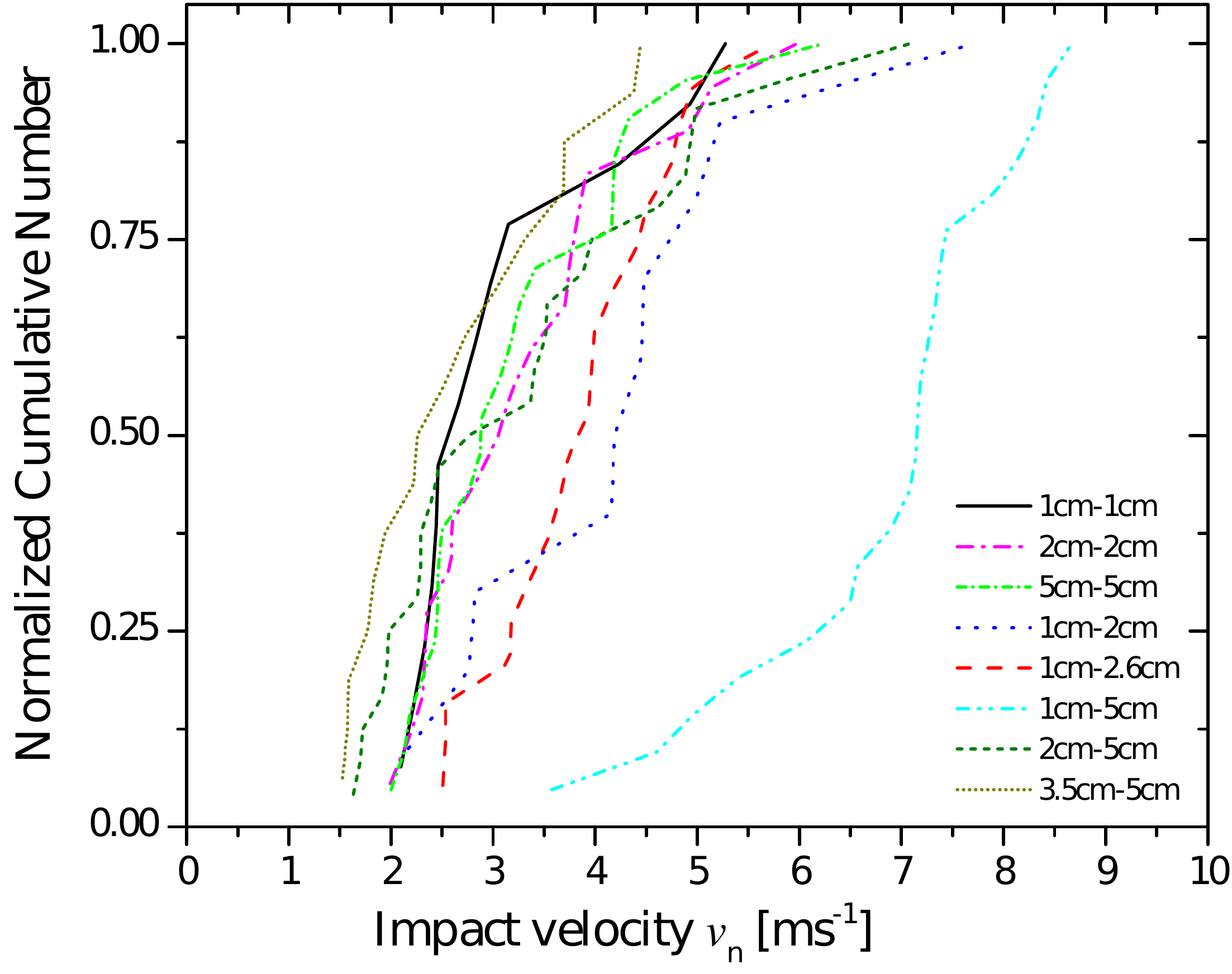}
		\caption{Normalised cumulative velocity distributions of the 8 experimental series listed in Table \ref{tab:parameter}. The mean collision velocity in this study is  $ \overline{v}_{n} = 4.51 \unit{m~s^{-1}}$. \label{fig:VD}}
	\end{center}
	
\end{figure}

The velocity distributions of the individual collision experiments in the 8 series listed in Table \ref{tab:parameter} are shown in Figure \ref{fig:VD} in a cumulative way. One can see that the chosen velocities are distributed evenly in the respective ranges given in Table \ref{tab:parameter}. The rather exceptionally high velocities of the 1 cm -- 5 cm series are due to the fact that the large targets possess higher fragmentation threshold velocities in collisions with small projectiles. This is further discussed below.

\begin{figure}[h!]
	\begin{center}
		\includegraphics[width= 150mm,  height = 116mm]{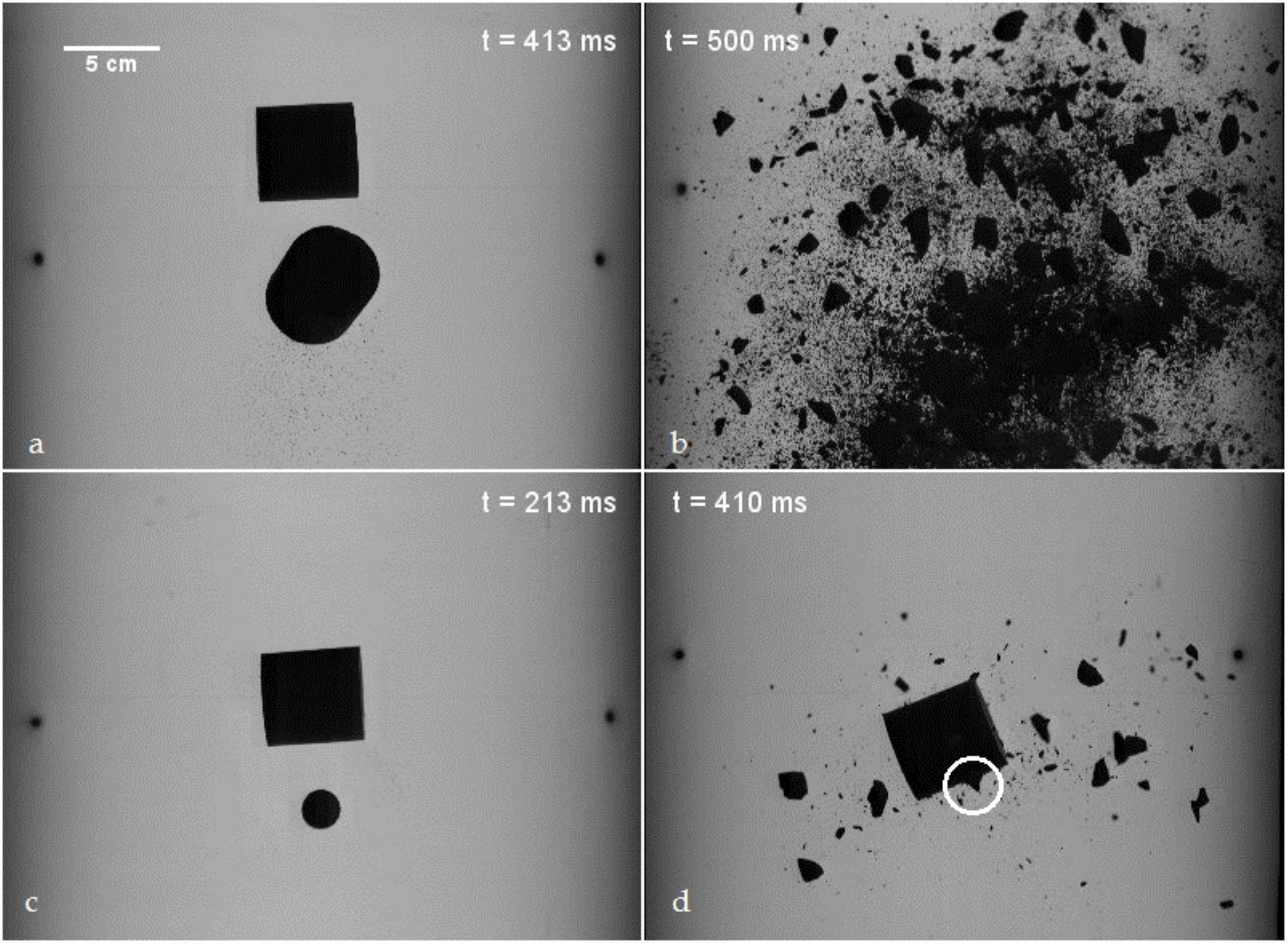}
		\caption{Examples of pre-collision and post-collision images. (a) and (b): Projectile and target aggregates of 5 cm each, colliding at 6.2 $\unit{m~s^{-1}}$ and resulting in catastrophic fragmentation of both aggregates. (c) and (d): A 2-cm-sized projectile aggregate, colliding with a 5-cm-sized target aggregate at 2.8 $\unit{m~s^{-1}}$, resulting in the fragmentation of the projectile only. Part of the projectile mass has been visibly transferred to the target, which has typically a cone shape (highlighted by the circle in (d)). 	 \label{fig:mont2}}
	\end{center}
\end{figure}

Figure \ref{fig:mont2} shows two examples of pre-collision and post-collision images taken with one of the high-speed cameras. The pre-collision images demonstrate the geometry of the collision, which has been set in such a way that the symmetry axis of the projectile was rotated by $90^{o}$ with respect to the symmetry axis of the target. The target aggregate, which is dropped from the top, thus projects a rectangular shape onto the field of view of the cameras, while the projectile aggregate, shot from the bottom, appears as a circle (see Figure \ref{fig:mont2}c). This geometry provides minimum contact area between the aggregates at first contact and is representative for collisions between spherical aggregates as shown by \citet{BeitzMeisneretal:2011}. However, in practice it has been challenging to guarantee these ideal geometrical conditions, first due to the slightly inherent nonalignment along the line joining the centres of release mechanism and accelerator, and second due to sometimes unavoidable rotation of the projectile aggregate. For instance, the 5-cm-sized projectile in Figure \ref{fig:mont2}a is slightly tilted due to rotation. Figures \ref{fig:mont2}b and \ref{fig:mont2}d show examples for complete fragmentation of projectile and target (annotated CF in Table \ref{tab:parameter}) and mass transfer from the fragmented projectile to the non-fragmented target (annotated FM in Table \ref{tab:parameter}), respectively.

The imperfect alignment between projectile and target also results in not perfectly central collisions, which we describe using a one-dimensional impact parameter. Due to the limitation of the experimental design, i.e. both cameras observing from the same direction, only one component of the two-dimensional impact parameter is accessible and taken into account. The velocities used in our data analysis have been corrected for impact parameter such that only the normal component of the relative velocity between projectile and target was taken into account. Formally the normal component of the collision velocity, $v_{\unit{n}}$, is derived from the relative collision speed, $v_{\unit{rel}}$, by
\begin{equation}\label{eq:imppar}
v_{\unit{n}} = v_{\unit{rel}} \cdot \sin\left(\arctan\frac{b}{r_{\unit{p}} + r_{\unit{t}}} \right) ,
\end{equation}
with $b$, $r_{\unit{p}}$ and $r_{\unit{t}}$ being the impact parameter, the radius of the projectile and the radius of the target aggregate, respectively. For our cylindrical aggregates with length $l$ and diameter $d$, we get $r_{\unit{p,t}} = l_{\unit{p,t}}/2 = d_{\unit{p,t}}/2$, with the indices p and t denoting the projectile and target aggregate, respectively.

\subsection{\label{sect:mt}Survival of the target aggregate and mass transfer}
Survival of the target aggregate combined with mass transfer from the fragmenting projectile to the intact target was a non-negligible experimental outcome in the 1 cm -- 2.6 cm, 1 cm -- 5 cm and 2 cm -- 5 cm series, with single events present in the 1 cm -- 1 cm and 2 cm -- 2 cm series.

\begin{figure*}[h!]
	\begin{center}
	\includegraphics[width= 80mm,  height = 80mm]{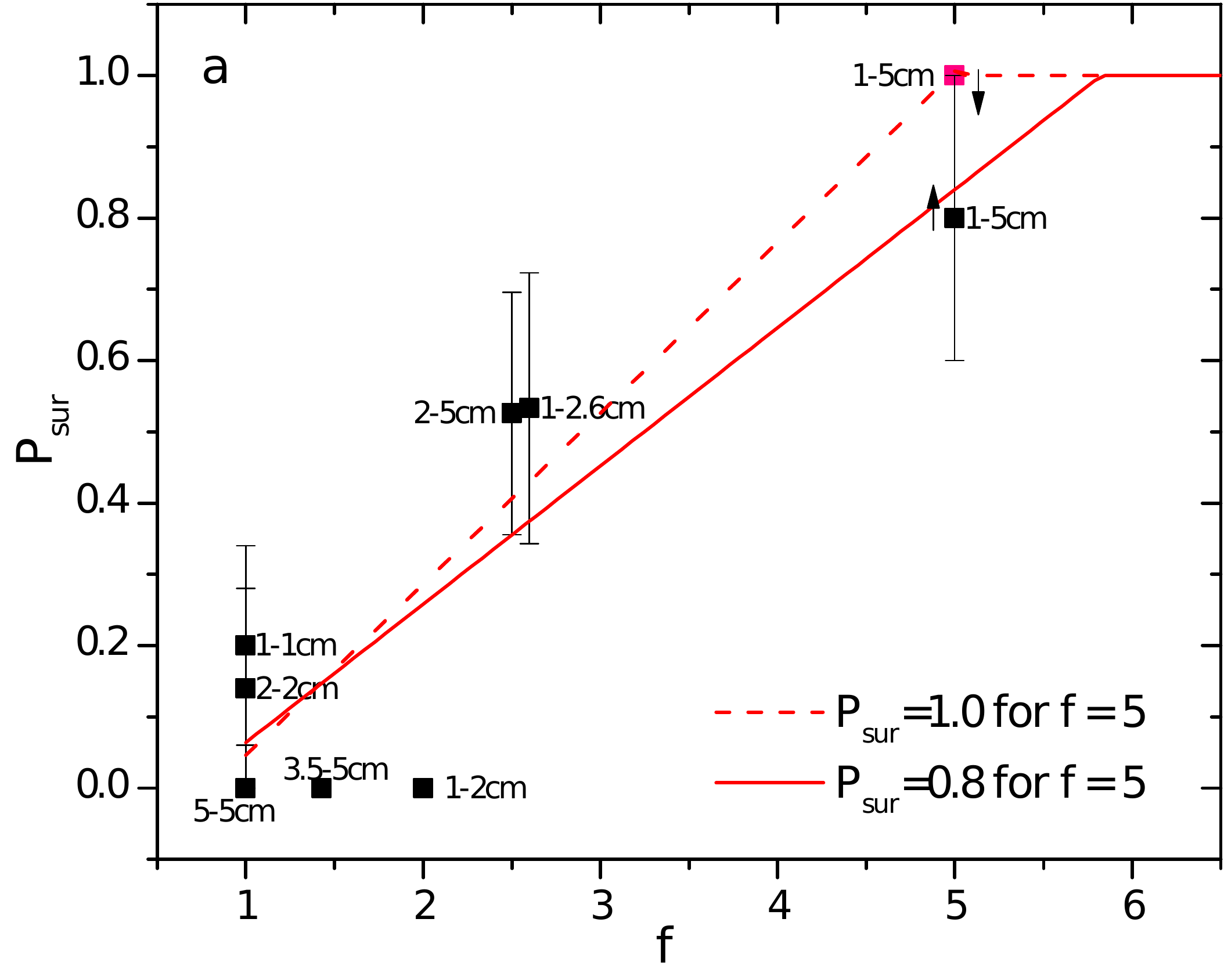}
	\includegraphics[width= 80mm,  height = 80mm]{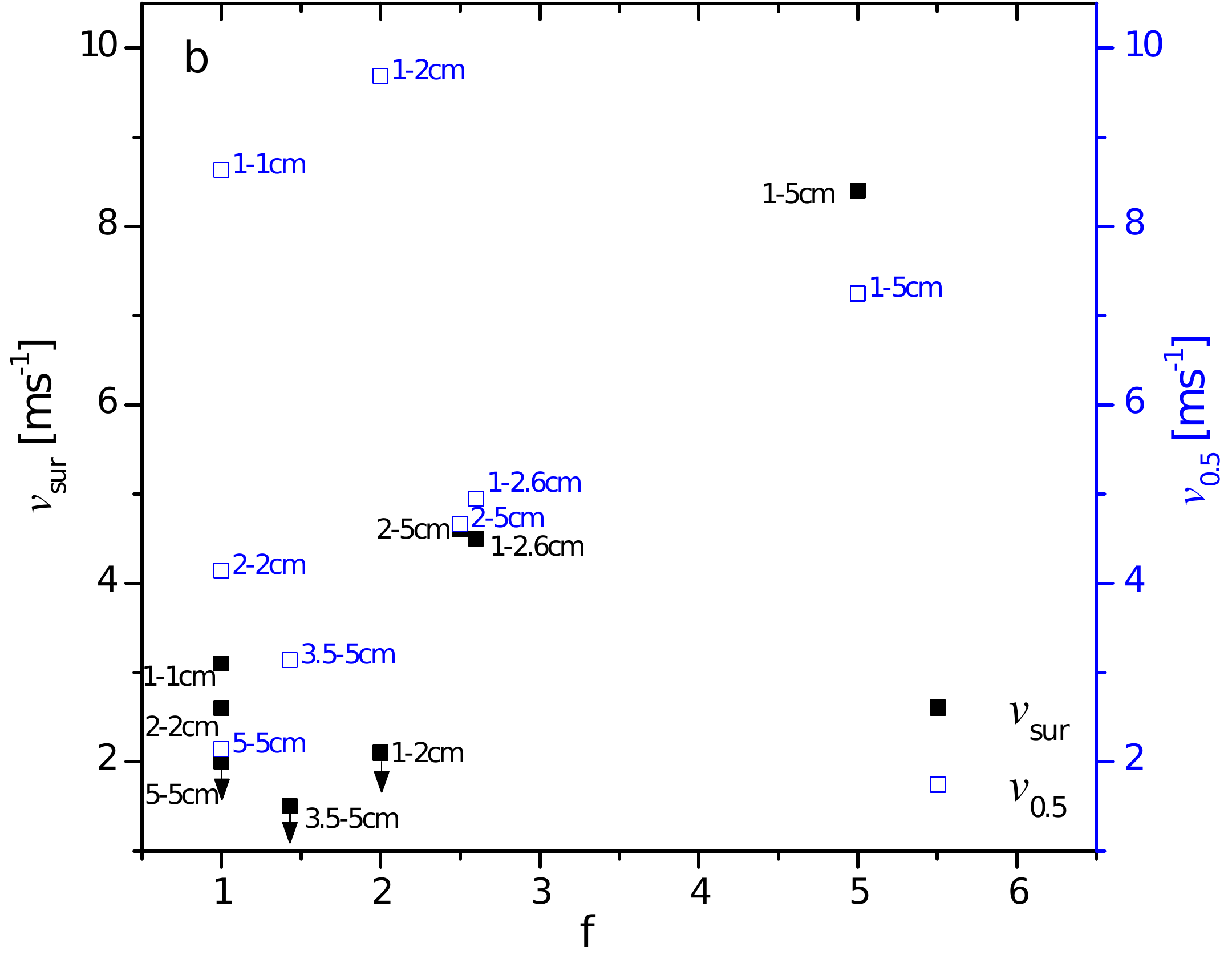}
		\caption {(a). Probability $p_{\unit{sur}}$  for the target aggregate to survive the impact of a projectile aggregate in the velocity range $v < v_{\unit{sur}}$, with $p_{\unit{sur}}$ and $v_{\unit{sur}}$ shown in Table \ref{tab:parameter}, as a function of the size ratio of target and projectile, $f$. The red solid line is a linear fit to the data, including the black square at $f=5$; the red dashed line is the same for the red square (see text) .
			(b). Highest collision velocity for which the target stayed intact, $v_{\unit{sur}}$ , as a function of the target-to-projectile size ratio, $f$ (black filled squares). In addition, we also show the critical fragmentation velocity of the target, $v_{\unit{0.5}}$ (blue open squares). In both figures the aggregates of similar $f$ tend to have similar values.
			\label{fig:mtprob}}
	\end{center}
\end{figure*}

As both the fragmentation of the target and its survival were possible outcomes in the same velocity range, we analysed the probability for the occurrence of mass transfer and, thus, for the intact survival of the target more closely by first determining the highest velocity $v_{\unit{sur}}$ for which mass transfer and intact survival of the target were observed. The sixth column of Table \ref{tab:parameter} shows this velocity. Then, we defined the probability for target survival, $p_{\unit{sur}}$, by the ratio between the number of mass-transfer events and the total number of experiments in the velocity range $v < v_{\unit{sur}}$. The seventh column of Table \ref{tab:parameter} lists the corresponding results. Plotting these probabilities in Figure \ref{fig:mtprob}a as a function of the size ratio $f$ between target and projectile shows that $p_{\unit{sur}}$ steadily increases with increasing $f$ values.

In the case of the 1 cm -- 5 cm series, i.e. for $f = 5$, the formal mass-transfer probability is $p_{\unit{sur}} = 0.8$ as shown by the black square in Figure \ref{fig:mtprob}a. However, the collision velocities in this series had been chosen systematically higher than for all the other series (see Figure \ref{fig:VD}), to achieve fragmentation of the target at all. Thus, $p_{\unit{sur}} = 0.8$ is most likely a lower limit to the true mass-transfer probability. As an upper limit, we chose $p_{\unit{sur}} = 1.0$ for the size ratio $f \stackrel{>}{\sim} 5.83$ (see red square in Figure \ref{fig:mtprob}a). For $f=1$, the mass-transfer probability is rather low and slightly decreases from $p_{\unit{sur}} = 0.2$ to $p_{\unit{sur}} = 0$ when the aggregate size increases from 1 cm to 5 cm. As the mass-transfer probabilities for the 1 cm -- 2.6 cm and the 2 cm -- 5 cm series are very similar, we conclude that these probabilities are merely dependent on the size (or mass) ratio between target and projectile and not on their absolute values. Thus, we approximated the mass-transfer probability by

\begin{equation}\label{eq:mtprob}
p_{\unit{sur}} = \left\{
\begin{array}{lcrcl}
0.19 f-0.13 & \unit{for} & 1 \le f \le 5.83 & \unit{and} & p_{\unit{sur}}(f=5) = 0.8\\
1 & \unit{for} &  f > 5.83 & \unit{and} & p_{\unit{sur}}(f=5) = 0.8\\
0.24 f - 0.20 & \unit{for} & 1 \le f \le 5 & \unit{and} & p_{\unit{sur}}(f=5) = 1\\
1 & \unit{for} &  f > 5 & \unit{and} & p_{\unit{sur}}(f=5) = 1\\
\end{array}
\right.
\end{equation}
(see the red solid and dashed lines in Figure \ref{fig:mtprob}a).

In Figure \ref{fig:mtprob}b the survival velocity  $v_{\unit{sur}}$ (on left y-axis in black squares) and critical fragmentation velocity $v_{\unit{0.5}}$ (on right y-axis in blue open squares) have been analyzed as a function of size ratio $f$. Here we see $v_{\unit{sur}}$ systematically increases with increasing $f$, which indicates an implicit correlation between $P_{\unit{sur}}$ and $v_{\unit{sur}}$.
Moreover it should be noticed that once again the series of similar size ratios i.e. 1 cm -- 2.6 cm and 2 cm -- 5cm, have similar values of $v_{\unit{sur}}$ and $v_{\unit{0.5}}$, which supports the hypotheses that it is the relative size of the target and projectile that matters.

For the three experiment series, where $f \geq 2.5$ and mass transfer was a common outcome (1 cm -- 2.6 cm, 1 cm -- 5 cm, 2 cm -- 5 cm, see Figure \ref{fig:MuPlt} below), we can also state that mass transfer always occurs down to the smallest collision velocities investigated, but possesses an upper velocity limit in the cases of 1 cm -- 2.6 cm and 2 cm -- 5 cm. Thus, for $v > v_{\unit{sur}}$, fragmentation is the only outcome. However, for 1 cm -- 5 cm, there is no such upper limit although in this series we extended the investigated velocity range up to 8.7 $\unit{m~s^{-1}}$. Moreover, for velocities $v < 5.6 \unit{m~s^{-1}}$, mass transfer is the only outcome. The sixth column in Table \ref{tab:parameter} summarizes our findings for $v_{\unit{sur}}$ and Figure \ref{fig:mtprob}b shows the data as a function of $f$. One can recognise that $v_{\unit{sur}}$ increases with increasing target-to-projectile size ratio so that dust-evolution models need to take mass transfer into account, particularly in cases when small projectiles hit large targets. These latter cases have been studied extensively in previous works (\citet{WurmEtal:2005a}, \citet{TeiserWurm:2009a}, \citet{GuettlerEtal:2010}, \citet{Kotheetal:2010}, \citet{Teiseretal:2011a}, and \citet{DeckersTeiser:2014}).

The target survival velocity $v_{\unit{sur}}$ can then be compared with the critical fragmentation velocity $v_{\unit{0.5}}$ (which will be derived in Section \ref{sect:Qstar} below), for which the mass of the largest fragment of the target equals half the initial target mass.
%%The latter can be derived from the target strength $Q^*$ (which will be derived in Section \ref{sect:Qstar} below) by $v_{\unit{0.5}} = \sqrt{2 Q^* \left( 1 + \frac{m_{\unit{t}}}{m_{\unit{p}}} \right)}$.
In Figure \ref{fig:mtprob}b we also show $v_{\unit{0.5}}$ for comparison. As can be seen, both velocities are very similar in those series in which mass transfer is a regular outcome (1 cm - 2.6 cm, 1 cm - 5 cm, 2 cm - 5 cm) so that we can conclude that mass transfer is only possible (and will occur with a probability $p_{\unit{sur}}$ as shown above) as long as the projectile is unable to destroy the target (for more details, please refer to Sections \ref{sect:e05} and \ref{sect:Qstar}).

The typical mass-transfer efficiencies, here defined by $\Delta m / m_{\unit{p}}$, with $\Delta m$ being the mass transferred from projectile to target, varied from $\Delta m / m_{\unit{p}} \approx 0.08$ to $\Delta m / m_{\unit{p}} \approx 0.30$ and are in the same range as previously found in collisions among cm-sized spherical dust agglomerates (\citet{BeitzMeisneretal:2011}, their Figure 8) or in collisions with varying mass ratios (\citet{WurmEtal:2005b}, their Figure 6; \citet{Kotheetal:2010}, their Figure 5; \citet{DeckersTeiser:2014}, their Figure 8).

\begin{figure*}[h!]
	\begin{center}
		\includegraphics[width= 80mm,  height = 80mm]{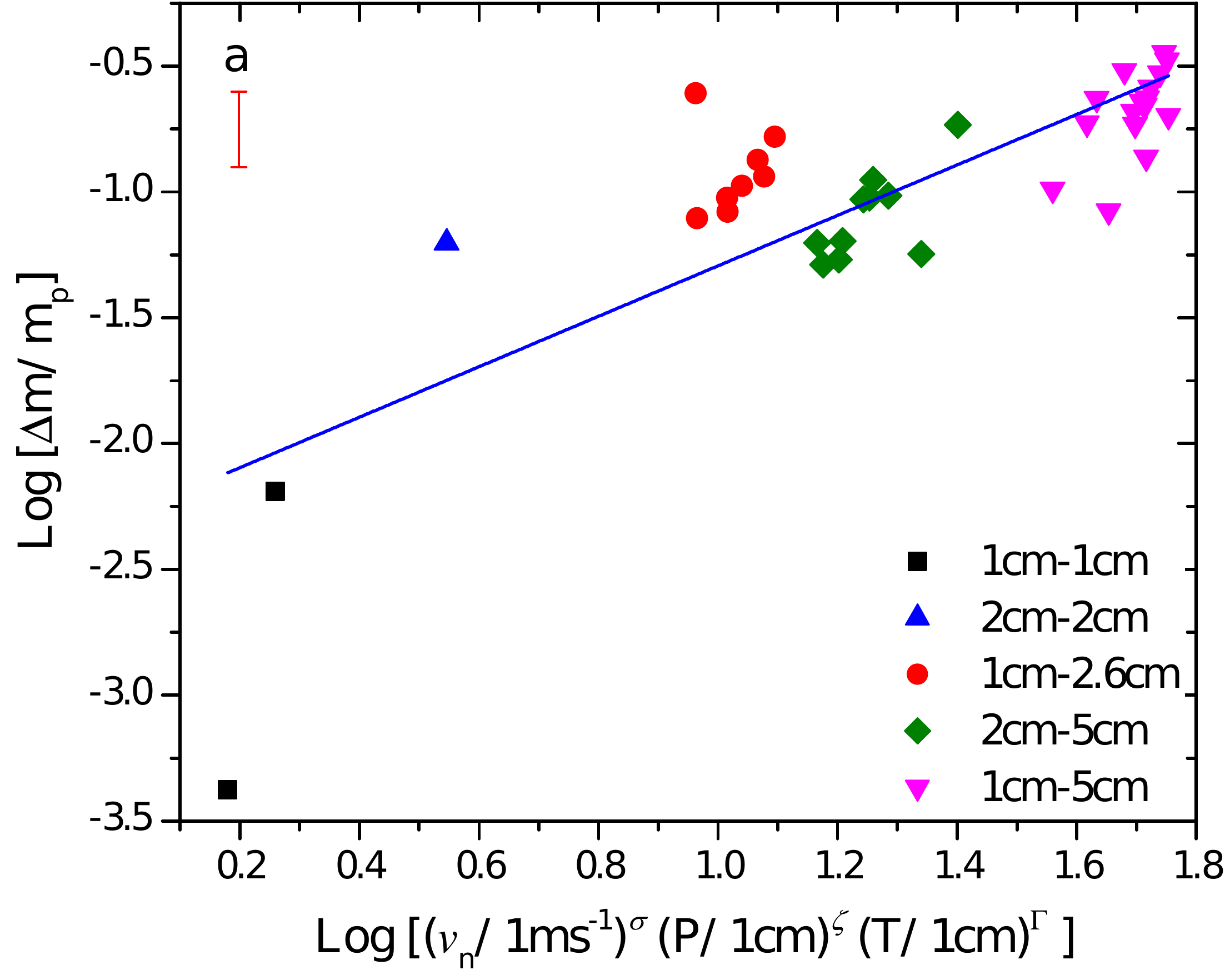}
		\includegraphics[width= 80mm,  height = 80mm]{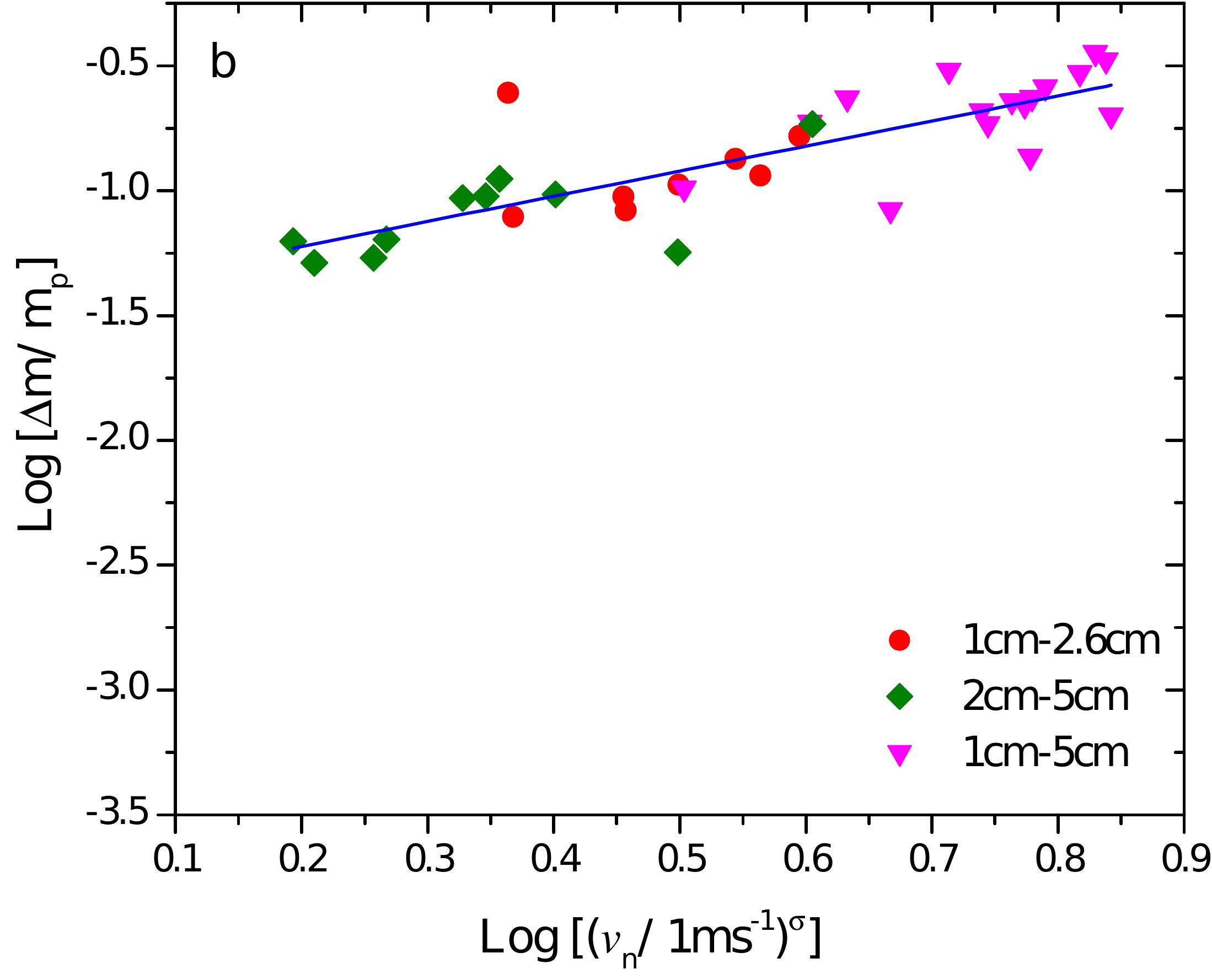}
		\caption {(a) Dependence of the mass-transfer efficiency $\Delta m / m_{\unit{p}}$ in all 37 events as a function of a power-law combination of the collision parameters, i.e. $v_{\unit{n}}^{\sigma} P^{\zeta} T^{\Gamma}$, with the respective coefficients given in first row of Table \ref{tab:RedChi}. Since the mass gain was determined by measuring the projected area of the transferred mass (as encircled in Figure \ref{fig:mont2}), we estimate the mean error as being a factor 2 in mass, which is shown by the red error bar at top left.(b) Removing the three data points with $f = 1$ reduces $\chi_{\unit{red}}^2$ considerably and shows that the dependence on $P$ and $T$ vanishes (see second row of Table \ref{tab:RedChi}). Thus, we re-fitted the 34 remaining data points with a power-law dependence on velocity only (see third row of Table \ref{tab:RedChi}).
			\label{fig:mtfit}}
	\end{center}
\end{figure*}

One novelty of our experimental data over previous work is that we now are able to derive the dependency of the mass-transfer efficiency on velocity, projectile size and target size. Earlier work either used equal-sized aggregates or studied impacts of dust aggregates into semi-infinite targets. The events of mass transfer have been observed in 5 series, providing 37 data points on the relative mass transfer $\Delta m / m_{\unit{p}}$ (shown in Figure \ref{fig:mtfit}a), which we analysed according to their (assumed power-law) dependency on velocity $v_{\unit{n}}$ and projectile/target sizes $P$ and $T$, respectively, i.e.,
\begin{equation}\label{eq:mtpowerlaw}
\log \left( \frac{\Delta m}{m_{\unit{p}}} \right) =
C_{\unit{mt}} + \sigma \log{\left(\frac{v_{\unit{n}}}{\unit{1~m~s^{-1}}}\right)} + \zeta \log \left(\frac{P}{\unit{1~cm}}\right) + \Gamma \log \left(\frac{T}{\unit{1~cm}}\right).
\end{equation}

In order to derive the coefficients $C_{\unit{mt}}$, $\sigma$, $\zeta$ and $\Gamma$ in the above equation, we minimised the reduced chi-squared value $\chi_{\unit{red}}^2$ for two cases. In the first case, all 37 events of mass-transfer were taken into account. In the second case, the three events from the series with $f = 1$ were dropped due to their low mass-transfer efficiency (see Figure \ref{fig:mtprob}a) so that only 34 events from the series with $f > 1$ were considered. The respective values of the coefficients and $\chi_{\unit{red}}^2$ in both cases are given in Table \ref{tab:RedChi}, where one can see that the restriction to $f>1$ significantly reduces the $\chi_{\unit{red}}^2$ value. The resulting correlations and fits in both cases are shown in Figure \ref{fig:mtfit}.

\begin{table}[h!]
	\caption{Derivation of the coefficients in Eq. \ref{eq:mtpowerlaw} by minimizing $\chi_{\unit{red}}^2$. The first row shows the results when all 37 data points are taken into account. The second row gives the results for 34 events of mass-transfer in the series with $f > 1$ (see text for details). The third row shows the fit values when projectile and target size are neglected for the same data set.
		\label{tab:RedChi}}
	\begin{center}
		\begin{tabular}{ |c|c|c|c|c|c|}	
			\hline
			Data Points. & $C_{\unit{mt}}$ & $\sigma$ & $\zeta$ & $\Gamma$ & $\chi_{\unit{red}}^2$ \\
			&& (exponent of $v_{\unit{n}}$)  & (exponent of \textit{P})& (exponent of \textit{T}) & \\
			\hline
			37 & $-2.30 \pm 0.24$ & $0.52 \pm 0.44$ & $-0.72 \pm 0.69$ & $1.82 \pm 0.40$&  $ 0.111 $ \\
			34 & $-1.37 \pm 0.14 $ & $0.81 \pm 0.23 $ & $-0.22 \pm 0.39$ & $0.05 \pm 0.32$ & $0.022 $ \\
			34 & $ -1.42\pm 0.07 $ & $0.91 \pm 0.11 $ & $ 0 $ & $ 0 $  & $0.021$ \\
			\hline
		\end{tabular}
	\end{center}
\end{table}

When we remove the 3 events from the series where $f = 1$ and fit the remaining 34 data points to Eq. \ref{eq:mtpowerlaw}, we get the coefficients shown in the second row of Table \ref{tab:RedChi}. It can be seen that $\chi_{\unit{red}}^2$ is reduced by a factor $\sim 5$, while the velocity dependence becomes significant. Surprisingly, the previous strong dependence on $T$ is now negligible, suggesting no role of the target in the case of $f > 1$. As the error of the exponent of $P$ is almost twice as large as the exponent itself, we also argue that the $P$ dependence of the mass-transfer is negligible. Therefore we neglect the coefficients of $P$ and $T$ by setting $\zeta=0$ and $\Gamma=0$, which leads to Eq. \ref{eq:mtvel}. The resulting values are shown in third row of Table \ref{tab:RedChi}. One can see that the omission of $P$ and $T$ further strengthens the dependence on velocity and even slightly reduces the value of $\chi_{\unit{red}}^2$. Figure \ref{fig:mtfit}(b) is the graphical representation of mass-transfer for size ratios $f > 1$ as a function of velocity only. Thus, we re-write Eq. \ref{eq:mtpowerlaw}, such that
\begin{equation}\label{eq:mtvel}
\log \left( \frac{\Delta m}{m_{\unit{p}}} \right) =
C_{\unit{mt}} + \sigma \log{\left(\frac{v_{\unit{n}}}{\unit{1~m~s^{-1}}}\right)}
\end{equation}
and get $C_{\unit{mt}} = -1.42 \pm 0.07$ and $\sigma = 0.91 \pm 0.11$, respectively, with $\chi_{\unit{red}}^2 = 0.021$. Obviously, if the aggregates are intrinsically different in size, i.e. $f > 1$, the dependence of mass-transfer on the size of individual aggregates becomes negligible and only the impact velocity $v_{\unit{n}}$ is the primary factor on which mass-transfer depends.

Formally, Eq. \ref{eq:mtvel} breaks down when $\Delta m / m_{\unit{p}} > 1$, i.e. for $v > 10^{-C_{\unit{mt}} / \sigma} = 36.34 \unit{m~s^{-1}}$. However, at these high impact velocities, other processes, like cratering, are important, which are not the subject of this study. One should, thus, be careful to use extrapolations of Eq. \ref{eq:mtvel} to too high velocities.

%% It should be mentioned that our values for the mass transfer efficiency are comparable to those found by \citet{BeitzMeisneretal:2011}.

\subsection{\label{sect:fragstren}Fragmentation strength $\mu$}
For each collision, we measured the fragmentation strength $\mu$, which we define as the mass ratio of the largest fragment $m_{\unit{l}}$ observed after the collision to the initial target mass $m_{\unit{t}}$, i.e.
\begin{equation} \label{eq:mul}
\mu = \frac {m_{\unit{l}}}{m_{\unit{t}}} .
\end{equation}

With this definition, we can use $\mu$ to distinguish between different collisional outcomes, i.e.
\begin{equation}
\mu \left\{\begin{array}{@{}l@{\quad}l}
> 1 & \textrm{mass transfer from projectile to target} \\[\jot]
= 1 & \textrm{bouncing}\\
< 1 & \textrm{fragmentation of projectile and target} \\
\end{array}\right.
% \textrm{1) In this study bouncing has never been observed.}
\end{equation}

\begin{figure*}[htp]
	\includegraphics[width= 150mm,  height = 180mm]{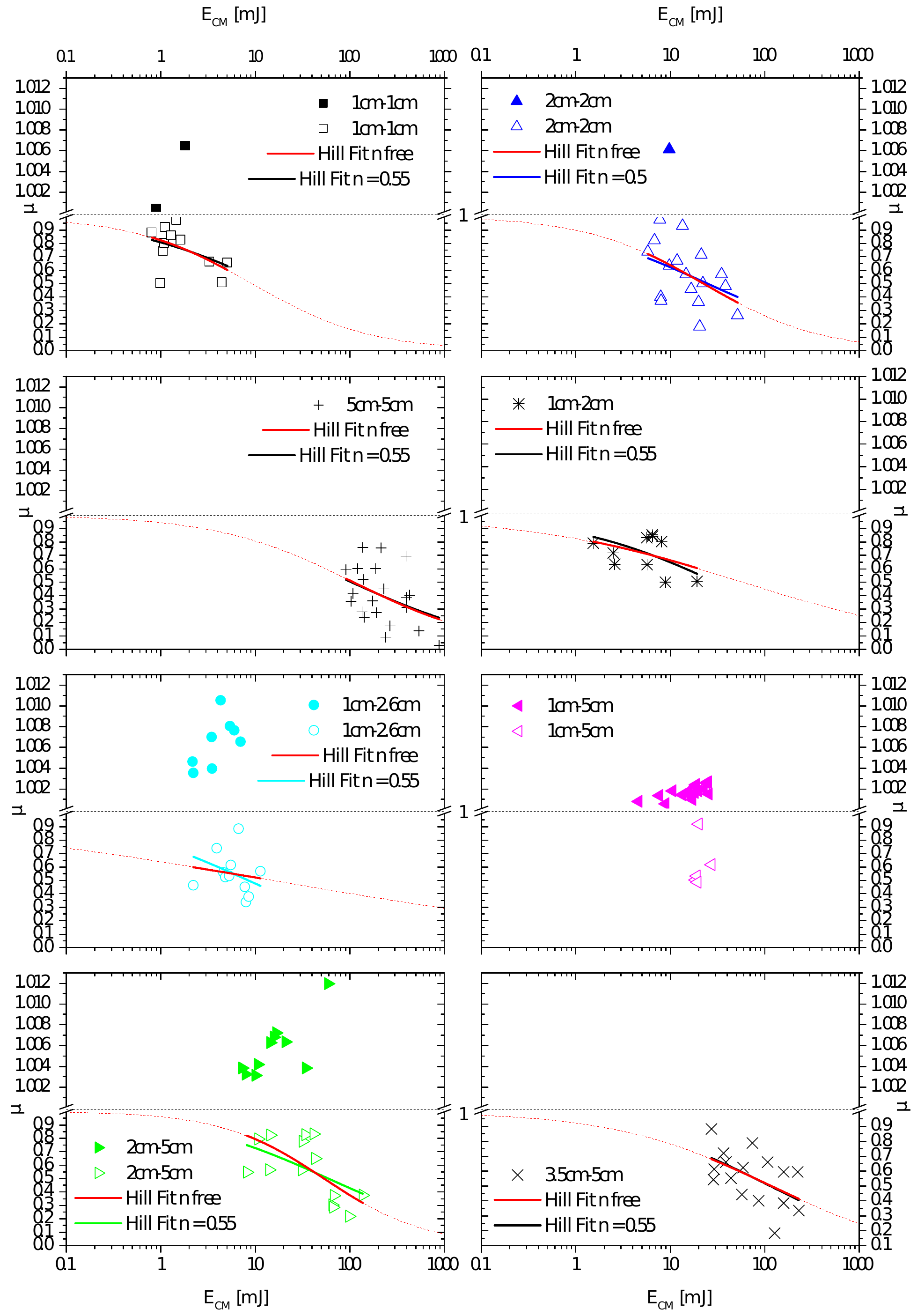}
	\caption {The fragmentation strength $\mu$ as a function of centre-of-mass kinetic energy. The dotted horizontal line at $\mu = 1$ is the line of bouncing, which separates the region of mass transfer ($\mu > 1$, represented by the filled data points) from the region of fragmentation ($\mu < 1$, represented by the open data points). Please mind the break in scale between $\mu < 1$ and $\mu > 1$. The curves in the $\mu < 1$ region follow Eq. \ref{eq:hill} and were fitted to the $\mu < 1$ data points only. Due to the small number of data points with $\mu < 1$ in the 1 cm -- 5 cm series, Eq. \ref{eq:hill} could not be fitted to this data set. \label{fig:MuPlt}}
\end{figure*}

Contrary to many previous studies in which $\mu$ has been investigated as a function of impact velocity, here we analyse it as a function of kinetic energy $E_{\unit{cm}}$ in the centre-of-mass system of projectile and target, i.e.
\begin{equation}
E_{\unit{cm}} = \frac {1}{2} m v_{\unit{n}}^{2} ,
\end{equation}
where $m$ is the reduced mass of the projectile-target system, given by $m^{-1} = m_{\unit{p}}^{-1} + m_{\unit{t}}^{-1}$, with $m_{\unit{p}}$ and $m_{\unit{t}}$ being the projectile and target mass, respectively. We use the kinetic energy, because we later intend to derive the collision strength $Q^*$ (see Sect. \ref{sect:Qstar}), and the reduced mass, because only this value has a contribution to the mass loss (the remainder of the kinetic energy refers to the motion of the center of mass). Figure \ref{fig:MuPlt} compiles the results for the fragmentation strength of all 8 collision series listed in Table \ref{tab:parameter}.

\subsection{\label{sect:e05}The catastrophic threshold energy $E_{0.5}$}
In previous studies, power-law dependencies between the fragmentation strength and the collision energy have frequently been used to determine the catastrophic threshold energy $E_{0.5}$ for which $\mu = 0.5$ (see, e.g., \citet{RyanEtal:1991}). Here, we rather use a Hill function \citep{Hill:1910}, given by
\begin{equation}
\label{eq:hill}
\mu(E_{\unit{cm}}) = 1 - \frac{E_{\unit{cm}}^{n}}{E^{n}_{0.5}+E_{\unit{cm}}^{n}} = \frac{E^{n}_{0.5}}{E^{n}_{0.5}+E_{\unit{cm}}^{n}},
\end{equation}
with the exponent $n$ as a free parameter to describe the dependency of the fragmentation strength (i.e. the relative mass of the largest fragment; see Sect. \ref{sect:fragstren}) on impact energy. The functional form of Eq. \ref{eq:hill} has the advantage of approaching the natural limit of $\mu \rightarrow 1$ (i.e. bouncing) for small impact energies, $E_{\unit{cm}} \ll E_{0.5}$, and a power law $\mu \propto E_{\unit{cm}}^{-n}$ for high energies, $E_{\unit{cm}} \gg E_{0.5}$, and $n > 0$.

Table \ref{tab:results} lists the results for $E_{0.5}$ and $n$; the data on the fragmentation strength and corresponding fit curves are shown in Figure \ref{fig:MuPlt}. Due to the small number of data points in the 1 cm -- 5 cm series, $E_{0.5}$ was estimated to be $E_{0.5} \approx 19$ mJ.

\begin{table}[h!]
	\caption{Results of the aggregate-aggregate collision experiments. The data shown in Figure \ref{fig:MuPlt} were fitted with the function shown in Eq. \ref{eq:hill} and deliver the catastrophic threshold energy $E_{0.5}$ and the exponent $n$. Mind that owing to the small number of data points, the catastrophic threshold energy for the 1.0 cm -- 5.0 cm series was estimated. $E_{0.5}(n=0.55)$ is the catastrophic threshold energy for a fixed exponent of $n=0.55$. The collision strength $Q^*$ is the catastrophic threshold energy (for $n=0.55$) per target mass, which inherits its error from the error in $E_{0.5}(n=0.55)$.
		\label{tab:results}	}
	\begin{center}
		\begin{tabular}{|l|l|l|l|l|l|}	
			\hline
			No. & projectile size & $E_{0.5}$ & Exponent $n$ & $E_{0.5}$ & $Q^*$ \\
			& -- target size & [mJ] & in Eq. \ref{eq:hill} & $(n=0.55)$ & [$\unit{J~kg^{-1}}$] \\
			&&&& [mJ] &\\
			\hline
			1 & 1.0 cm -- 1.0 cm & $9.20\pm 7.70$  & $0.69 \pm 0.36$ & $13.58 \pm 5.57$ & $18.61 \pm 7.63$  \\
			2 & 2.0 cm -- 2.0 cm & $22.52 \pm 7.84$ & $0.70 \pm 0.35$ & $24.82\pm 9.10$ & $4.28 \pm 1.57$ \\
			3 & 5.0 cm -- 5.0 cm & $109.43 \pm 50.21$   & $0.60 \pm 0.34$ & $104.10 \pm 34.06$ & $1.14 \pm 0.37$ \\
			4 & 1.0 cm -- 2.0 cm & $58.24 \pm109.63$ & $0.38 \pm 0.31$ & $30.42 \pm 11.08$   & $5.24 \pm 1.91 $ \\
			5 & 1.0 cm -- 2.6 cm & $15.27 \pm 37.32$  & $0.21 \pm 0.47$ & $8.43 \pm 3.02$ & $0.67 \pm 0.24$ \\
			6 & 1.0 cm -- 5.0 cm & $\sim 19.0 $     & --   & $\sim 19.0 $  & $\sim 0.21$ \\
			7 & 2.0 cm -- 5.0 cm & $53.42 \pm 14.97$    & $0.80 \pm 0.30$ & $59.40 \pm 21.65$ & $0.65 \pm 0.29$\\
			8 & 3.5 cm -- 5.0 cm & $ 117.29 \pm 43.88$    & $0.52 \pm 0.23$ & $114.24 \pm 33.20$ & $1.26 \pm 0.36 $ \\
			\hline
		\end{tabular}
	\end{center}
\end{table}

However, from Table \ref{tab:results}, it can be seen that the exponent $n$ is constrained to values between $n \sim 0.2$ and $n \sim 0.8$ without no obvious dependence on the projectile/target size. Thus, we determined the weighted average of the exponents shown in Table \ref{tab:results} and got $\bar{n} = 0.55 \pm 0.11$. It should be remarked that all values for $n$ in Table \ref{tab:results} are within their individual errors consistent with $n = 0.55$. For this fixed exponent, the resulting energy values $E_{0.5}(n=0.55)$ are also shown in Table \ref{tab:results} and will hereafter be used.

The catastrophic threshold energy varies systematically between $E_{0.5} \sim 10$ mJ for the smaller projectiles/targets and $E_{0.5} \sim 117$ mJ for the larger projectiles/targets. This will be analysed in more detail in Section \ref{sect:Qstar}.

\subsection{\label{sect:Qstar}The collision strength $Q^*$}
If the catastrophic threshold energy $E_{0.5}$ is known, the collision strength $Q^*$, which we define through
\begin{equation}
\label{eq:q*}
Q^* = \frac{E_{0.5}}{m_{\unit{t}}} ,
\end{equation}
can be calculated. Here, we use the target mass for normalisation rather than the total mass of the system, $m_{\unit{t}}+m_{\unit{p}}$, as used by \citet{StewartLeinhardt:2009} and \citet{BeitzMeisneretal:2011}. The reasons for doing this are (1) that in the previous studies the variation in mass ratio between projectile and target was not so extreme, but here it varies by more than two orders of magnitude and (2) that the largest fragment always stems from the target (in the case of equal-mass dust aggregates, the target is the one that delivers the largest fragment). As we are interested in systematically following the fate of the more massive of the colliding dust aggregates, we normalise the catastrophic fragmentation energy to the target mass. Another advantage of doing this is the comparison with different projectile sizes for a constant target size, which can be interpreted as the fragmentation efficiency of the projectile as a function of target mass. The higher $Q^*$, the lower is the fragmentation efficiency of the projectile. The absolute values of $Q^*$ are given in Table \ref{tab:results}.

\begin{figure}[h!]
	\begin{center}
		\includegraphics[width= 180mm,  height = 80mm]{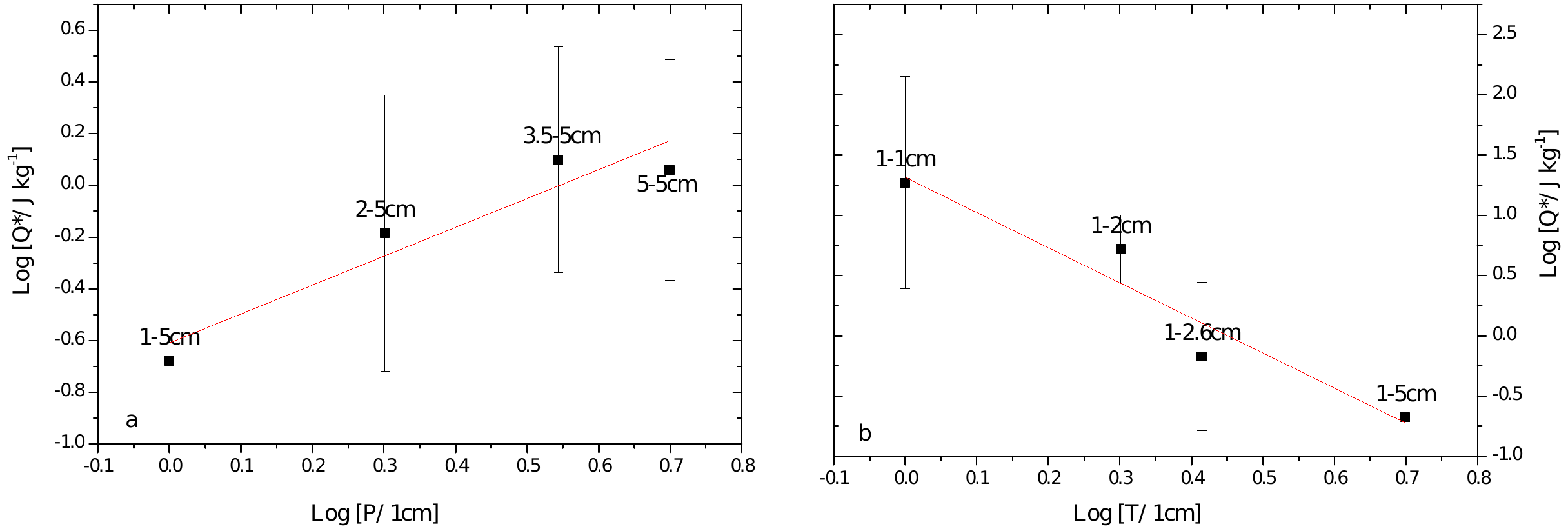}
		\caption{a. Collision strength $Q^*$ as a function of projectile size $P$ for 5-cm targets. The straight line is a linear fit to the data with a slope of 1.12.  b. Collision strength $Q^*$ as a function of target size $T$ for 1-cm projectiles. The straight line is a linear fit to the data with a slope of -2.92. The labelling of the data points represents the series of experiments referring to Table \ref{tab:parameter}.
			\label{fig:pt1}}
	\end{center}
\end{figure}

\begin{figure}[h!]
	\begin{center}
		\includegraphics[width= 120mm,  height = 100mm]{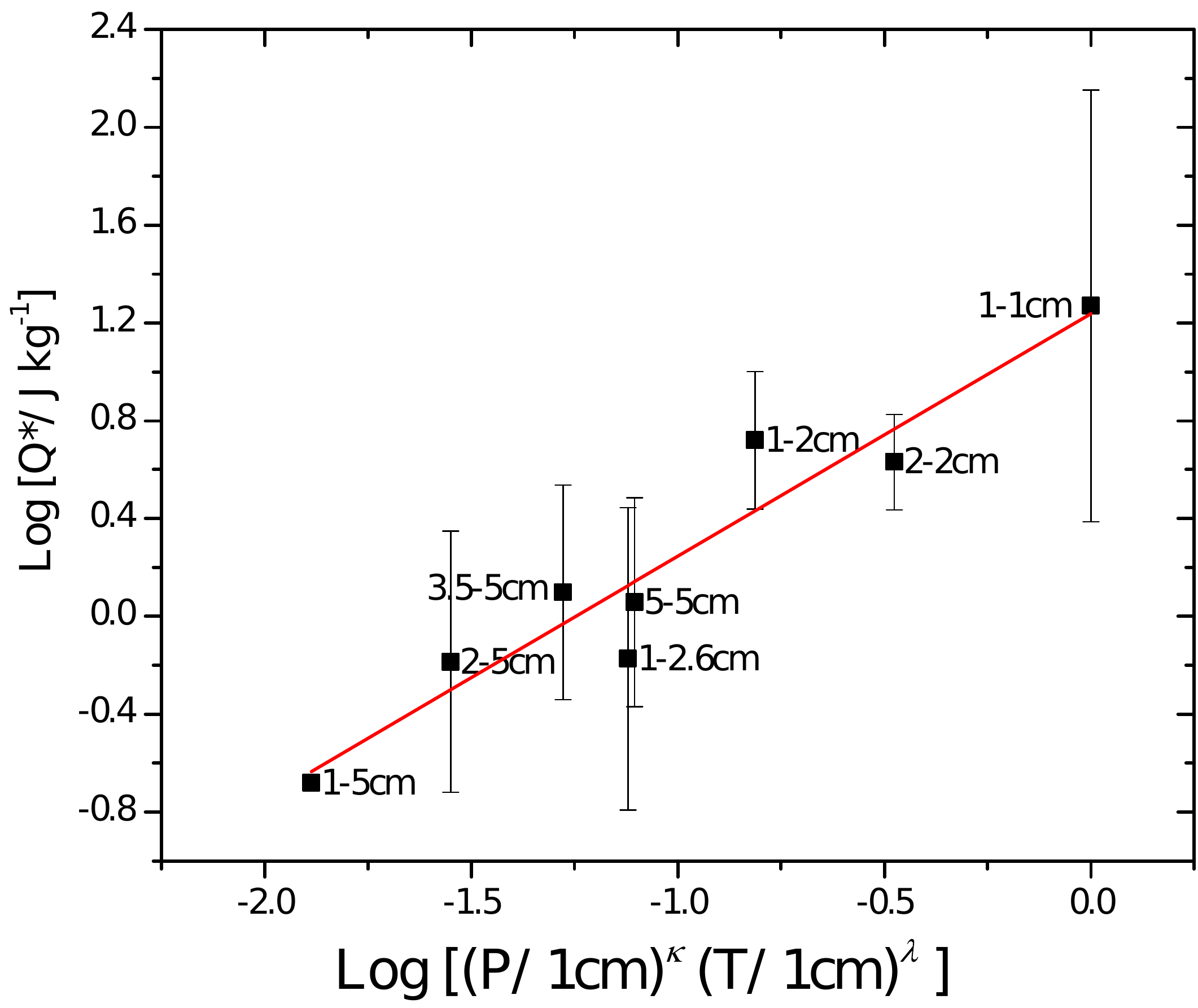}
		\caption{Correlation between collision strength $Q^*$ and projectile size $P$ and target size $T$ in the combination of $P^{\kappa} \cdot T^{\lambda}$ with $\kappa = 1.12$ and $\lambda = -2.70$, which yield the least squares error. Here it should be noticed that inclusion of the 2 cm -- 2 cm series did not have impact on the values of the exponents.	
			\label{fig:pt2}}
	\end{center}
\end{figure}

Figure \ref{fig:pt1} shows $Q^*$ as a function of the projectile size $P$ for a fixed target size of $T = 5$ cm (Figure \ref{fig:pt1}a) as well as a function of the target size $T$ for a fixed projectile size of $P = 1$ cm (Figure \ref{fig:pt1}b). The $Q^*$ value of the 2 cm -- 2 cm series is not shown, as it belongs neither to the fixed projectile ($P = 1$ cm) nor to the fixed target ($T = 5$ cm) parameter space. However this series is included in the collective analysis (Figure \ref{fig:pt2}). As can be seen in Figure \ref{fig:pt1}, the collision strength follows roughly a power law of the projectile size with an exponent of $\sim 1.12 \pm 0.25$ and a power law of the target size with an exponent of $\sim -2.92 \pm 0.57$. Varying the projectile size by a factor of 5 changes the value of $Q^*$ by a factor of $\sim 6$, whereas a variation of the target size by the same factor 5 changes $Q^*$ by a factor of $\sim 100$. The collision strength of a target for a given projectile size thus becomes considerably weaker for increasing target sizes.

Although the kind of data analysis as shown in Figure \ref{fig:pt1} is very intuitive, we favor the simultaneous correlation of $Q^*$ with $P$ and $T$, which also has the advantage that all 8 available $Q^*$ data can be used. Thus, we used the ansatz
\begin{equation}
\label{eq:q*fit}
\log \left(\frac{Q^*(P,T)}{\unit{1~J~kg^{-1}}}\right) = C_{\unit{Q}} + \kappa \log \left(\frac{P}{\unit{1~cm}}\right) + \lambda \log \left(\frac{T}{\unit{1~cm}}\right).
\end{equation}
Least-squares-fitting of all 8 data points of $Q^*$ to the respective $(P,T)$ pairs delivers the coefficients $C_{\unit{Q}} = 1.24 \pm 0.16$, $\kappa = 1.12 \pm 0.35$ and $\lambda = -2.70 \pm 0.37$, slightly off the previously independently determined values, but within the standard errors. The data and fit are shown in Figure \ref{fig:pt2}. We here remind the reader that the series 2 cm -- 2 cm is included. However the values of $\kappa$ and $\lambda$ did not vary significantly, when $Q^*$ was analysed as a function of $P$ and $T$ separately. Figure \ref{fig:pt2} shows the goodness of the fit, with $\log \left(\frac{Q^*(P,T)}{\unit{1~J~kg^{-1}}}\right)$ as y-values and $\log \left( \left(\frac{P}{\unit{1~cm}}\right)^\kappa \cdot  \left(\frac{T}{\unit{1~cm}}\right)^\lambda \right)$ as x-values.

Our results can be applied to the calculation of the catastrophic fragmentation velocity (for which $\mu = 0.5$) if we recall (see Eq. \ref{eq:q*}) that $Q^* = E_{0.5}/m_{\unit{t}} = \frac{1}{2} \frac{m_{\unit{p}} \cdot m_{\unit{t}}}{m_{\unit{p}} + m_{\unit{t}}} v_{0.5}^2 / m_{\unit{t}} \approx \frac{1}{2} m_{\unit{p}} v_{0.5}^2 / m_{\unit{t}} \propto v_{\unit{0.5}} P^3 / T^3$ so that we get, using Eq. \ref{eq:q*fit}
\begin{equation}\label{eq:vcf1}
v_{\unit{0.5}} \propto \left( Q^* \cdot \frac{T^3}{P^3} \right)^{1/2} \propto P^{\frac{\kappa - 3}{2}} \cdot T^{\frac{\lambda + 3}{2}} \propto P^{-0.94} \cdot T^{0.15} .
\end{equation}
For  equal sized dust aggregates, i.e. $P=T$, we get
\begin{equation}\label{eq:vcf2}
v_{\unit{0.5}} \propto T^{\frac{\kappa + \lambda}{2}} \propto T^{-0.79} \propto m_{\unit{t}}^{-0.26} .
\end{equation}

This is a rather weak dependence of the catastrophic fragmentation velocity on aggregate mass for equal-sized collision partners.

\begin{figure}[h!]
	\begin{center}
		\includegraphics[width= 100mm,  height = 80mm]{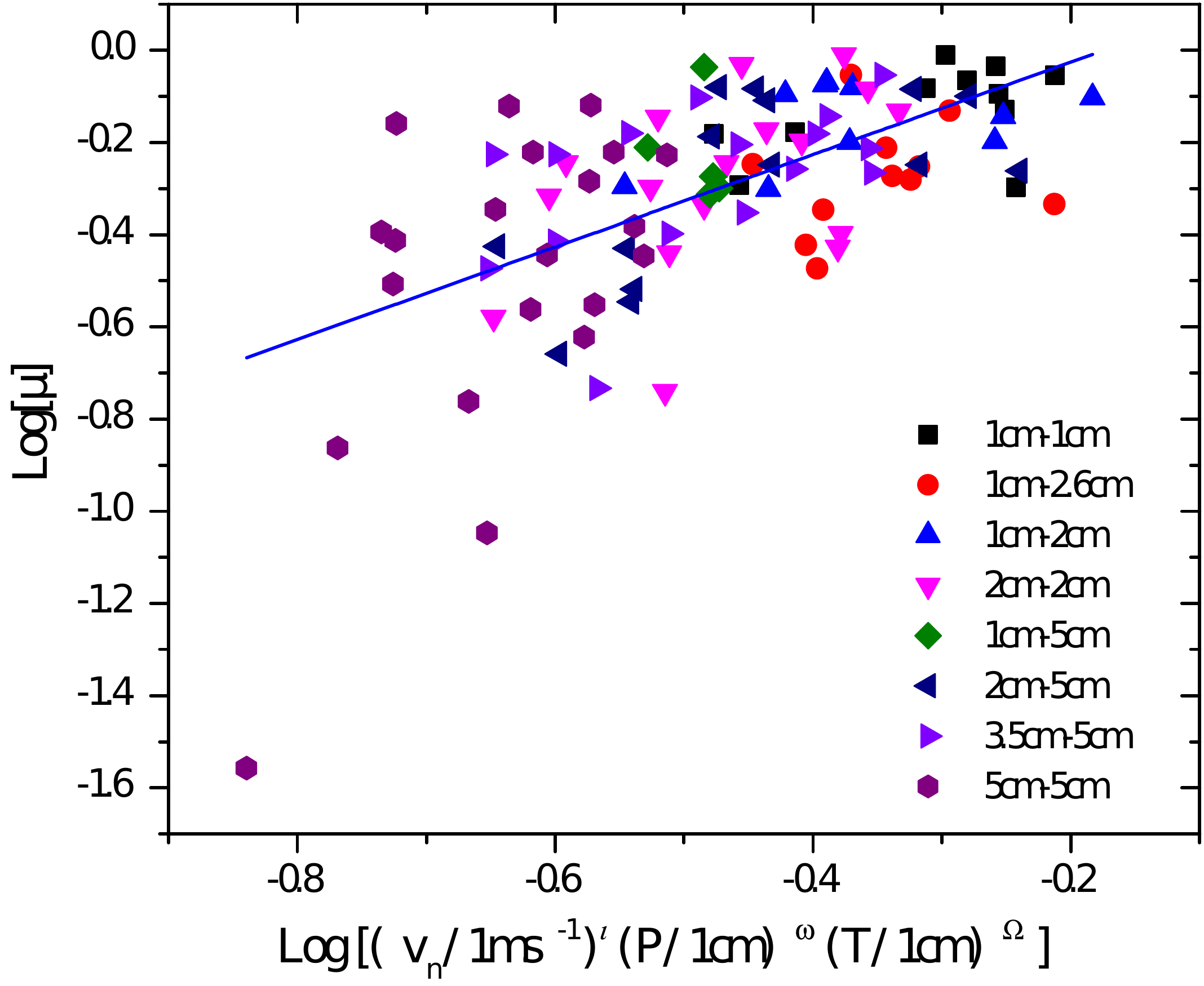}
		\caption{Correlation between the fragmentation strength $\mu$ and projectile/target size and impact velocity in the combination of $v_{\unit{n}}^{\iota} \cdot P^{\omega} \cdot T^{\Omega}$ with $\iota = -0.66$, $\omega = -0.58$ and $\Omega = 0.13$, which yield the least square error.	
			\label{fig:mupowlaw}}
	\end{center}
\end{figure}

As an alternative data analysis of the dependence of the fragmentation strength on impact velocity and projectile/target size, we assume a power-law relationship among these values (and thus circumvent the collision strength) of the following form
\begin{equation}\label{eq:mypowerlaw}
\log \mu = C_{\unit{\mu}} + \iota \log{\left(\frac{v_{\unit{n}}}{\unit{1~m~s^{-1}}}\right)} + \omega \log \left(\frac{P}{\unit{1~cm}}\right) + \Omega \log \left(\frac{T}{\unit{1~cm}}\right).
\end{equation}
Minimising the squared deviations between measured $\mu$ values and those calculated, we get a reasonable fit to Eq. \ref{eq:mypowerlaw} (shown in Figure \ref{fig:mupowlaw}) with $C_{\unit{\mu}} = 0.18 \pm 0.07$, $\iota = -0.66 \pm 0.12$, $\omega = -0.58 \pm 0.10$, and $\Omega = 0.13 \pm 0.11$, respectively, which is only valid as long as $\log \mu \le 0$. This can be applied to derive the general expression for the onset-velocity for fragmentation, for which $\mu = 1$, i.e.
\begin{equation}\label{vfron1}
v_{\unit{1}} \propto \left( P^\omega \cdot T^\Omega \right)^{-1/\iota} = P^{-0.88} \cdot T^{0.20},
\end{equation}
and for $T=P$ we get
\begin{equation}\label{vfron2}
v_{\unit{1}} \propto T^{-0.68} \propto m_{\unit{}t}^{-0.23}.
\end{equation}
As these results also hold for $\mu=0.5$ and, thus, for $v_{0.5}$, this is comparable with the results shown in Eq \ref{eq:vcf2}. It should, however, be mentioned that a power-law description of $\mu$ as shown in Eq. \ref{eq:mypowerlaw} does not adequately describe the asymptotic behavior of the largest fragment mass for small velocities, $\mu \rightarrow 1$. As can be seen in Figure \ref{fig:mupowlaw}, the high-velocity behavior of $\mu$ is not well represented by Eq. \ref{eq:mypowerlaw}. This can also be seen by comparing the velocity dependence $\mu \propto v^{-0.66}_{\unit{n}}$ in Eq. \ref{eq:mypowerlaw} with that of the Hill function (Eq. \ref{eq:hill}). Using $E_{\unit{cm}} \propto v^2_{\unit{n}}$, we get for the asymptotic velocity behavior of the Hill function (Eq. \ref{eq:hill}) $\mu \propto v^{2 n}_{\unit{n}} = v^{-1.1}_{\unit{n}}$. Thus, we think that the velocity dependence of the largest fragment is better described by Eq. \ref{eq:hill}.

It should be mentioned that Eq. \ref{eq:q*fit} can be rewritten for $P=T$ as $\left(\frac{Q^*}{\unit{1~J~kg^{-1}}}\right) = 10^{C_{\unit{Q}}} \left(\frac{T}{\unit{1~cm}}\right)^{\kappa + \lambda} = 17.38 \left(\frac{T}{\unit{1~cm}}\right)^{-1.58}$. This is about one order of magnitude higher than the data given by \citet{BeitzMeisneretal:2011} but with a comparable slope, which \citet{BeitzMeisneretal:2011} give as $-0.95 \pm 0.38$. However, our new results (see Figure \ref{fig:pt2} and Eq. \ref{eq:q*fit}) indicate that $Q^*$ independently depends on both, the projectile and the target size. As far as we know, this aspect has not been described before and has thus not been considered yet in collisional evolution models.

As shown in Eq. \ref{eq:vcf1}, the catastrophic threshold velocity scales with projectile and target size as $v_{\unit{0.5}} \propto P^{\frac{\kappa - 3}{2}} \cdot T^{\frac{\lambda + 3}{2}} \propto P^{-0.94} \cdot T^{0.15}$. This means that smaller projectiles require higher impact velocities to achieve the same collisional result with the same target. On the other hand, for a given projectile size, larger targets require higher impact speeds according to $v_{\unit{0.5}} \propto T^{0.15}$, but with a much shallower size dependence. As a result, in Figure \ref{fig:mtprob} (b) we see that the series where $f = 1$, tend to have lower catastrophic threshold velocity than that of the series where $f > 1$. 
However, as far as the relative strength $Q^{*}$ is concerned, the larger aggregates (of same filling factor) are intrinsically weaker.

\subsection{The fragment size distribution}
A physical model for the fragmentation in aggregate-aggregate collisions requires more than the knowledge of the mass of the largest fragment or the catastrophic fragmentation energy. Thus, we also measured the fragment size distribution in all collisions.

Fragment size distributions of colliding dust aggregates have previously been observed to be composed of two parts, with a high count of smaller fragments following a power law in size-frequency distribution and fewer counts of the largest ones (see, e.g., \citet{BlumMuench:1993}; \citet{DeckersTeiser:2014}). Technically the largest fragment in such collisions is actually the remnant of the original target aggregate. \citet{DeckersTeiser:2014} showed that the mass fraction of the largest fragment decreases with increasing impact energy, in agreement with our findings (see Figure \ref{fig:MuPlt}) and discussions in Section \ref{sect:fragstren}.

In previous experiments, owing to technical limitations, the factors influencing the fragment size distribution have not been fully revealed. But thanks to the high frame rate of 7,500 frames per second of our high-speed cameras (i.e. a temporal resolution of $\sim 130 \unit{\mu s}$), we could trace back the trajectories of the distinguishable fragments and analyse the time evolution of the fragmentation process. In order to count the fragments, a time series of frames was generated with the following two conditions: (i) avoiding secondary collisions among the fragments and between fragments and the drop-tower walls; (ii) following the image sequence at least until the fragment cloud becomes optically thin and the largest fragments become visible, but without violating the first condition. The implementation of these criteria left a small window of time for the determination of the fragment size-frequency distribution, usually about $\sim 50 \unit{ms}$ after the collision.

After processing the selected frames (flat fielding and background subtraction), all distinguishable trajectories of the fragments were traced and counted. Parallel to that, the projected cross-sectional area of each fragment was registered in units of pixels, from the smallest discernable fragments of 1 pixel in cross section to the largest remnant. Here 24 pixels correspond to a linear size of 10 mm and 1 pixel cross section is equivalent to $0.174 \unit{mm}^2$. Hereafter, the fragments were size-sorted according to their cross sections. Thus, any fragment can be counted more than once up to a maximum equal to the number of frames selected. The assumptions which have been made while implementing this method are the following:

\begin{enumerate}
	\item In principle, the fragment size distribution is completed immediately after the collision energy is consumed and the last bond is broken. Secondary collisions are irrelevant.
	\item Small fragments move faster than large ones so that they potentially leave the field of view before the larger fragments become visible in the initially optically thick cloud. This is the reason for also selecting the later frames.
\end{enumerate}

In order to test whether the statistical analysis, based upon the above assumptions applied within the temporal window of the first $\sim 50$ ms, is a realistic approximation of the actual fragment size distribution, we analysed the time variation of the slope of the cumulative area-frequency distribution, $\alpha$. We assume a power-law relation between the cumulative number of fragments with area $x$, $N_{\unit{cum}}(x)$, and the cross-sectional area $x$ of the form
\begin{equation}
N_{\unit{cum}}(x) = \sum_{x'=x_{\unit{max}}}^{x} N(x') = C_{\unit{N}} x^{-\alpha}.
\label{eq:cumdist}
\end{equation}
Here, $C_{\unit{N}}$ is a normalization constant, the bin width of the summation is $\Delta x = 1$ pixel, and the summation starts with the largest discernible fragment of the continuous area-frequency distribution function, $x_{\unit{max}}$, in each image sequence. The latter is not necessarily equal to the projected area of the largest fragment (see Sect. \ref{sect:fragstren}), particularly in the cases where $\mu \approx 1$.

\begin{figure}[h!]
	\begin{center}
		\includegraphics[width= 80mm,  height = 80mm]{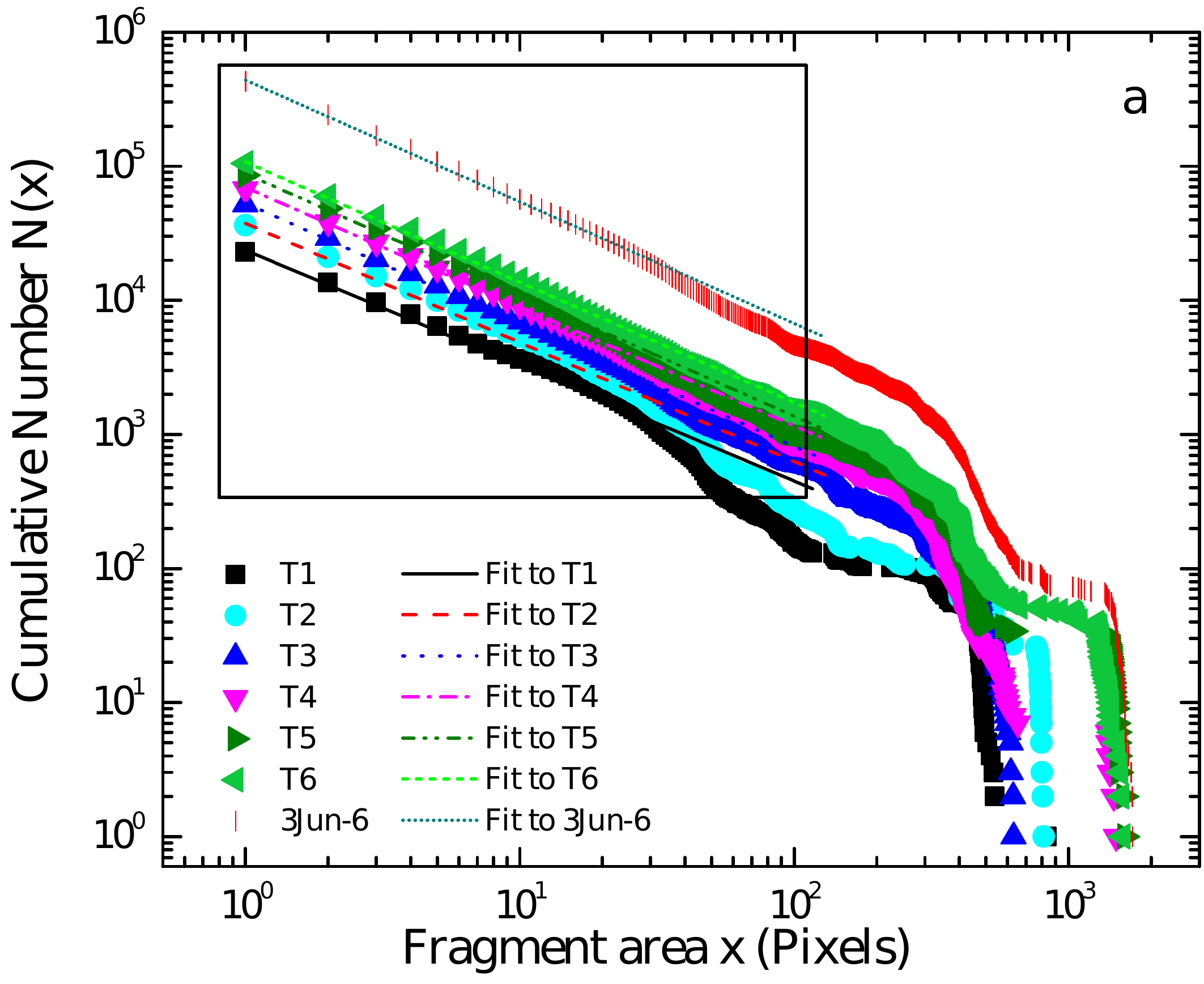}
		\includegraphics[width= 80mm,  height = 80mm]{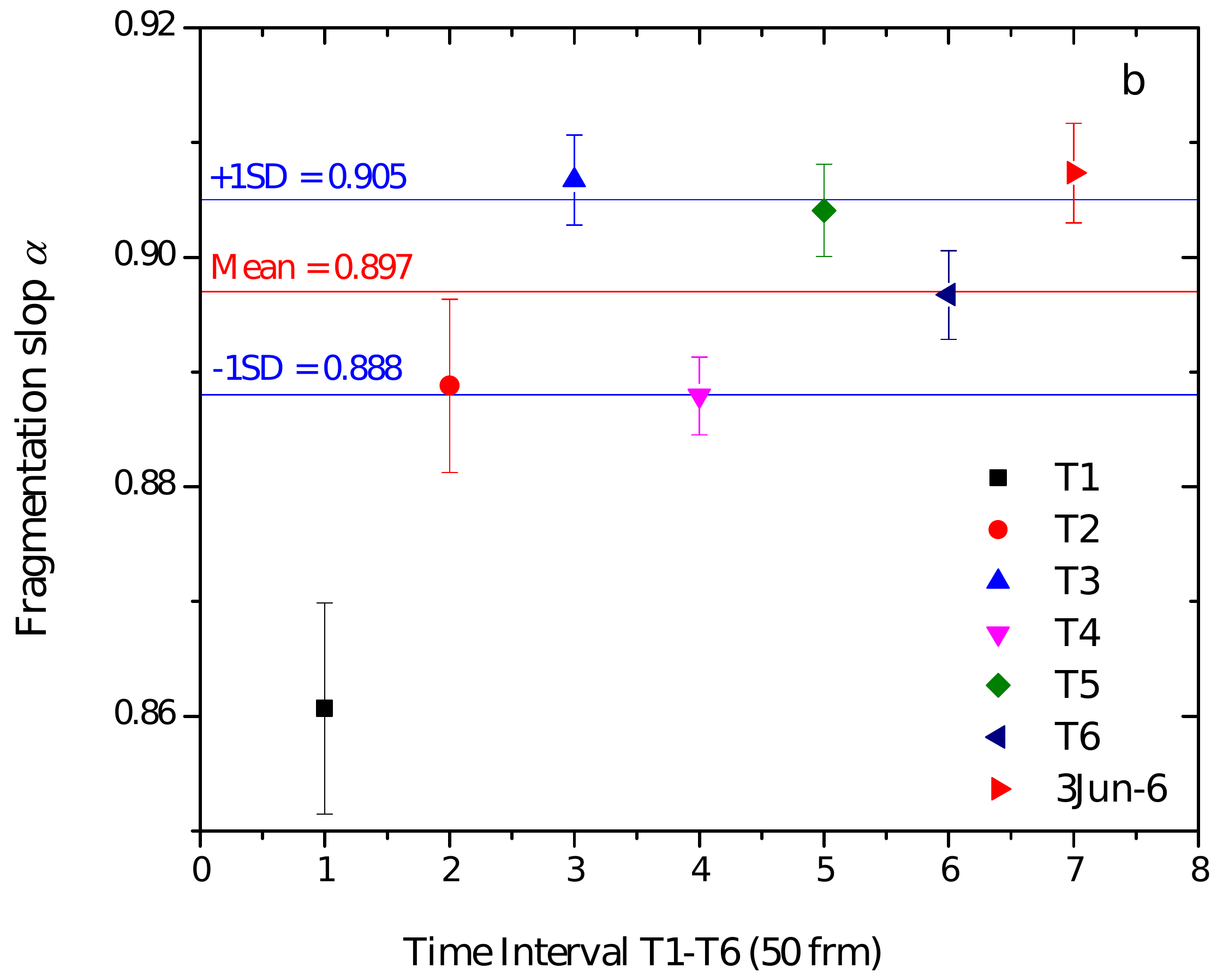}
		\caption{(a). Time series of the single fragmentation curve 3Jun-6, analysed in six equal and non-overlapping time intervals, T1 - T6, of 50 frames each. The red vertical dashes show the cumulative number of fragments of the total sequence. (b). Slopes and error bars of the six time-sorted and the total cumulative area distributions shown as fits to the data points in a. The mean slope of the six individual fits and its standard deviation are shown by the red and blue horizontal lines, respectively. \label{fig:ts1}}
	\end{center}
\end{figure}

For demonstration of our analysis, we randomly selected an experiment, named 3Jun-6 from the 3.5 cm -- 5 cm series. Out of the images of this experiment, six non-overlapping equal time intervals of 50 frames were selected, covering a total time of 40 ms after the collision. We applied a power law of the form presented in Eq. \ref{eq:cumdist} to the area range depicted in Figure \ref{fig:ts1}a with the rectangular box, in order to avoid small-number effects with the largest fragments. Here the curve of black squares T1 and the curve of green triangles T6 represent the first and the last time interval, respectively. For reference, the complete cross-section data of the 3Jun-6 experiment (comprising all data from the six sequences T1-T6) are also plotted, here represented by red vertical dashes. As the time after the collision elapses from T1 through T6, the cumulative count of each fragment bin increases, which shifts each subsequent curve upward. At the same time, the larger fragments get separated from the power-law tail of small fragments and become countable, so that the curves shift rightwards to higher cross sections. However, the slope $\alpha$ (see Eq. \ref{eq:cumdist}) does not change much over time as Figure \ref{fig:ts1}b demonstrates. With the exception of T1, the slopes remain within a narrow interval of $\bar{\alpha} = 0.897 \pm 0.0086$, determined by averaging the five $\alpha$ values for the intervals T2-T6. This mean slope corresponds very well to the reference slope $\alpha = 0.907$ of 3Jun-6 for full time interval. From this analysis, we conclude that the slope of the power-law part of the cumulative area-frequency distribution of fragments can be derived from the full interval of images. In other words, the slope of the fragment area distribution, observed at any instance after the collision, remains almost the same.

\begin{figure}[h!]
	\begin{center}
		\includegraphics[width= 140mm,  height = 120mm]{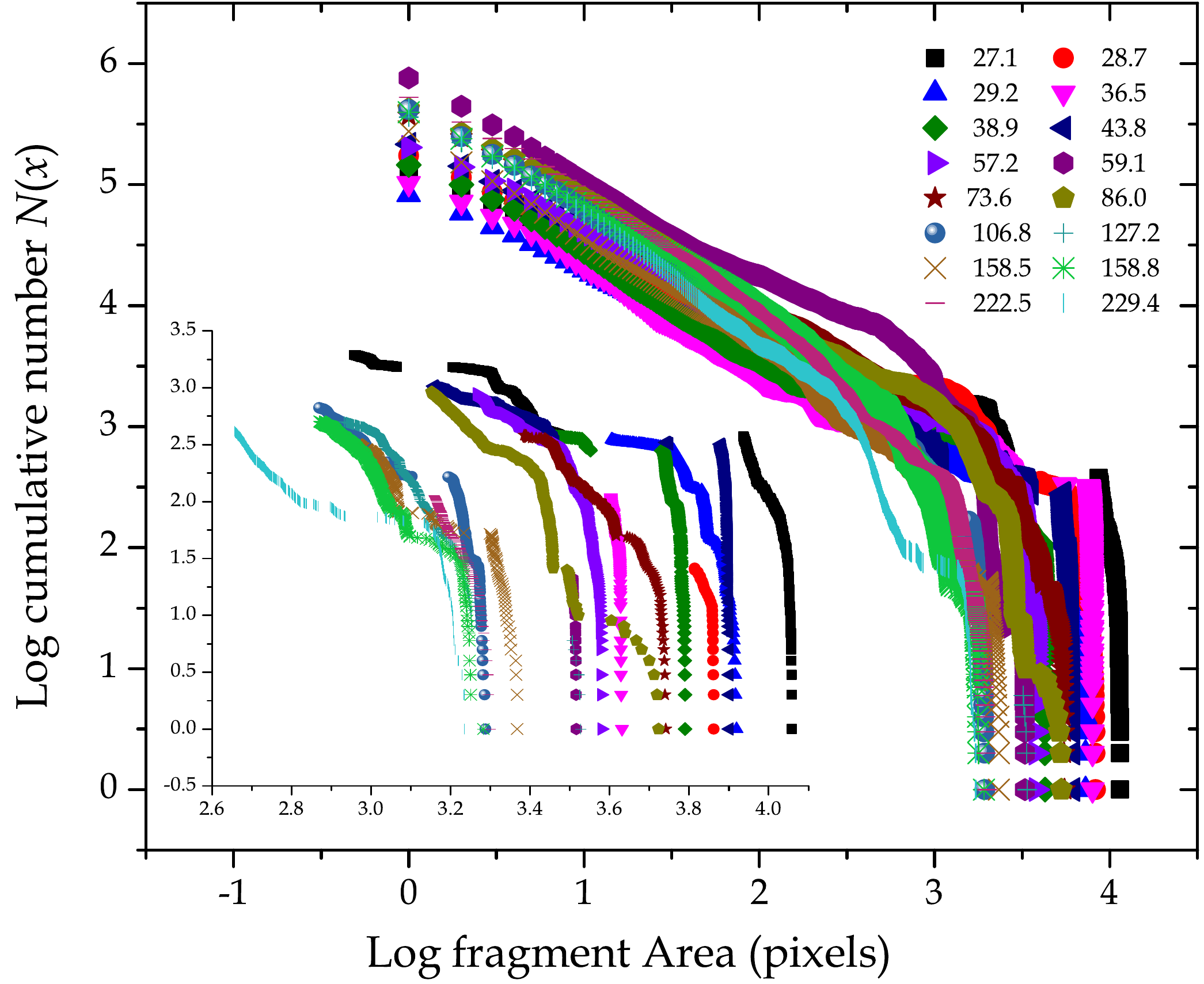}
		\caption{Cumulative area-frequency distributions of the fragments in all 16 experiments of the 3.5 cm -- 5 cm series. The larger fragments in the distributions are responsible for the vertical features shown in the inset, which is a zoom into the data at the lower right end of the cumulative area-frequency distribution. The symbols are sorted with respect to collision energy (in mJ) as shown in the legend. \label{fig:sd1}}
		\end{center}
	\end{figure}

In Figure \ref{fig:sd1}, we show all 16 size-frequency distributions in the 3.5 cm -- 5 cm series derived for full time intervals of $\sim 50$ ms and the assumptions mentioned on page 28. The numbers next to the symbols in the legend indicate the collision energy in units of mJ.

	We have seen above that the cumulative area-frequency distribution of the fragments are well represented by a power law for small fragments (see Eq. \ref{eq:cumdist}), with a rather sharp cut-off at the high-mass end. In order to mathematically describe the cumulative area-frequency distribution for the full range of fragment sizes, we used an exponential cut-off of the form
	\begin{equation}
	N_{\unit{cum}}(x) = \sum_{x'=x_{\unit{max}}}^{x} N(x') = C_{\unit{N}} x^{-\alpha} e^{-\left({\frac{x}{x_{\unit{i}}}}\right)^{\nu}}.
	\label{eq:cumdistexp}
	\end{equation}
	As the area of the fragments increases, $N_{\unit{cum}}(x)$ declines with a slope $-\alpha$, up to about the critical fragment area $x \approx x_{\unit{i}}$, the knee of the distribution, which indicates the end of the continuous regime of the power law. Above $x \approx x_{\unit{i}}$, the cumulative count drops exponentially with an exponent $-\left({\frac{x}{x_{\unit{i}}}}\right)^{\nu}$. Fitting the experimental data to Eq. \ref{eq:cumdistexp} using the four fit parameters $C_{\unit{N}}$, $\alpha$, $x_{\unit{i}}$, and $\nu$, respectively, shows that $\nu \approx 2$ for all data sets. The slope varies between $\alpha \approx 0.2$ and $\alpha \approx 2$ and the critical fragment area ranges between $x_{\unit{i}} \approx 50$ pixels and $x_{\unit{i}} \approx 10,000$ pixels. In the Electronic Appendix 1, we show all 142 size-frequency distributions of all eight series with their respective fit functions according to Eq. \ref{eq:cumdistexp}.

	As can be seen in Figure \ref{fig:sd1}, the curves are sorted with respect to the collision energy. The more energetic collisions tend to possess steeper slopes and smaller critical fragment areas. To derive the dependency of $\alpha$ on the projectile and target size as well as on the impact velocity, we fitted the 141 data points (1 point being an outlier was not included) to the function
	\begin{equation}
	\alpha = C_{\unit{\alpha}} + \delta \log{\left(\frac{v_{\unit{n}}}{\unit{1~m~s^{-1}}}\right)} + \epsilon \log{\left(\frac{P}{\unit{1~cm}}\right)} + \psi \log{\left(\frac{T}{\unit{1~cm}}\right)} .
	\label{eq:alphamin}
	\end{equation}
	Minimising the residuals, we get $C_{\unit{\alpha}} = 0.15 \pm 0.09$, $\delta=1.01 \pm 0.14$, $\epsilon=-0.02 \pm 0.11$, and $\psi=0.36 \pm 0.12$, respectively, with $\chi_{\unit{red}}^2 = 0.066$. Obviously, the dependence on $P$ is very weak and statistically not significant. A variation of $P$ by a factor 5 results in a maximum deviation of $\sim 0.01$ in $\alpha$, which is very small compared to the range of slopes.
	Here we shall also like to show that if only collision velocity is considered it increases the $\chi_{\unit{red}}^2$. For this we set $\epsilon$ and $\psi$ equal to zero in Eq. \ref{eq:alphamin} and we get
		\begin{equation}
		\alpha = C_{\unit{\alpha}} + \delta \log{\left(\frac{v_{\unit{n}}}{\unit{1~m~s^{-1}}}\right)} ,
		\label{eq:alphamin2}
		\end{equation}

%%	\begin{equation}
%%	\alpha = C_{\unit{\alpha}} + \delta \log{\left(\frac{v_{\unit{n}}}{\unit{1~m~s^{-1}}}\right)} +\psi \log{\left(\frac{T}{\unit{1~cm}}\right)} ,
%%	\label{eq:alphamin2}
%%	\end{equation}

which yielded $C_{\unit{\alpha}} = 0.30 \pm 0.07$, $\delta=1.06 \pm 0.12$ with $\chi_{\unit{red}}^2 = 0.071$, slightly higher than when using Eq. \ref{eq:alphamin}.
	
\begin{figure}[h!]
		\begin{center}
			\includegraphics[width= 160mm,  height = 80mm]{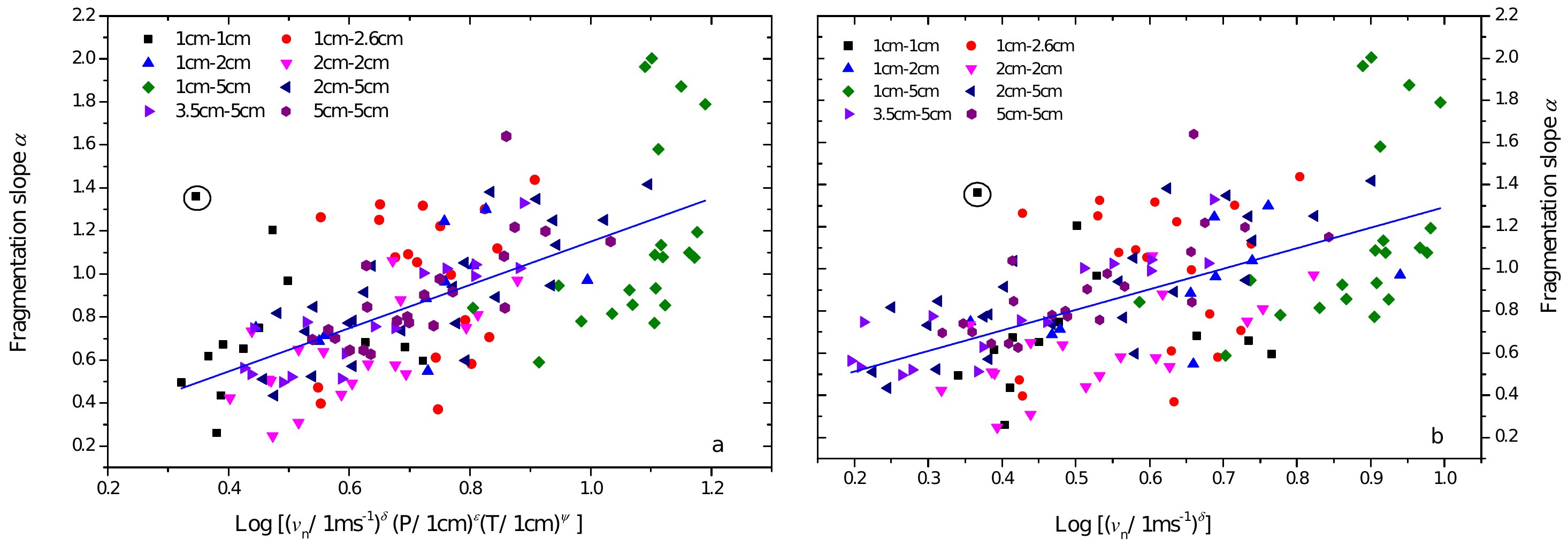}
			\caption{(a) Slope of the area-frequency distribution function,$\alpha$, introduced in Eq. \ref{eq:cumdistexp}, as a function of collision parameters. The blue solid line shows the fit to the data following Eq. \ref{eq:alphamin} with $\delta=1.02$, $\epsilon= -0.02$ and $\psi=0.36$. (b) Same as (a), but assuming that $\alpha$ depends only on the collision velocity. The blue line shows the fit to Eq. \ref{eq:alphamin2} with $\delta=1.06$. In both cases, one data point (encircled) from the series 1 cm -- 1 cm has been ignored during fitting process.
				\label{fig:av1}}
		\end{center}
	\end{figure}

In Figure \ref{fig:av1}(a), we show the best-fit $\alpha$ values of all 141 impact experiments as a function of collision velocity and target size according to Eq. \ref{eq:alphamin2}, whereas the dependence of $\alpha$ on velocity according to Eq. \ref{eq:alphamin2}, is shown Figure \ref{fig:av1}(b) (one outlier data point from the 1 cm -- 1 cm series was not used in both fits; this data point is circled).

\begin{figure}[h!]
	\begin{center}
	\includegraphics[width= 140mm,  height = 120mm]{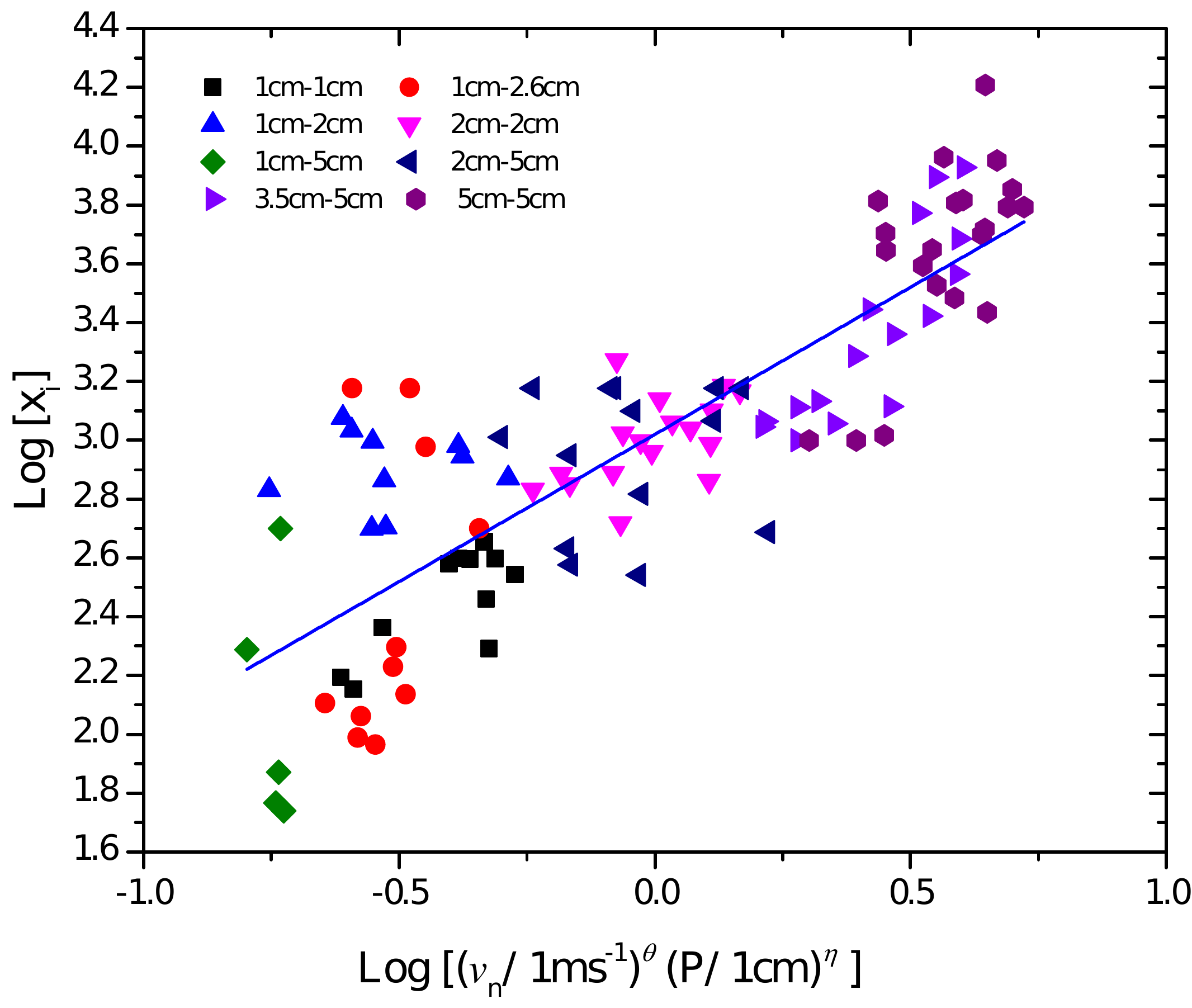}
	\caption{Correlation between the cut-off area $x_{\unit{i}}$ in the cumulative area-frequency distribution function as introduced in to Eq. \ref{eq:cumdistexp} and $v_{\unit{n}}^{\theta} P^{\eta}$, with $\theta = -0.85$ and $\eta = 1.40$. 	
	\label{fig:xifrag}}
	\end{center}
\end{figure}

	In addition to $\alpha$ we have also analysed the critical fragment size, the cut-off size of the area-frequency distribution function $x_{\unit{i}}$. This is the size of $x_{\unit{i}}$ which determines the boundary between the continuous and the discrete size distribution. Therefore we are interested in seeing its dependence on collision parameters. We realise that in the case of mass transfer (37 cases in which the target stayed intact and only the projectile had fragmented), $x_{\unit{i}}$ exclusively belongs to projectile, therefore we eliminate 37 events of mass-transfer and consider the remaining 105 events of complete fragmentation of projectile and target. We fitted the $x_{\unit{i}}$ data to a combined power law
		\begin{equation}
	\log{x_{\unit{i}}} = C_{\unit{x}} + \theta \log{\left(\frac{v_{\unit{n}}}{\unit{1~m~s^{-1}}}\right)} + \eta \log{\left(\frac{P}{\unit{1~cm}}\right)} + \tau \log{\left(\frac{T}{\unit{1~cm}}\right)}
	\label{eq:ximin}
	\end{equation}
	and got $C_{\unit{x}} = 3.03 \pm 0.11$, $\theta = -0.83 \pm 0.19$, $\eta = 1.44 \pm 0.17$, and $\tau = -0.05 \pm 0.17$, respectively, with $\chi_{\unit{red}}^2 = 0.087$. As the dependence on target size turns out to be negligible, we re-fitted Eq. \ref{eq:ximin} for $\tau = 0$, which leads to
		\begin{equation}
			\log{x_{\unit{i}}} = C_{\unit{x}} + \theta \log{\left(\frac{v_{\unit{n}}}{\unit{1~m~s^{-1}}}\right)} + \eta \log{\left(\frac{P}{\unit{1~cm}}\right)}
		\label{eq:ximin2}
		\end{equation}
and got new fit values $C_{\unit{x}} = 3.02 \pm 0.11$, $\theta = -0.85 \pm 0.17$ and $\eta = 1.40 \pm 0.11$ with $\chi_{\unit{red}}^2 = 0.085$, which are slightly better than before. Figure \ref{fig:xifrag} shows the dependence of $x_{\unit{i}}$ on $v_{\unit{n}}$ and $P$.
	
	Contrary to the results for the slope of the area-frequency distribution function, the largest fragment area of the continuous distribution is strongly dependent on the projectile size, but not at all on the target size. If we recall that $x_{\unit{i}}$ is the cross section of the cut-off fragment and $P$ is the diameter of the projectile, Eq. \ref{eq:ximin} shows that the size of the cut-off fragment almost linearly scales with the projectile size. Thus, the cut-off size of the continuous area-frequency distribution function is dominated by the contribution of the fragmenting projectile, which is also seen by the vanishing dependence on the target size. Figure \ref{fig:xifrag} demonstrates the excellent correlation of $x_{\unit{i}}$ with $v_{\unit{n}}^{\theta} P^{\eta} T^{\tau}$. As the collision velocity increases, the size of the cut-off area decreases, as expected.
	
	%% In contrast, taking into account only those cases in which the target stayed intact, applying Eq. \ref{eq:ximin} results in $C_{\unit{x}} = 2.60 \pm 0.21$, $\theta = -0.28 \pm 0.38$,  $\eta = 2.98\pm 0.60$, and $\tau = -1.15 \pm 0.34$, respectively, with $\chi_{\unit{red}}^2 = 0.082$ (see Figure \ref{fig:ximt}). Here, the dependence on impact velocity has virtually vanished, the rough proportionality between projectile size and cut-off fragment size is still present, but the target size is now negatively correlated to the size of the largest fragment of the continuous distribution.
	
%%\begin{figure}[h!]
%%	\begin{center}
%%	\includegraphics[width= 120mm,  height = 100mm]{Fig17-Xi-PT.pdf}
%%	\caption{Correlation between the cut-off area $x_{\unit{i}}$ in the cumulative area-frequency distribution function according to Eq. \ref{eq:cumdistexp}, and $v_{\unit{n}}^{\theta} P^{\eta} T^{\tau}$, with $\theta = -0.28$, $\eta = 2.98$ and $\tau = -1.15$ for all 37 experiments in which the target remained intact.	
%%				\label{fig:ximt}}
%%		\end{center}
%%	\end{figure}
	
%%	From these data, we can conclude that we have to distinguish between mass transfer (target remains intact) and fragmentation of the target, as concerns the cut-off area of the continuous part of the cumulative area-frequency distribution function. When mass transfer can be ignored and both, projectile and target, fragment in a collision, the influence of the projectile on the size of the largest fragment of the continuous distribution is much stronger than that of the target.
	
\subsection{Fragment mass distribution}
In the next step, we analysed the mass-frequency distribution of the fragments, which is required for the implementation of the model, which is discussed in Paper II. Here, we will derive a normalised distribution, required for the modelling (details in Section \ref{sect:fragmod}). In a first step, we derive the mass $m_{\unit{f}}(x)$ of an individual fragment with measured cross-sectional area $x$, assuming spherical aggregate-fragments with a mass density $\rho_{\unit{agg}}$ (see Section \ref{sect:samples}). Using this approach, we get $m_{\unit{f}} = 5.46 \cdot 10^{-5} \unit{cm^3} \rho_{\unit{agg}} x^{3/2}$, with $\rho_{\unit{agg}} = \phi \rho_{\unit{SiO_2}} = 0.91 \unit{g~cm^{-3}}$, $\phi = 0.35$, $x$ in pixels and $m_{\unit{f}}$ in grams. Assuming a simple power law for the cumulative area-frequency distribution, as shown in Eq. \ref{eq:cumdist}, we can derive for the cumulative mass distribution function
	
	\begin{equation}\label{eq:massdistpowlaw}
	M_{\unit{cum}}(m_{\unit{f}})  = \left\{
	\begin{array}{lcr}
	M_{\unit{tot}} \left( 1 - \left( \frac{m_{\unit{f}}}{m_{\unit{max}}} \right) ^{\beta} \right) & \unit{for} & \beta > 0\\
	M_{\unit{tot}} \left( \frac{m_{\unit{max}}}{m_0} \right)^{\beta} \left(\left( \frac{m_{\unit{f}}}{m_{\unit{max}}} \right) ^{\beta} - 1 \right) & \unit{for} & \beta < 0 ,
	\end{array}
	\right.
	\end{equation}
	
	with $\beta = 1 - \frac{2 \alpha}{3}$ and $m_{\unit{max}}$ being the maximum fragment mass of the continuous mass-frequency distribution function. The functional conversion of cross section into mass for the truncated power law as shown in Eq. \ref{eq:cumdistexp} will be presented in Paper II.
	
	As $\alpha < 3/2$ for the majority of the cases considered here (see Figure \ref{fig:av1}), and thus $\beta > 0$, $M_{\unit{cum}}(m)$ asymptotically approaches a constant total mass $M_{\unit{tot}}$ for $m \rightarrow 0$ with no need to insert a smallest fragment mass. Thus, Eq. \ref{eq:massdistpowlaw} is a good fit to the fragment mass distribution function as long as the deviations from the power law (compare Eqs. \ref{eq:cumdist} and \ref{eq:cumdistexp}) are small. In those cases for which $\alpha > 3/2$, particularly for the highest experimental velocities and for reasons of extrapolating to even higher impact energies, Eq. \ref{eq:massdistpowlaw} is still valid, but $M_{\unit{cum}}(m_{\unit{f}}) \rightarrow \infty$ for $m_{\unit{f}} \rightarrow 0$ so that a reasonable smallest fragment mass, $m_0$, needs to be inserted to keep the total mass finite. Thus, for $\beta < 0$, the fragment mass must be constrained to $m_0 \le m_{\unit{f}} \le m_{\unit{max}}$.
	
\begin{figure}[htp]
	\begin{center}
	\includegraphics[height = 80mm]{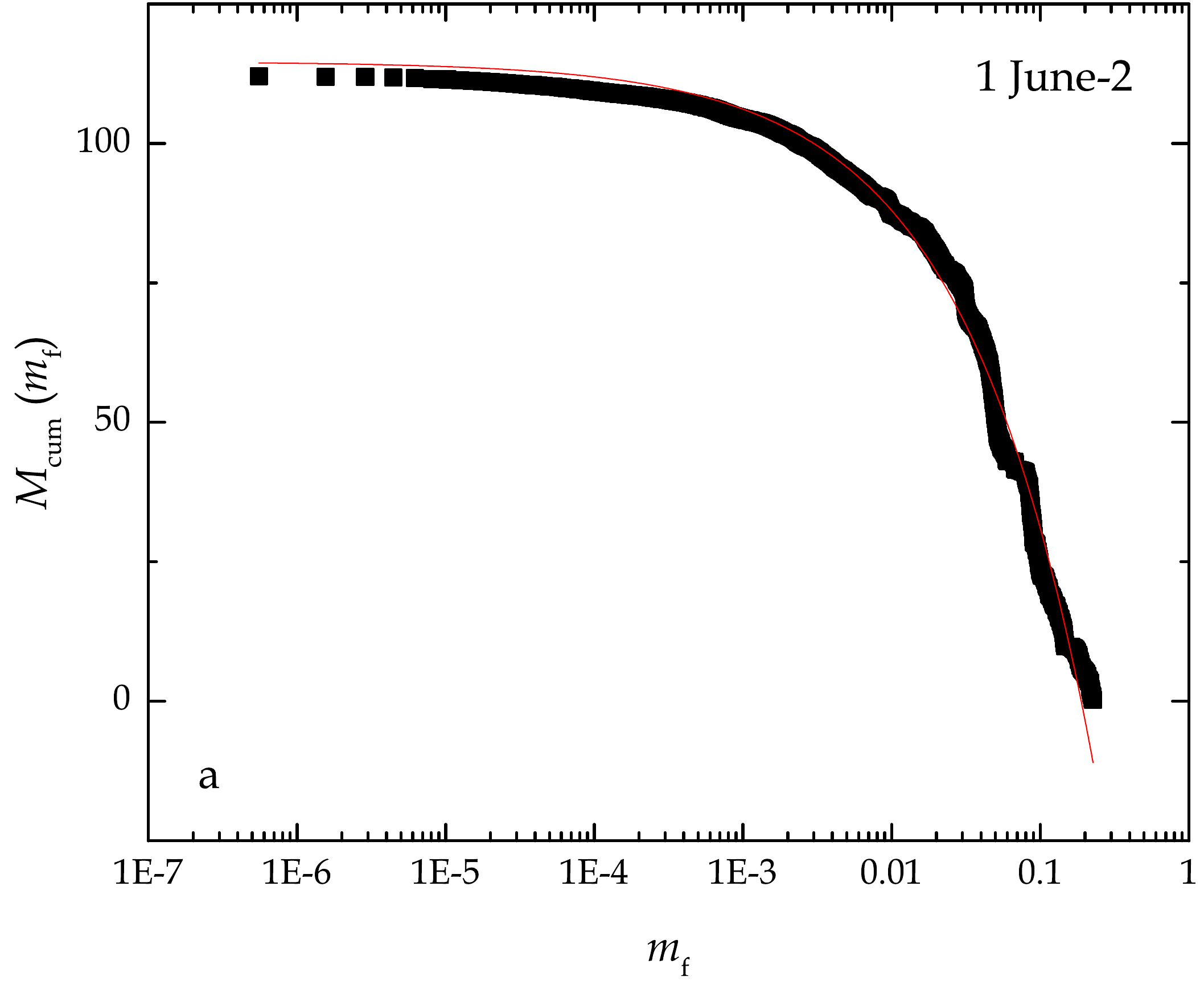}
	\includegraphics[height = 80mm]{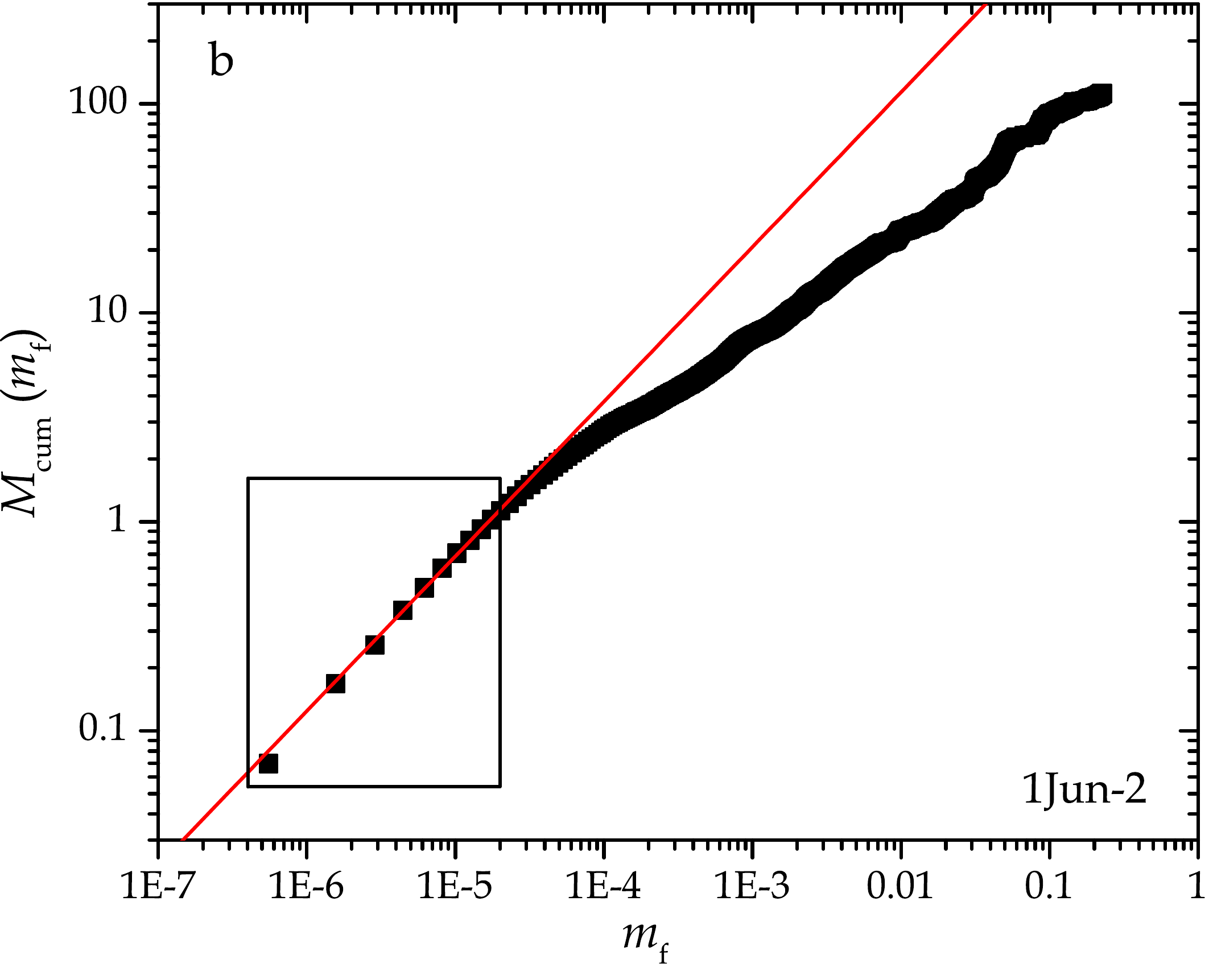}
	\caption{a. Example of a fit of Eq. \ref{eq:massdistpowlaw} to the measured cumulative mass-frequency distribution function. Here, a collision from the 3.5 cm -- 5 cm series is shown. The free parameters of the fit were $m_{\unit{max}}$ and $M_{\unit{tot}}$, with $\beta$ derived from the earlier determined slope $\alpha$ of the area-frequency distribution function by $\beta = 1 - \frac{2 \alpha}{3}$. b. Same as a. but with a cumulative mass-frequency distribution function started at the smallest mass and a log-log display. The apparent power-law start of the cumulative mass-frequency distribution function is shown by the solid line, which was fitted to the first ten data points (as shown by the box).
    \label{fig:exmassfit}}
	\end{center}
\end{figure}

\begin{figure}[h!]
	\begin{center}
	\includegraphics[width= 120mm,  height = 100mm]{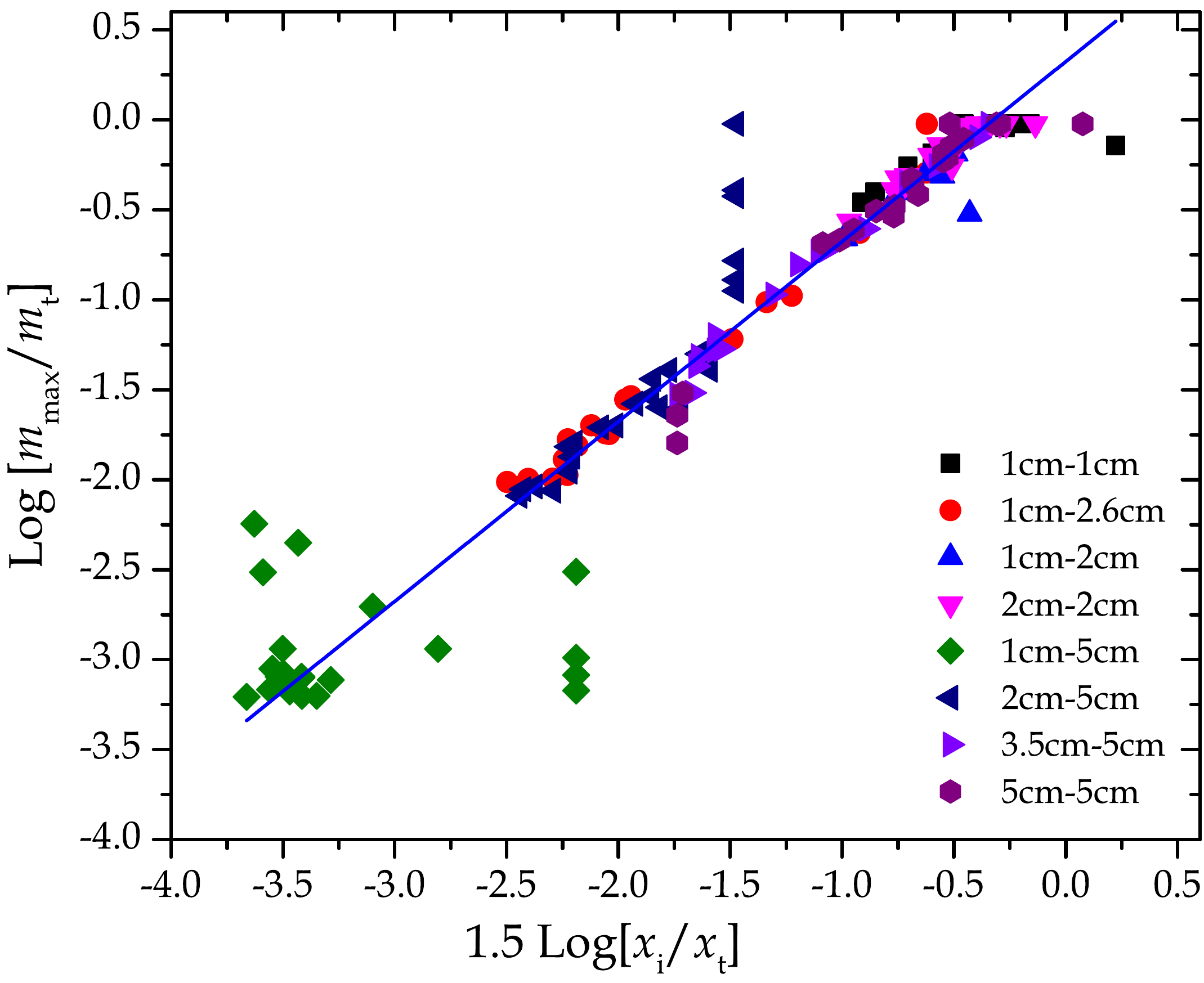}
	\caption{Correlation between $m_{\unit{max}}/m_{\unit{t}}$, derived from fitting the experimental cumulative mass-frequency distribution to Eq. \ref{eq:micorr}, and the cut-off area $x_{\unit{i}}$ from Eq. \ref{eq:cumdistexp}, normalised to the corresponding target value, $x_{\unit{t}}$.
	\label{fig:xmxi}}
    \end{center}
\end{figure}

To determine $m_{\unit{max}}$ in Eq. \ref{eq:massdistpowlaw} and correlate it with $x_{\unit{i}}$ in Eq. \ref{eq:cumdistexp}, we fitted Eq. \ref{eq:massdistpowlaw} to the cumulative mass-frequency distribution of all experiments, which we derived from the corresponding cumulative area-frequency distributions, with only $m_{\unit{max}}$ and $M_{\unit{tot}}$ as fit parameters and $\beta$ derived from the respective $\alpha$ through $\beta = 1 - \frac{2 \alpha}{3}$. In Figure \ref{fig:exmassfit}a, we show an example of the fit of Eq. \ref{eq:massdistpowlaw} to the data and in the Electronic Appendix 2, we present all 142 experiments with their corresponding fit functions. Figure \ref{fig:xmxi} shows the relation between the fit value of $m_{\unit{max}}$ and the cut-off area $x_{\unit{i}}$ from Eq. \ref{eq:cumdistexp}, with both values having been normalised to their respective target values $m_{\unit{t}}$ and $x_{\unit{t}}$, respectively. As expected, these values correlate such that we can state that
\begin{eqnarray}
	\log{\left(\frac{m_{\unit{max}}}{m_{\unit{t}}} \right)} & = & C_{\unit{m}} + 1.5 \log{\left(\frac{x_{\unit{i}}}{x_{\unit{t}}} \right)} \\
	\frac{m_{\unit{max}}}{m_{\unit{t}}} & = & 10^{C_{\unit{m}}} \cdot \left(\frac{x_{\unit{i}}}{x_{\unit{t}}} \right)^{1.5}
	\label{eq:micorr}
	\end{eqnarray}
where $C_{\unit{m}} = 0.32 \pm 0.02$.

\begin{figure}[h!]
	\begin{center}
	\includegraphics[width= 140mm,  height = 120mm]{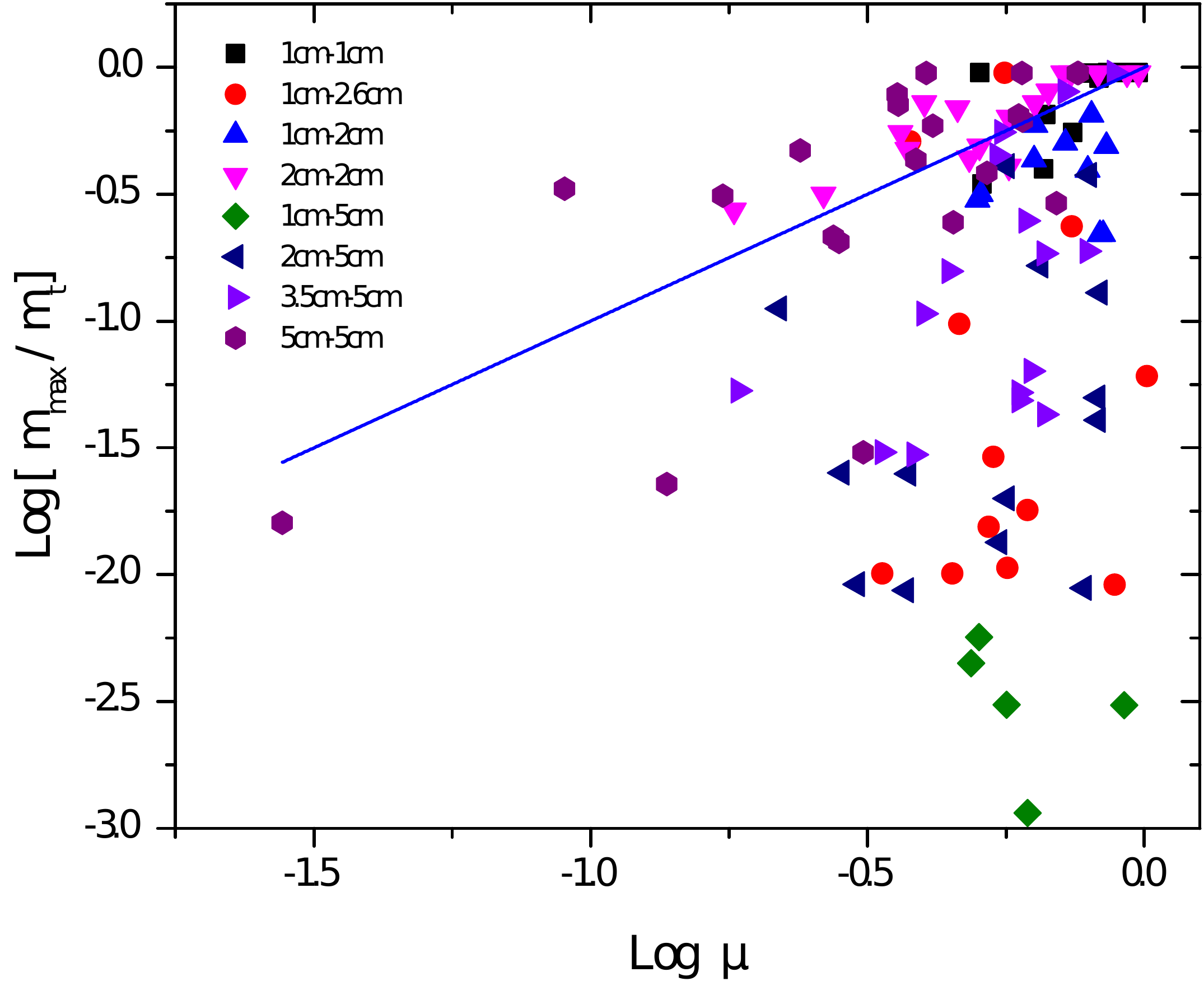}
	\caption{Absence of correlation between $m_{\unit{max}}/m_{\unit{t}}$ and the fragmentation strength $\mu$ for all the cases in which $\mu < 1$. The solid line denotes $m_{\unit{max}}/m_{\unit{t}} = \mu$.
	\label{fig:mmaxmu}}
	\end{center}
\end{figure}

Finally, we investigated the correlation between the normalised cut-off value of the cumulative fragment mass distribution,  $m_{\unit{max}}/m_{\unit{t}}$, derived from Eq. \ref{eq:micorr} and the mass of the largest fragment $\mu$ from Eq. \ref{eq:mul} and found that there is no such correlation (see Figure \ref{fig:mmaxmu}). Thus, the cut-off mass and the mass of the largest fragment are independent and should be treated separately. As generally $m_{\unit{max}}/m_{\unit{t}} \le \mu$, this also means that the continuum part of the fragment mass-frequency distribution, whose upper mass is $m_{\unit{max}}$, and the largest fragment, whose mass is $\mu \cdot m_{\unit{t}}$, are distinct for the projectile/target sizes and impact velocities treated in this study. From their scaling behaviour, i.e. $m_{\unit{max}} \propto \left( x_{\unit{i}} \right)^{3/2} \propto v_{\unit{n}}^{-1.23}$ (Eq. \ref{eq:ximin}) and $\mu \propto v_{\unit{n}}^{-1.1}$ (Eq. \ref{eq:hill} with $n=0.55$) or $\mu \propto v_{\unit{n}}^{-0.66}$ (Eq. \ref{eq:mypowerlaw}), respectively, we can argue that this is also the case for higher impact speeds.

\begin{figure}[h!]
	\begin{center}
	\includegraphics[width= 120mm,  height = 100mm]{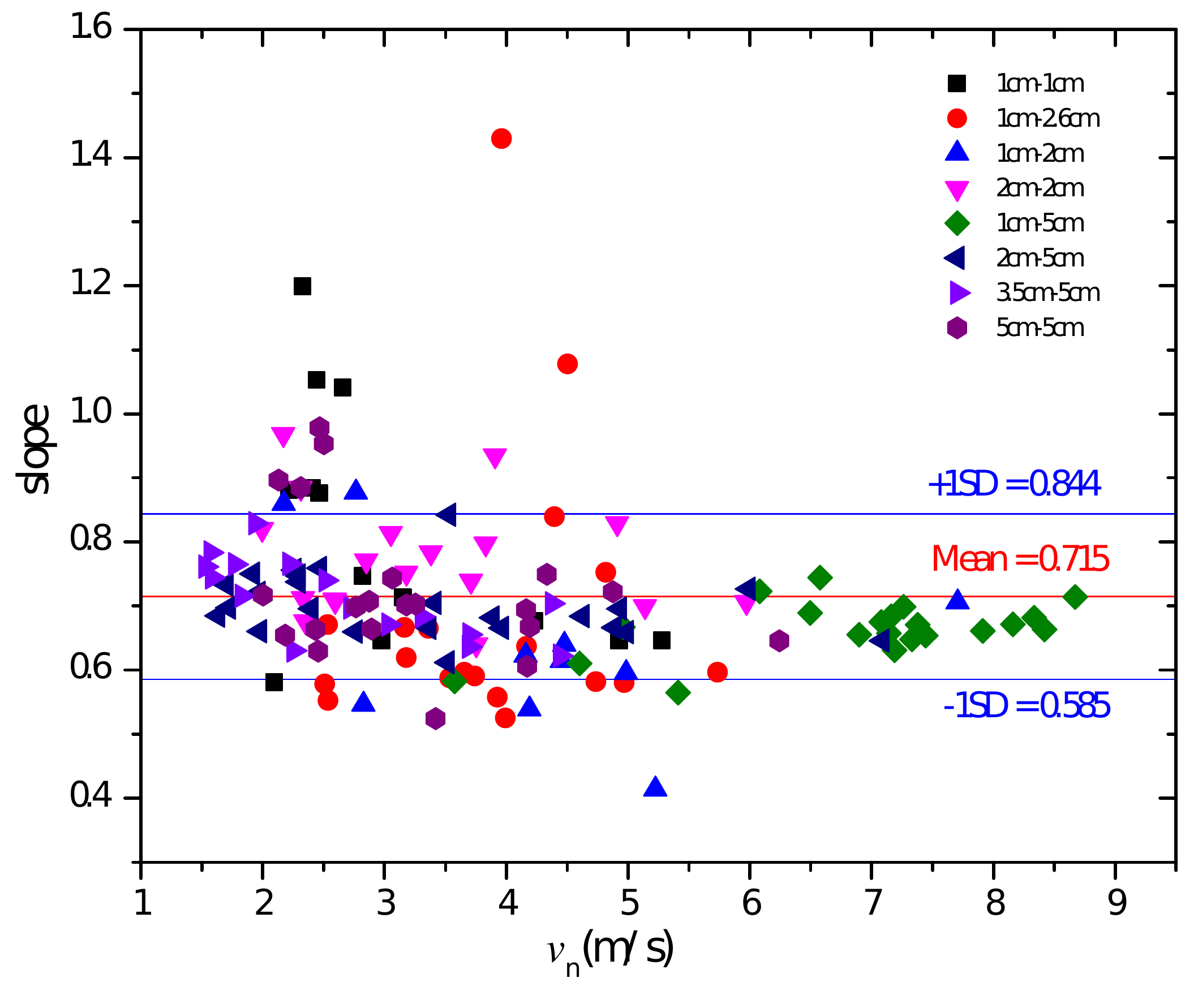}
	\caption{Slope of the power law at the low-mass end of the cumulative mass-frequency distribution function, if started at the smallest fragments for all eight collision series as a function of the normal component of the collision velocity. The horizontal red and blue lines show the mean value and standard deviations, respectively.
	\label{fig:cmslw}}
	\end{center}
\end{figure}

Figure \ref{fig:exmassfit}b shows another representation of the cumulative mass distribution function, now displayed in log-log form with the cumulation started at the low-mass end. It is obvious that for the smallest fragment masses the cumulative mass-frequency distribution function follows a power law but flattens for higher masses. This is generally the case for all collisions. In Figure \ref{fig:cmslw}, we show the slope of the initial power law, as determined in the example shown in Figure \ref{fig:exmassfit}b, as a function of the normal collision velocity. Except for a few outliers, these slopes fall within a narrow range around the average of 0.72 with a standard deviation of 0.13 (see Figure \ref{fig:cmslw}), independent of impact velocity.

\section{\label{sect:fragmod}A Utilitarian Fragmentation Model for Aggregate-Aggregate Collisions}
	Following the results of our extensive study on aggregate-aggregate collisions, which include the determination of the mass of the largest fragment as a function of the normal component of the impact velocity and the projectile/target size as well as a full description of the mass distribution function of the escaping fragments, we suggest the following recipe for a complete description of the fragmentation event:
	\begin{enumerate}
		\item {\bf Determination whether mass transfer or fragmentation of the target occurs.}\\ The outcome in aggregate-aggregate collisions is not unique. In the velocity range above the fragmentation threshold of $\sim 1 \unit{m~s^{-1}}$, mass transfer or complete fragmentation can coexist. In the former case, the target agglomerate stays intact and acquires part of the mass of the projectile. In the latter case, both colliding aggregates lose mass. Based on our experimental results, we propose the following approach:
		\begin{itemize}
			\item Determine whether mass transfer can occur. For given sizes (i.e., diameters) of projectile and target aggregate, $P$ and $T$, and their masses $m_{\unit{p}}$ and $m_{\unit{t}}$, respectively, the necessary condition for the occurrence of mass transfer is that the collision velocity must not exceed the critical fragmentation velocity $v_{\unit{0.5}}$. This can be determined through the strength of the target agglomerate $Q^*$ as a function of $P$ and $T$ using Eq. \ref{eq:q*fit}, $\log \left(\frac{Q^*(P,T)}{\unit{1~J~kg^{-1}}}\right) = C_{\unit{Q}} + \kappa \log \left(\frac{P}{\unit{1~cm}}\right) + \lambda \log \left(\frac{T}{\unit{1~cm}}\right)$, with $C_{\unit{Q}} = 1.24$, $\kappa = 1.12$ and $\lambda = -2.70$. Then, the critical fragmentation velocity is given by $v_{\unit{0.5}} = \sqrt{2 Q^* \left( 1 + \frac{m_{\unit{t}}}{m_{\unit{p}}} \right)}$ (see discussion before Eq. \ref{eq:vcf1}). For $v_{\unit{n}} > v_{\unit{0.5}}$, mass transfer is not possible, because the target aggregate necessarily fragments.
			\item If mass transfer can occur, determine the probability for mass transfer against the probability for fragmentation of the target. According to our analysis in Section \ref{sect:mt}, the probability for mass transfer (and the intact survival of the target aggregate) is given by Eq. \ref{eq:mtprob}, $p_{\unit{sur}} = 0.194 f-0.13$ for $1 \le f \le 5.83$ and $p_{\unit{sur}} = 1$ for $f > 5.83$. Thus, the probability for the fragmentation of the target is $p_{\unit{frag}}= 1 - p_{\unit{sur}}$.
			\item In case of mass transfer, determine the mass gain of the target. According to Eq. \ref{eq:mtvel}, the mass-transfer efficiency $\Delta m / m_{\unit{p}}$ is given by $\log \left( \frac{\Delta m}{m_{\unit{p}}} \right) = C_{\unit{mt}} + \sigma \log{\left(\frac{v_{\unit{n}}}{\unit{1~m~s^{-1}}}\right)}$, with $C_{\unit{mt}} = -1.50$ and $\sigma = 0.99$.
		\end{itemize}
		\item {\bf Determination of the relative mass of the largest fragment $\mu$.}\\ As the sizes of projectile and target, $P$ and $T$, their masses $m_{\unit{p}}$ and $m_{\unit{t}}$, the reduced mass $m = \left( m_{\unit{p}}^{-1} + m_{\unit{t}}^{-1} \right) ^{-1}$ and the normal component of the collision velocity, $v_{\unit{n}}$, are known, we can
		\begin{itemize}
			\item first determine the strength of the target agglomerate $Q^*$ as a function of $P$ and $T$ using Eq. \ref{eq:q*fit}, $\log \left(\frac{Q^*(P,T)}{\unit{1~J~kg^{-1}}}\right) = C_{\unit{Q}} + \kappa \log \left(\frac{P}{\unit{1~cm}}\right) + \lambda \log \left(\frac{T}{\unit{1~cm}}\right)$, with $C_{\unit{Q}} = 1.24$, $\kappa = 1.12$ and $\lambda = -2.70$,
			\item then calculate $E_{0.5}$ using Eq. \ref{eq:q*}, $Q^* = \frac{E_{0.5}}{m_{\unit{t}}}$,
			\item and finally derive the relative mass of the largest fragment using Eq. \ref{eq:hill}, $\mu(E_{\unit{cm}}) = 1 - \frac{E_{\unit{cm}}^{n}}{E^{n}_{0.5}+E_{\unit{cm}}^{n}}$, with $n=0.55$. Here, $E_{\unit{cm}}$ is the centre-of-mass kinetic energy given by $E_{\unit{cm}} = \frac {1}{2} m v_{\unit{n}}^{2}$.
			\item In the case of mass transfer, the mass of the largest fragment is $\mu = 1 + \frac{\Delta m}{m_{\unit{t}}}$ (determination of ${\Delta m}$, see above).
	\end{itemize}
	\item {\bf Determination of the exponent $\beta$ in the fragment mass-frequency distribution function.}\\ The exponent in the continuous part of the fragment mass-frequency distribution function Eq. \ref{eq:massdistpowlaw}, $M_{\unit{cum}}(m_{\unit{f}}) = M_{\unit{tot}} \left( 1 - \left( \frac{m_{\unit{f}}}{m_{\unit{max}}} \right) ^{\beta} \right)$, can be calculated using its relation to the slope of the area-frequency distribution function, $\beta = 1 - \frac{2 \alpha}{3}$. The latter is solely a function of the collision velocity, as expressed in Eq. \ref{eq:alphamin2}, $\alpha = C_{\unit{\alpha}} + \delta \log{\left(\frac{v_{\unit{n}}}{\unit{1~m~s^{-1}}}\right)} +\psi \log{\left(\frac{T}{\unit{1~cm}}\right)}$, with $C_{\unit{\alpha}} = 0.14$, $\delta = 1.02$ and $\psi = 0.34$.
	\item {\bf Determination of the largest relative fragment mass of the continuous distribution $m_{\unit{max}}$.}\\ The continuous fragment mass distribution function, Eq. \ref{eq:alphamin2}, requires an upper mass limit $m_{\unit{max}}$, which we propose to equate to the cutoff-mass of the area-frequency distribution function Eq. \ref{eq:cumdistexp}. Following Eq. \ref{eq:micorr}, we get $m_{\unit{max}} = 2.1 m_{\unit{t}} \cdot \left( x_{\unit{i}} / x_{\unit{t}} \right)^{3/2}$, with (see Eq. \ref{eq:ximin})  $\log{x_{\unit{i}}} = C_{\unit{x}} + \theta \log{\left(\frac{v_{\unit{n}}}{\unit{1~m~s^{-1}}}\right)} + \eta \log{\left(\frac{P}{\unit{1~cm}}\right)} + \tau \log{\left(\frac{T}{\unit{1~cm}}\right)} $, and coefficients $C_{\unit{x}} = 3.01$, $\theta = -0.82$, $\eta = 1.44$ and $\tau = -0.04$ in the case of complete fragmentation of projectile and target, and $C_{\unit{x}} = 2.60$, $\theta = -0.28$, $\eta = 3.00$ and $\tau = -1.15$, respectively, in the case of mass transfer.

\item {\bf Total-mass scaling.}\\  With this information, the functional behavior of the cumulative mass distribution of the fragments, $M_{\unit{cum}}(m_{\unit{f}})$, is fully determined, following Eq. \ref{eq:massdistpowlaw}, except for the scaling parameter $M_{\unit{tot}}$. However, the latter can easily be derived by acknowledging the fact that the full mass distribution function consist of two parts, (i) a continuous fragment mass distribution function and (ii) isolated from that the largest fragment mass. That the latter is really distinct from the former can be seen by the fact that $m_{\unit{max}}/m_{\unit{t}}$ is practically always smaller than $\mu$ (see Figure \ref{fig:mmaxmu} and scaling behavior $m_{\unit{max}} \propto v_{\unit{n}}^{-1}$ (Eq. \ref{eq:ximin}) and $\mu \propto v_{\unit{n}}^{-0.6}$ (Eq. \ref{eq:mypowerlaw})). Thus, we can calculate $M_{\unit{tot}}$ by normalising Eq. \ref{eq:massdistpowlaw} by $M_{\unit{cum}}(m_{\unit{ll}}) = (1 - \mu)m_{\unit{t}} + m_{\unit{p}}$, with $m_{\unit{ll}}$ being the lower limit of the fragment mass distribution. We recognize that the lower fragment size/mass limit in this work was due to the finite spatial resolution of our imaging system. In reality, fragments as small as the monomer size might appear. As long as $\beta > 0$, the lower fragment-mass limit may be set to $m_{\unit{ll}} = 0$, without causing mass-conservation problems. However, if $\beta < 0$, a reasonable lower mass limit for the fragments needs to be found.

\end{enumerate}

\section{\label{sect:simul}Simulating the Collapse}

The results of our experiments can be used to study the formation of planetesimals. In a protoplanetary disk, gravitationally bound pebble clouds can form through the interaction between pebbles and the gas in the disk by e.g. by the streaming instability (see Sect. \ref{sect:intro}). Such a cloud will collapse into a solid planetesimal thanks to the negative heat capacity property of gravitationally bound systems and energy dissipation in pebble-pebble collisions. \citet{Wahlberg:2014} studied the collapse process of such a cloud to find the internal structure of the resulting planetesimal. They find that the density of a planetesimal formed increases with planetesimal mass. More massive clouds have more fragmenting collisions and a wide range of particle sizes in the resulting planetesimal leading to better packing capabilities. In their numerical simulations, however, they use a simplified model of fragmenting collisions, treating fragmentation as erosion. In Paper II, the collapse process is investigated with an updated model. The new model includes the results of our experiments (critical speeds, fragment size distribution and mass transfer probability), to get more physically realistic simulations. For low-mass planetesimals ($R_\textnormal{solid}\lesssim 40$ km) the results are similar (they end up as porous pebble-piles). For more massive planetesimals, however, the internal structure show a strong dependence on both fragmentation model and pebble composition (silicates vs. ice).

\section{\label{sect:conclusions}Conclusion and Discussion}

We developed a new experimental setup dedicated to the study of the low-velocity fragmentation behavior of porous dust aggregates. Aggregates consisted of micrometer-sized $\rm SiO_2$ grains and possessed volume filling factors of $\phi = 0.35$, i.e. porosities of 65\%. The sizes of the dust aggregates ranged between 1 cm and 5 cm, with collision velocities in the range from $1.5 \unit{m~s^{-1}}$ and $8.7 \unit{m~s^{-1}}$ (see Figure \ref{fig:VD}).

In all cases studied, the smaller (or equal) sized projectile aggregate fragmented. The larger (or equal sized) target aggregate survived impact when the target-to-projectile size ratio was large and the impact velocity rather small (see Figure \ref{fig:mtprob}b). However, we found that the outcome in these cases is probabilistic between target survival and target fragmentation, with a probability for target survival given by Eq. \ref{eq:mtprob} (see Figure \ref{fig:mtprob}a).

We described the fragmentation of the colliding dust aggregates by the mass of the largest fragment and a continuous area-frequency distribution function of the smaller fragments. When we express the mass of the largest fragments in units of the target-aggregate mass, we can describe its dependence on impact energy with a Hill function (see Eq. \ref{eq:hill}) with two free parameters, the energy $E_{0.5}$ for which the largest fragment is $\mu = 0.5$ and an exponent $n$ for which we find that $n=0.55$. Following our recipe summarized in Sect. \ref{sect:fragmod}, a full description of the fragmentation process in collisions between arbitrary dust aggregates is possible.

Besides the application of our high-velocity dust-aggregation collision model in the description of the fate of dust aggregates in collapsing pebble clouds (see Sect. \ref{sect:simul} and Paper II), it will also be useful for mass-transfer based formation models of planetesimals \citep{WindmarkEtal:2012a,WindmarkEtal:2012b,Garaudetal:2013}. The successive growth of dust aggregates beyond the bouncing barrier by mass transfer in catastrophic collisions between dissimilar-sized dust aggregates is an essential part of these models. With the data and formal descriptions of the collision outcomes presented in this paper, the validity of models for the formation of planetesimals by direct sticking via mass transfer can be assessed with more realistic collision outcomes.

\section*{Acknowledgments}
We are thankful to Kathrin Gebauer, Bernd Stoll, Ingo von Borstel  for their technical assistance in constructing the drop tower. We are also grateful to Christropher Perschke, Stefan Kothe and Rainer Schr\"apler for fruitful discussions and support during the data analysis. We thank Alexander Diener from IPAT, TU Braunschweig, for performing the XRT measurements on our dust-aggregate samples. This work has been funded through the DFG project SFB 963 ``Astrophysical Flow Instabilities and Turbulence'' in sub-project A07. KWJ and AJ were supported by the European Research Council under ERC Starting Grant agreement 278675-PEBBLE2PLANET. AJ was also supported by the Swedish Research Council (grant 2014-5775) and by the Knut and Alice Wallenberg Foundation.

\bibliographystyle{abbrvnat}
\bibliography{literatur}

\begin{appendix}

\section{Electronic Appendix 1}

\begin{figure}[htp]
\includegraphics[height= 200mm]{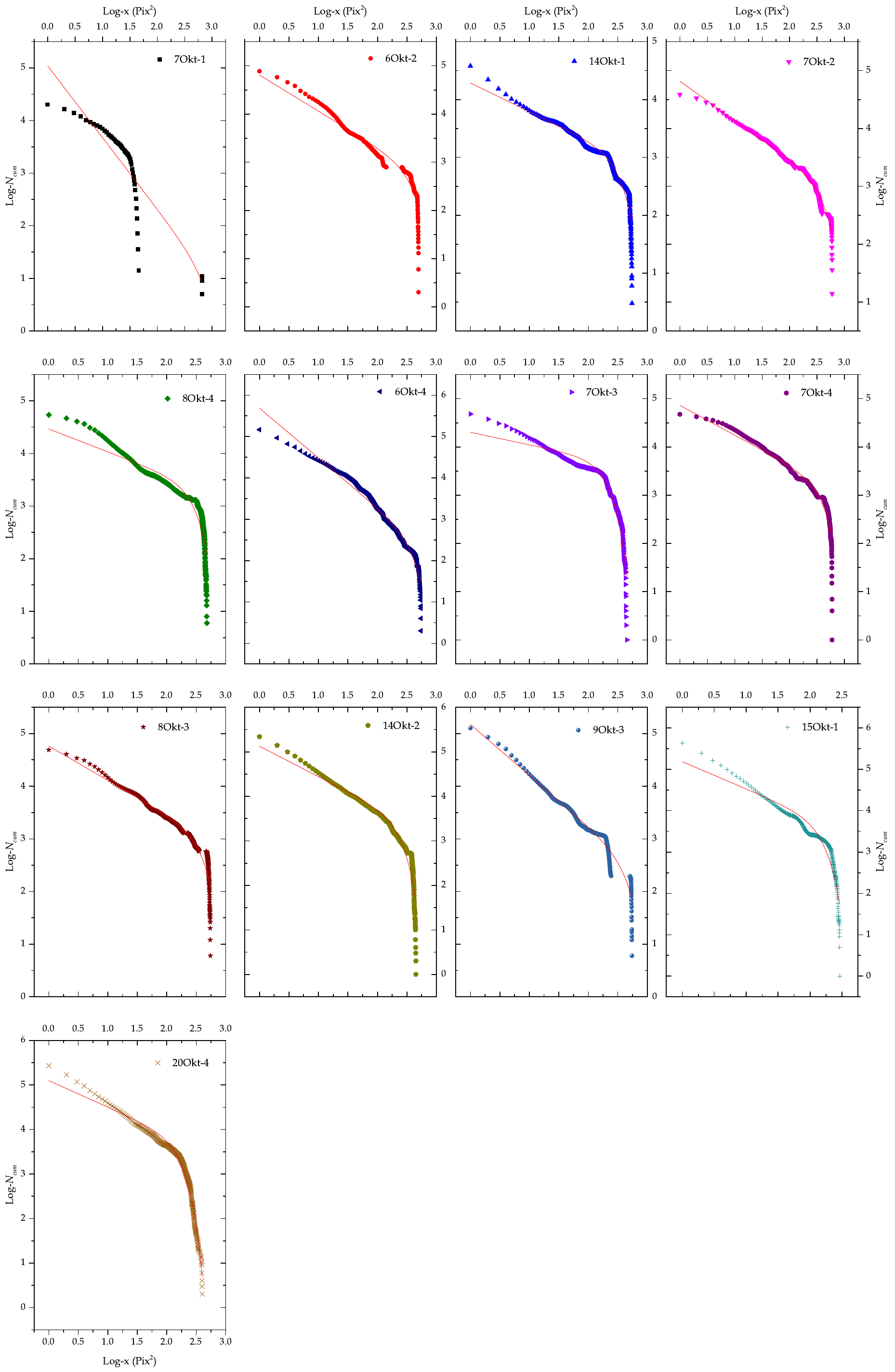}
	\caption{Area-frequency distribution of all individual experiments in the 1 cm - 1 cm series.
\label{app1-1-1}}
\end{figure}

\begin{figure}[htp]
\includegraphics[height= 200mm]{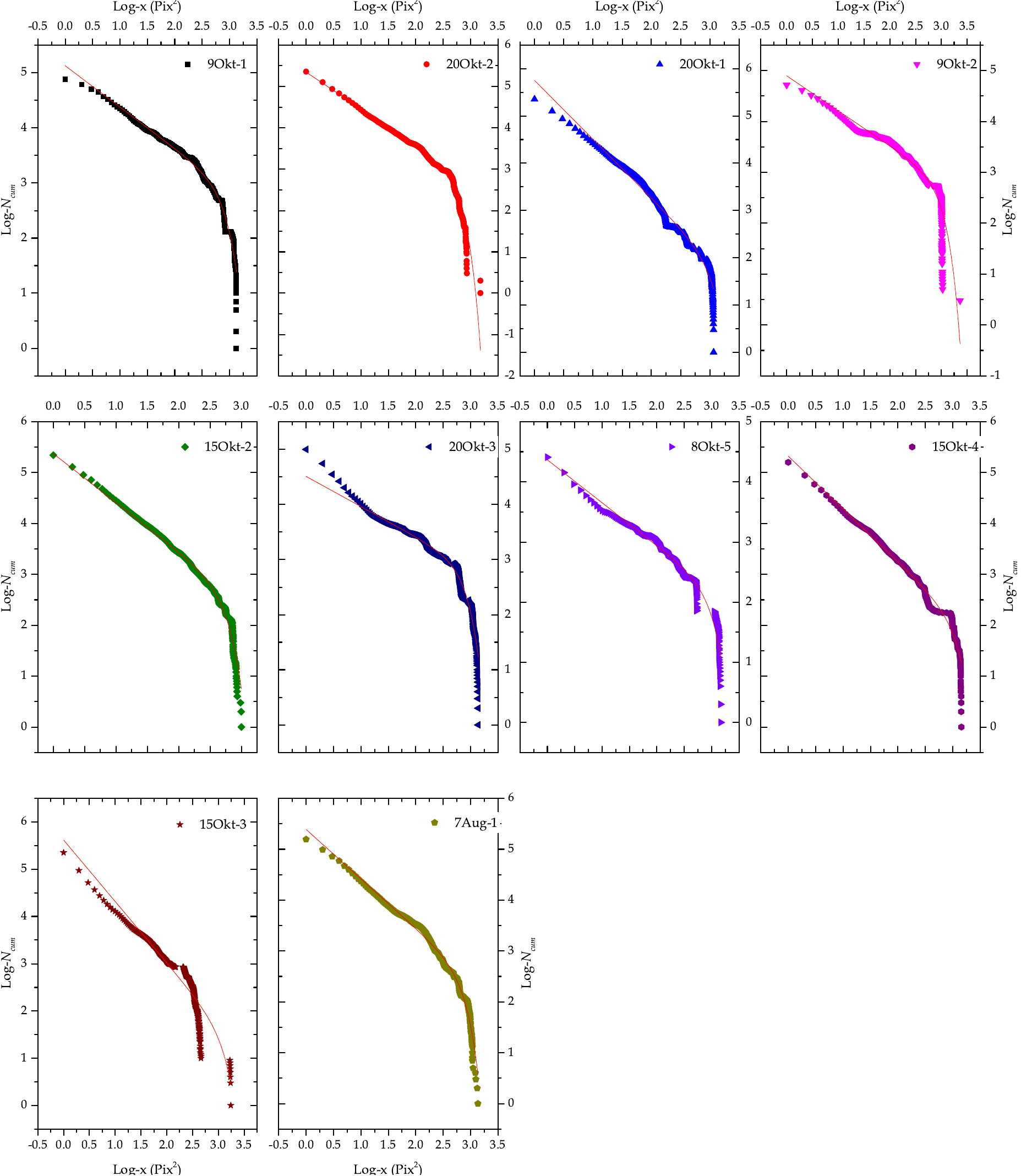}
	\caption{Area-frequency distribution of all individual experiments in the 1 cm - 2 cm series.
\label{app1-1-2}}
\end{figure}

\begin{figure}[htp]
\includegraphics[height= 200mm]{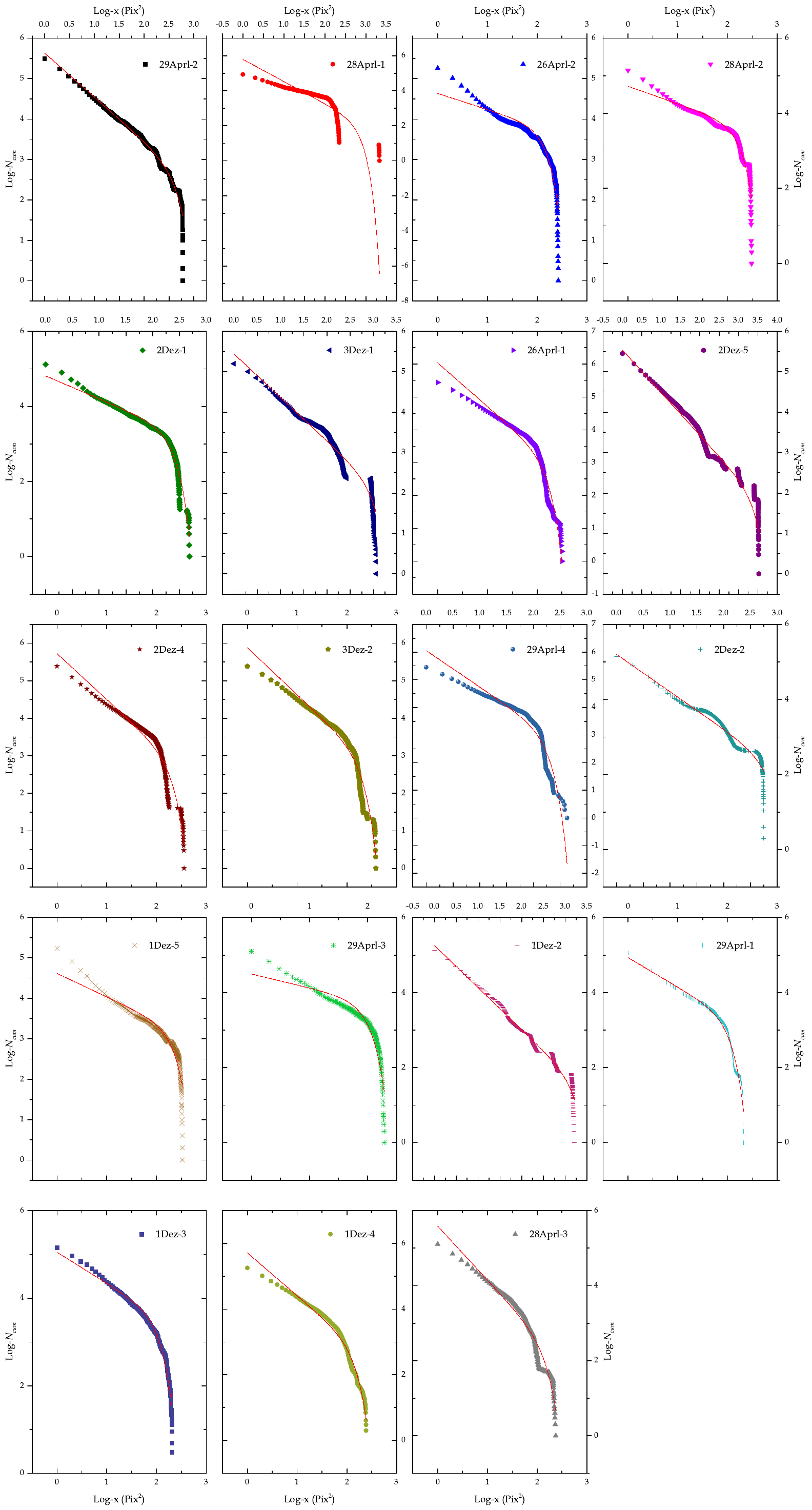}
	\caption{Area-frequency distribution of all individual experiments in the 1 cm - 2.6 cm series.
\label{app1-1-2.6}}
\end{figure}

\begin{figure}[htp]
\includegraphics[height= 200mm]{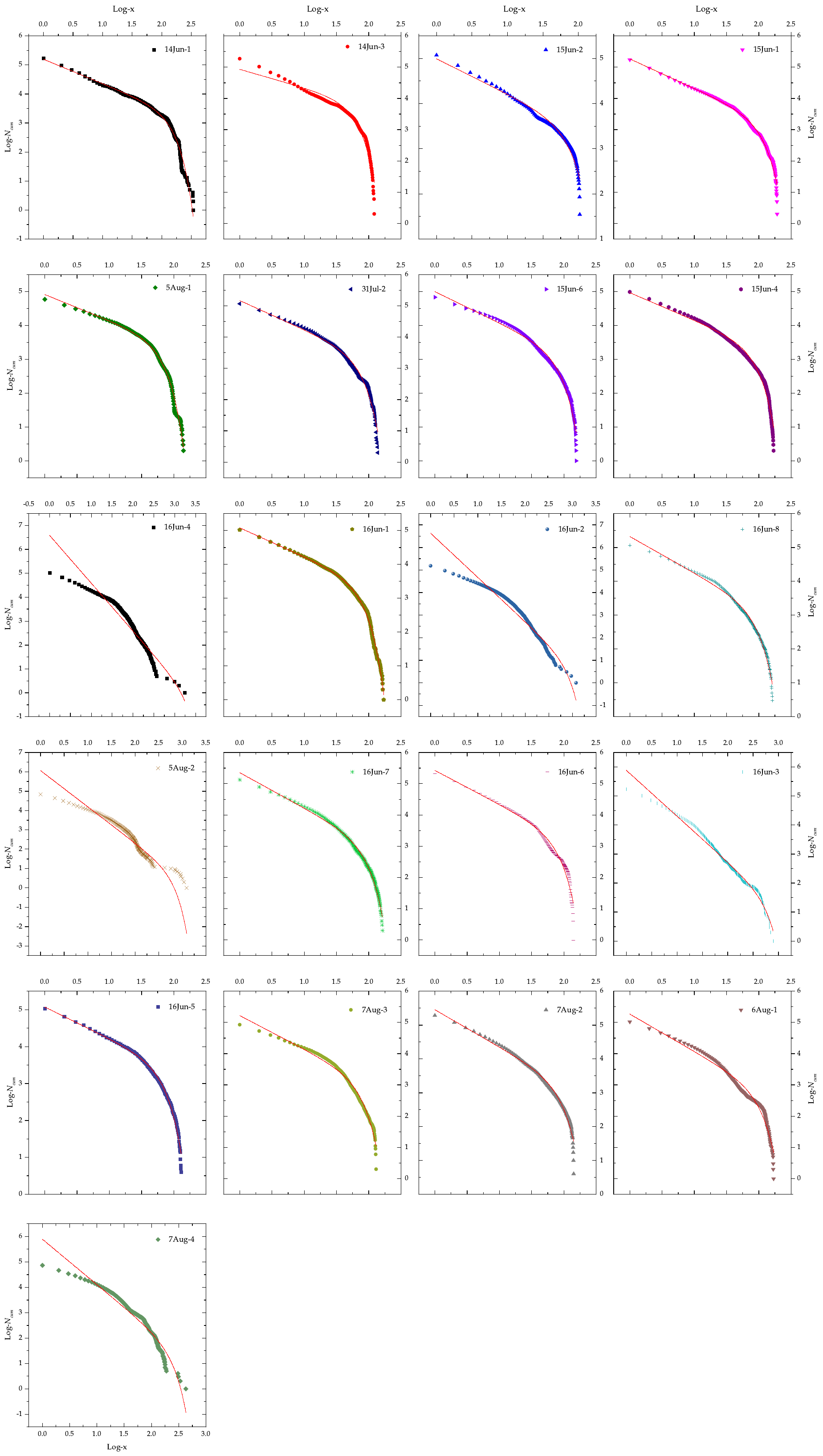}
	\caption{Area-frequency distribution of all individual experiments in the 1 cm - 5 cm series.
\label{app1-1-5}}
\end{figure}

\begin{figure}[htp]
\includegraphics[height= 200mm]{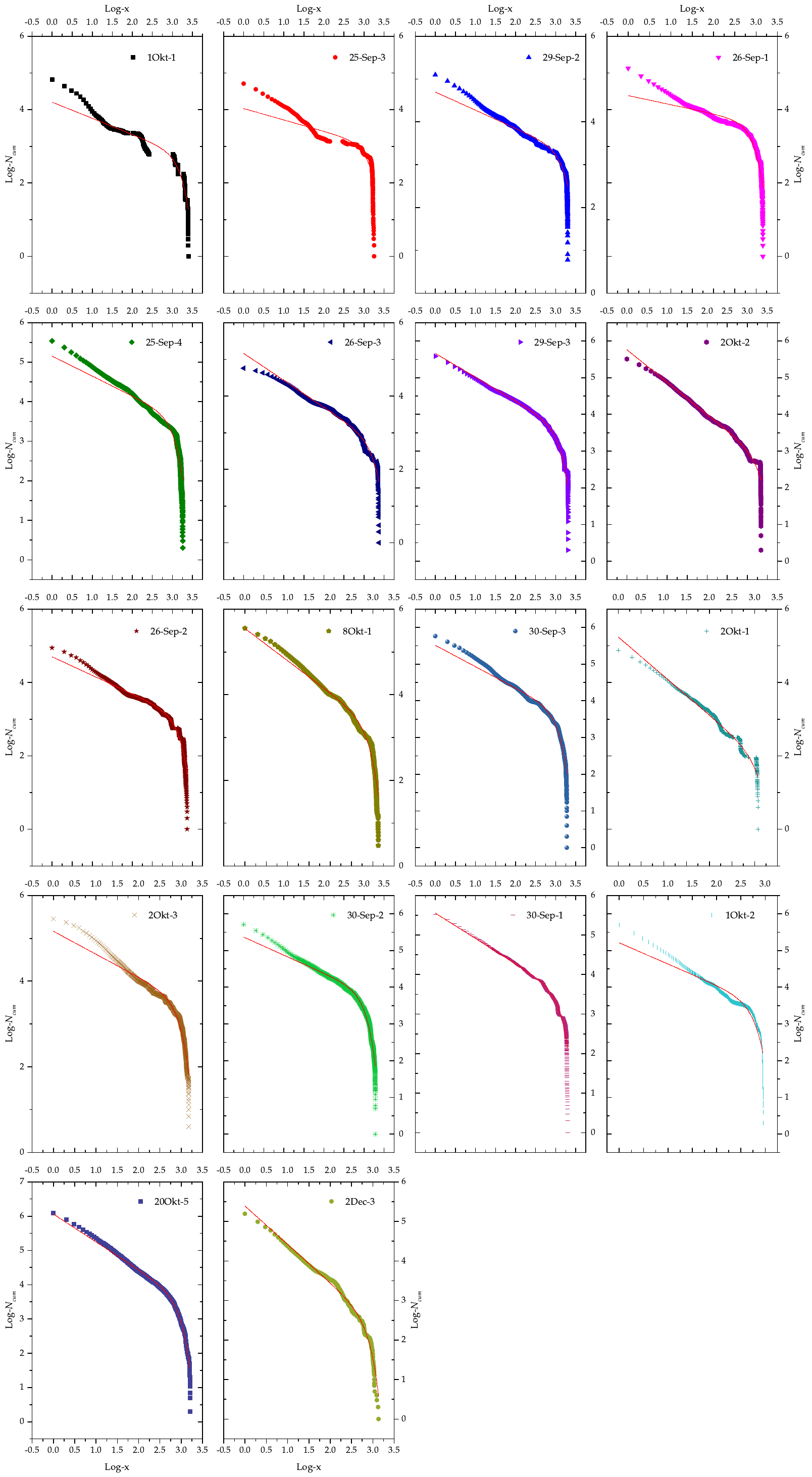}
	\caption{Area-frequency distribution of all individual experiments in the 2 cm - 2 cm series.
\label{app1-2-2}}
\end{figure}

\begin{figure}[htp]
\includegraphics[height= 200mm]{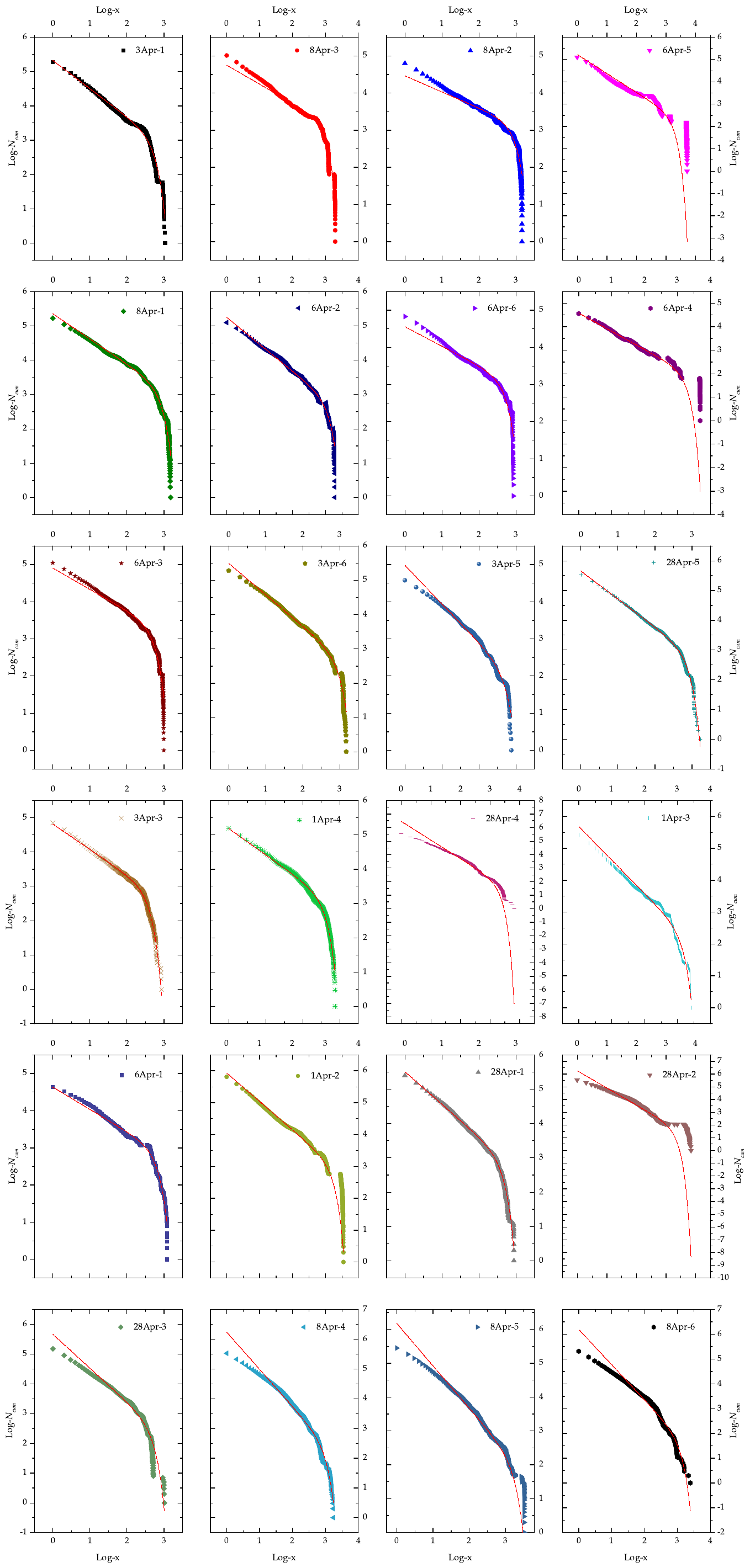}
	\caption{Area-frequency distribution of all individual experiments in the 2 cm - 5 cm series.
\label{app1-2-5}}
\end{figure}

\begin{figure}[htp]
\includegraphics[height= 200mm]{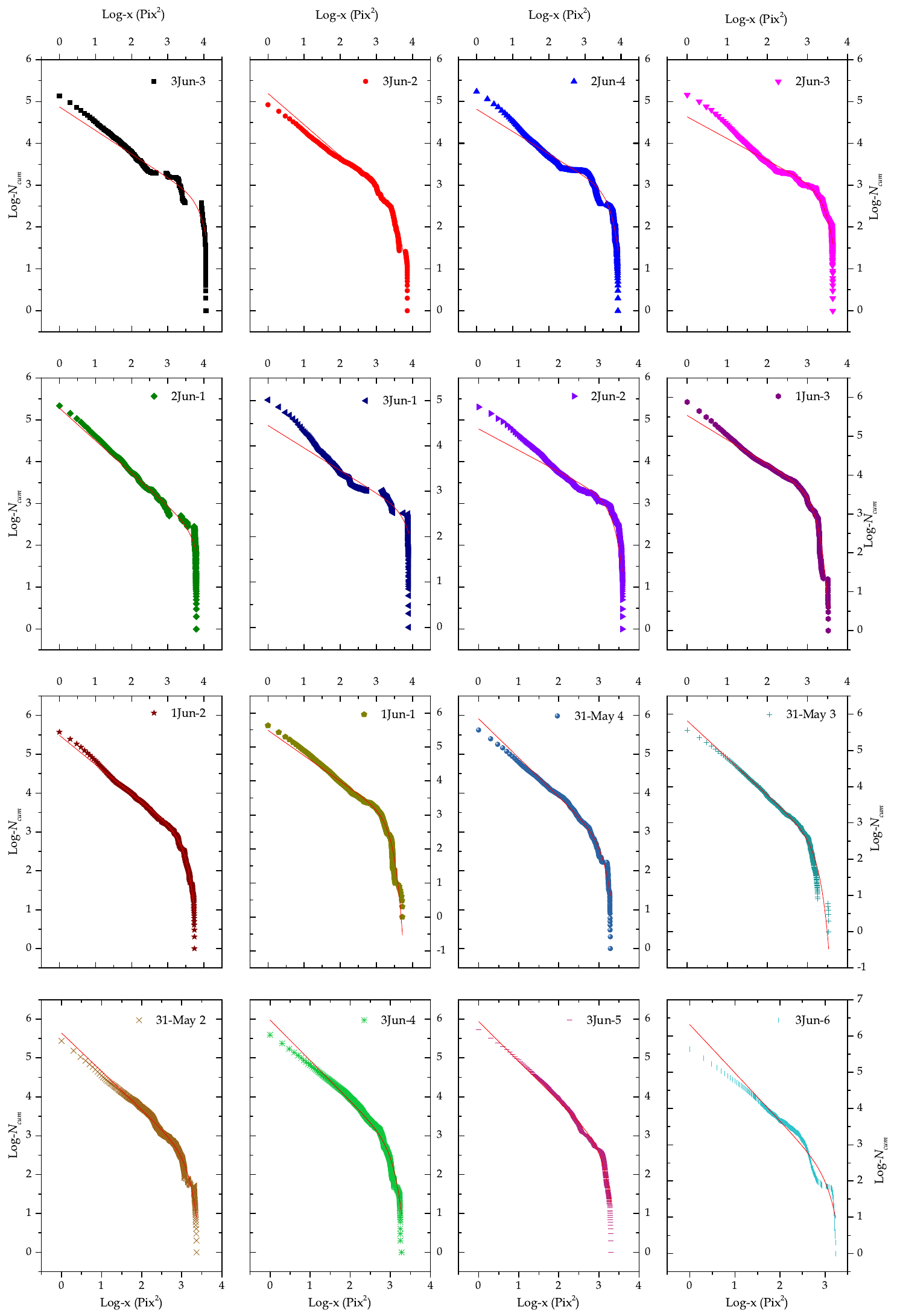}
	\caption{Area-frequency distribution of all individual experiments in the 3.5 cm - 5 cm series.
\label{app1-3.5-5}}
\end{figure}

\begin{figure}[htp]
\includegraphics[height= 200mm]{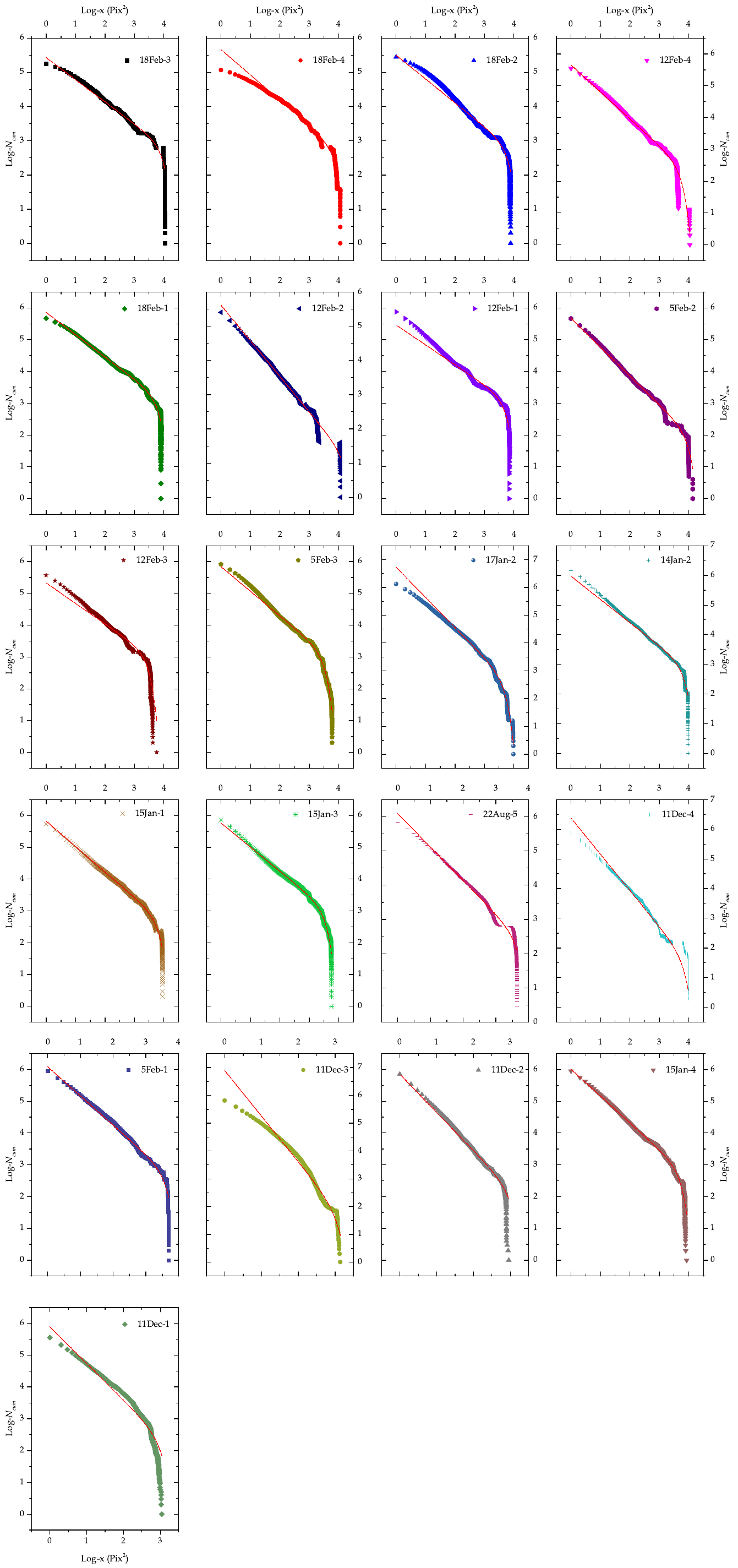}
	\caption{Area-frequency distribution of all individual experiments in the 5 cm - 5 cm series.
\label{app1-5-5}}
\end{figure}

Figures \ref{app1-1-1} - \ref{app1-5-5}

\section{Electronic Appendix 2}

\begin{figure}[htp]
\includegraphics[height= 200mm]{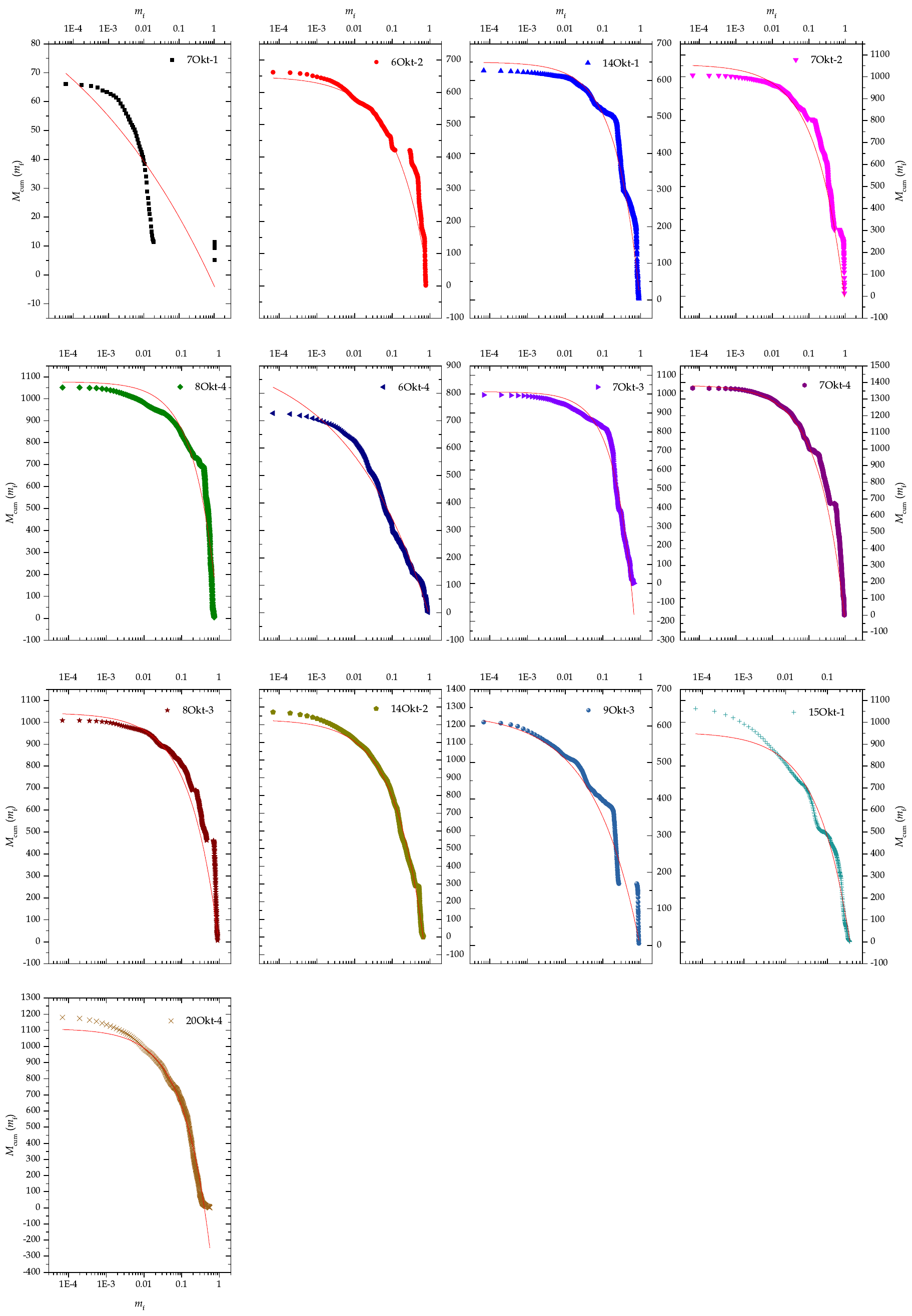}
	\caption{Mass-frequency distribution of all individual experiments in the 1 cm - 1 cm series.
\label{app2-1-1}}
\end{figure}

\begin{figure}[htp]
\includegraphics[height= 200mm]{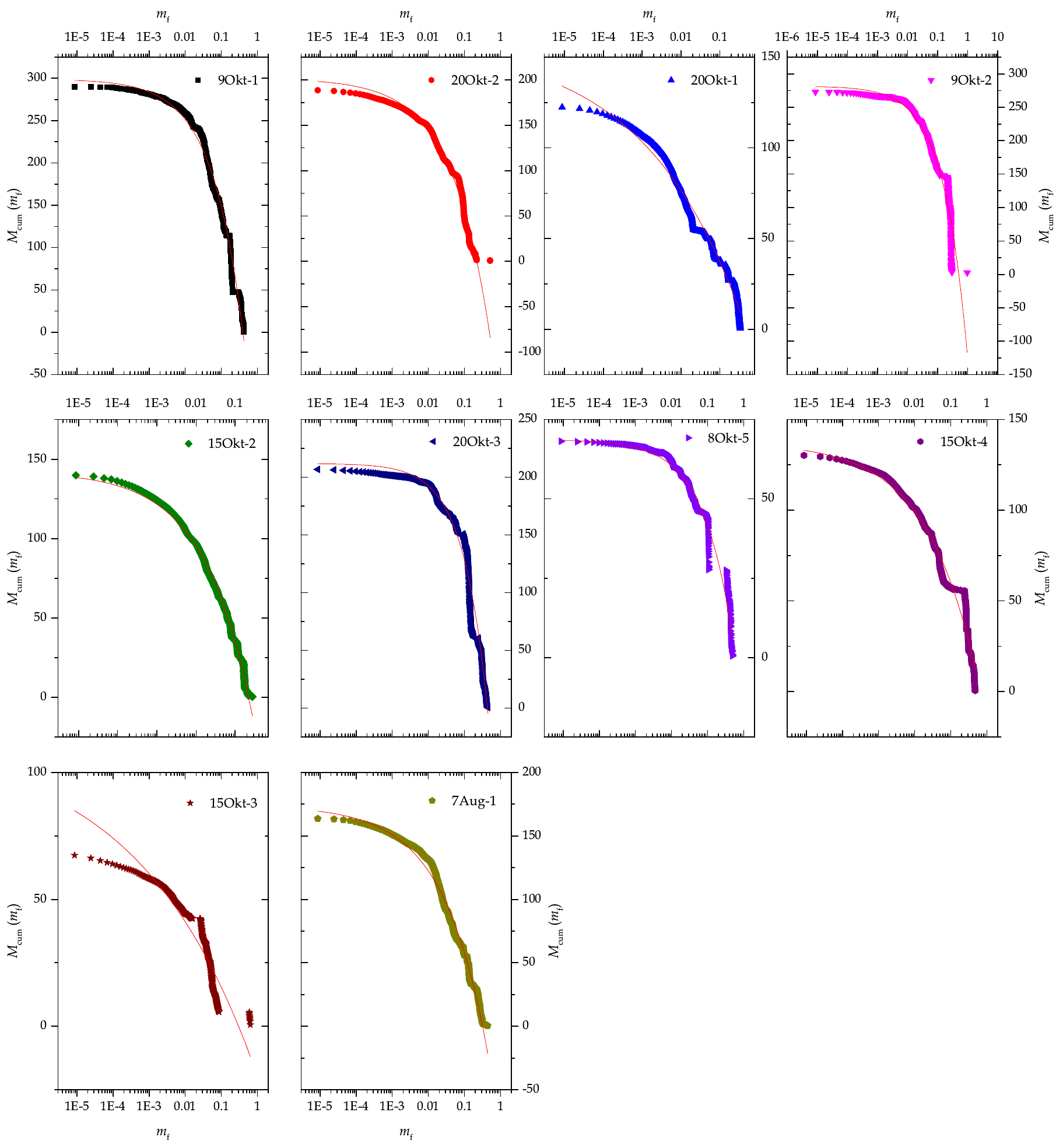}
	\caption{Mass-frequency distribution of all individual experiments in the 1 cm - 2 cm series.
\label{app2-1-2}}
\end{figure}

\begin{figure}[htp]
\includegraphics[height= 200mm]{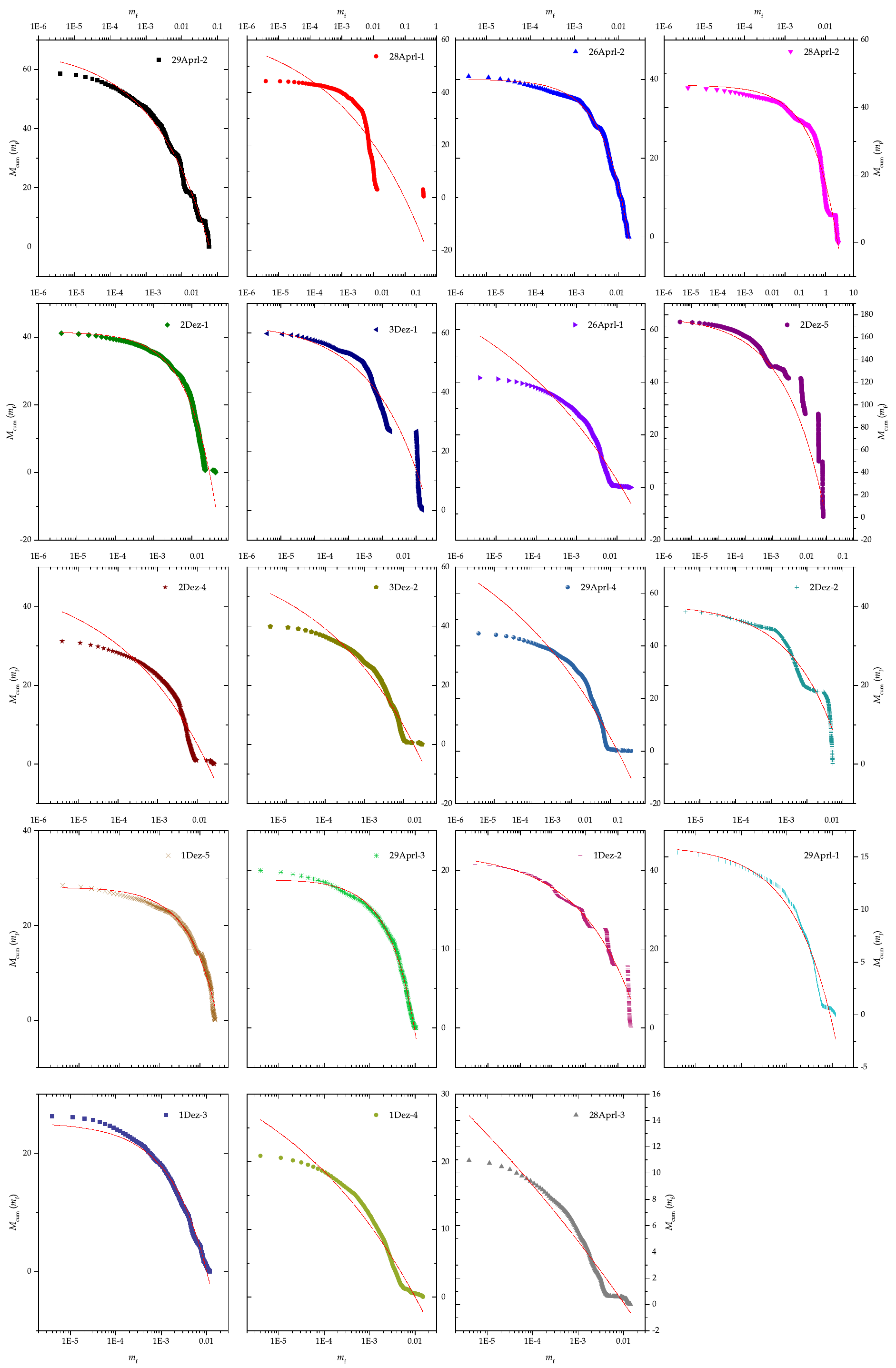}
	\caption{Mass-frequency distribution of all individual experiments in the 1 cm - 2.6 cm series.
\label{app2-1-2.6}}
\end{figure}

\begin{figure}[htp]
\includegraphics[height= 200mm]{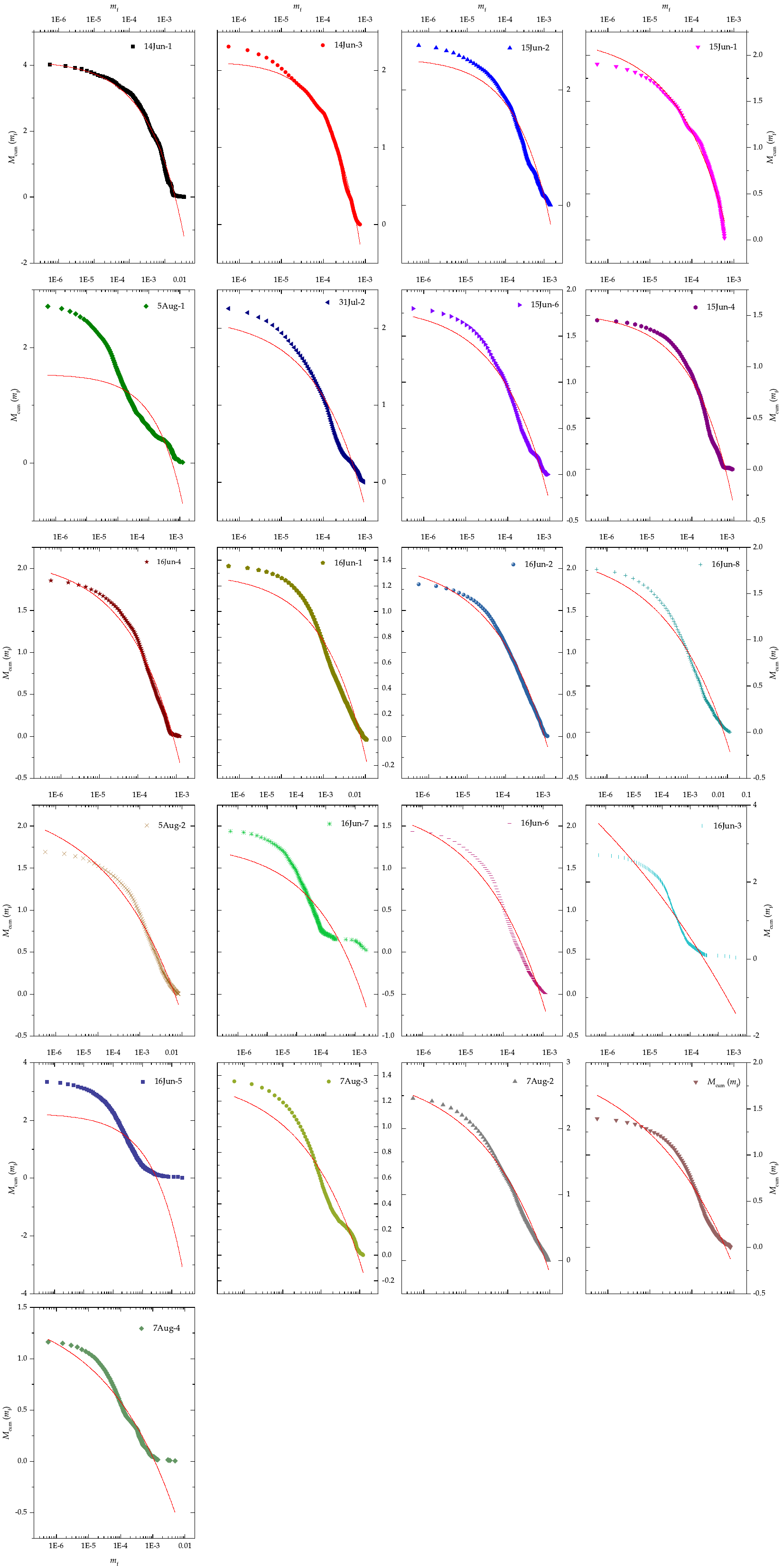}
	\caption{Mass-frequency distribution of all individual experiments in the 1 cm - 5 cm series.
\label{app2-1-5}}
\end{figure}

\begin{figure}[htp]
\includegraphics[height= 200mm]{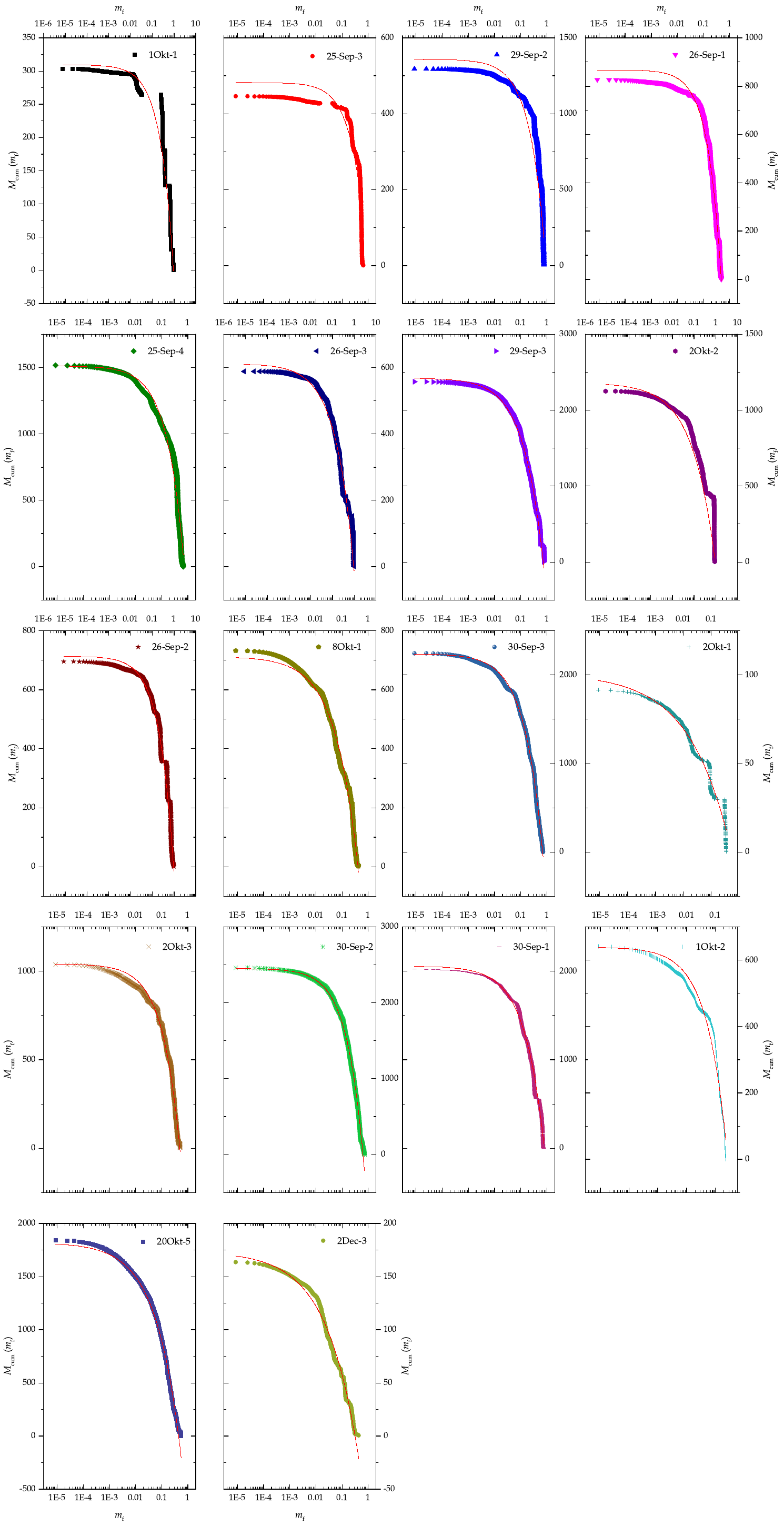}
	\caption{Mass-frequency distribution of all individual experiments in the 2 cm - 2 cm series.
\label{app2-2-2}}
\end{figure}

\begin{figure}[htp]
\includegraphics[height= 200mm]{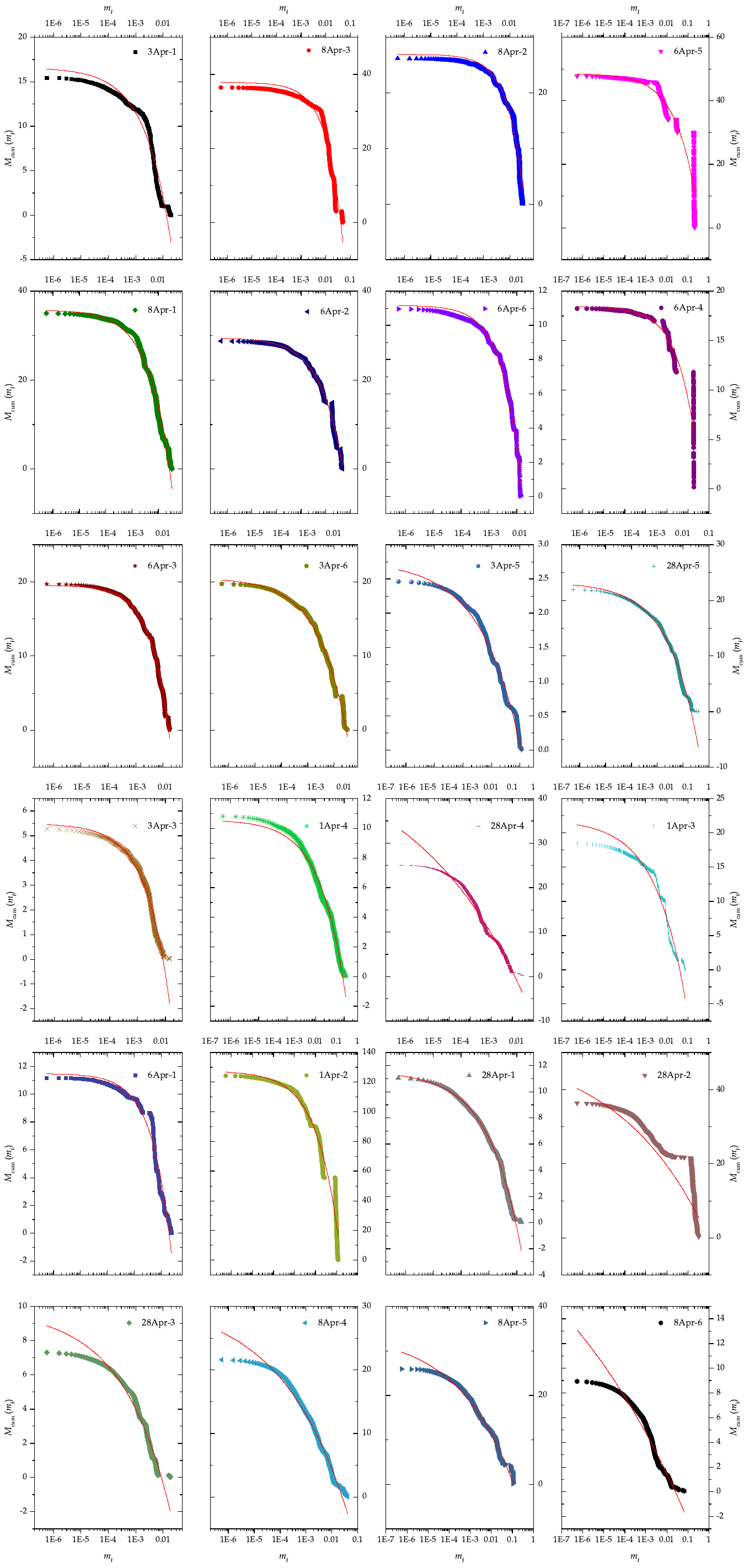}
	\caption{Mass-frequency distribution of all individual experiments in the 2 cm - 5 cm series.
\label{app2-2-5}}
\end{figure}

\begin{figure}[htp]
\includegraphics[height= 200mm]{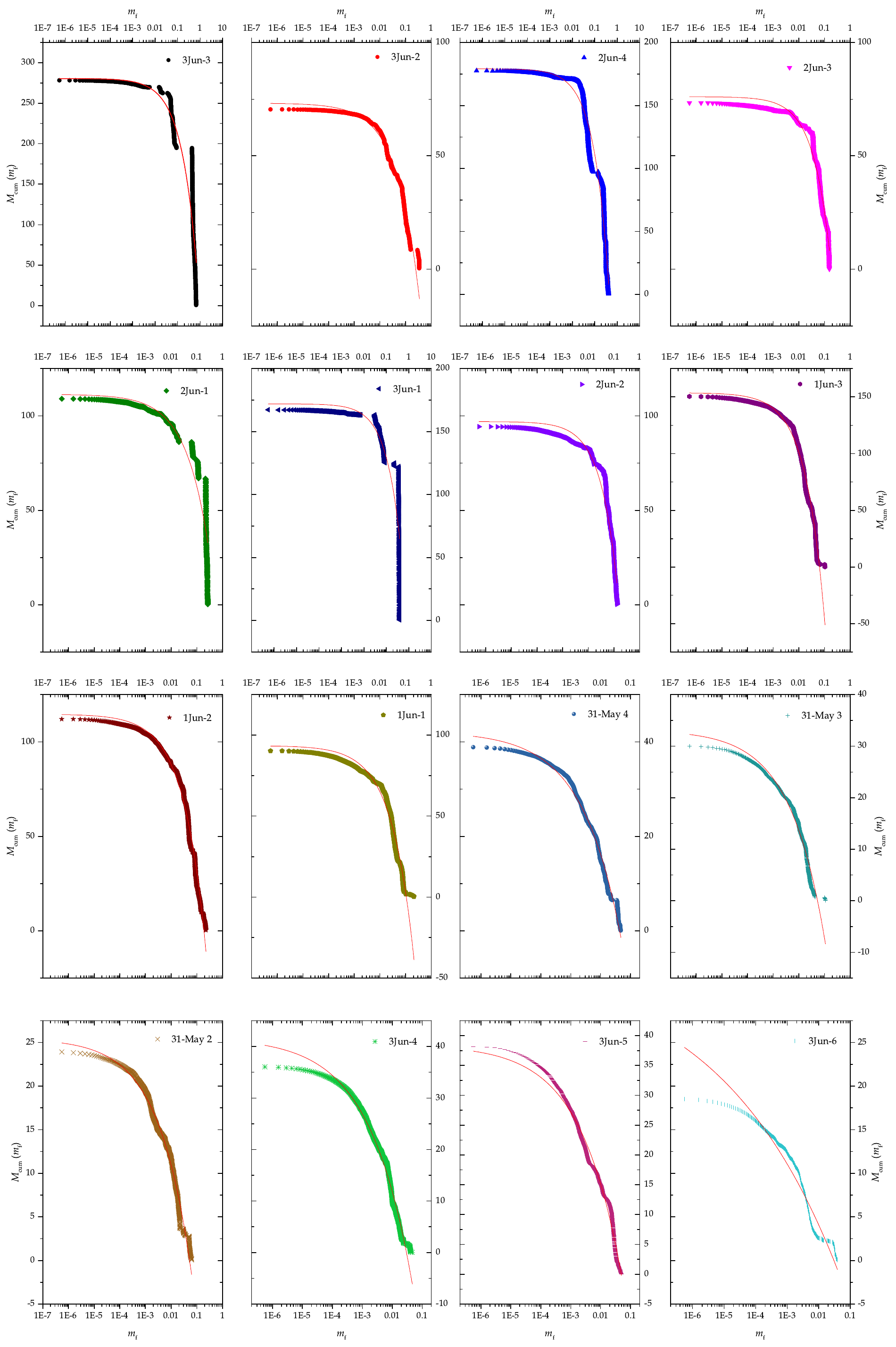}
	\caption{Mass-frequency distribution of all individual experiments in the 3.5 cm - 5 cm series.
\label{app2-3.5-5}}
\end{figure}

\begin{figure}[htp]
\includegraphics[height= 200mm]{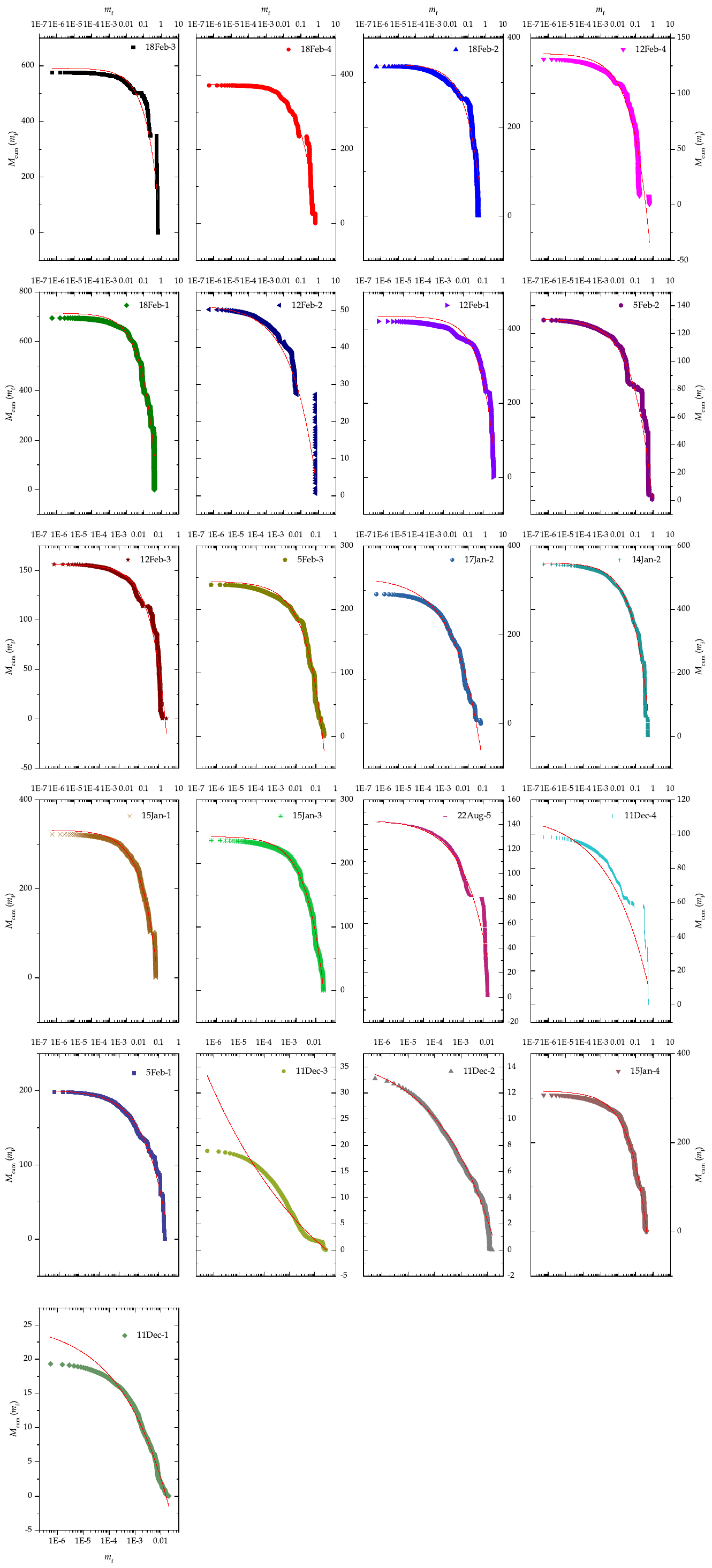}
	\caption{Mass-frequency distribution of all individual experiments in the 5 cm - 5 cm series.
\label{app2-5-5}}
\end{figure}

Figures \ref{app2-1-1} - \ref{app2-5-5}

\end{appendix}

\end{document}